\documentclass[a4paper]{article}
\pdfoutput=1
\usepackage{pdfpages}

\usepackage{pos}
\usepackage{amsmath}
\usepackage{placeins}
\usepackage{subcaption}
\usepackage{epstopdf}
\usepackage{epsfig}
\usepackage{pdflscape}
\usepackage[utf8]{inputenc}
\usepackage{graphicx}
\usepackage{float}
\usepackage{xspace}

\begin{document}

\title{A Proposed Forward Silicon Tracker for the Future Electron-Ion Collider and Associated Physics Studies}
\ShortTitle{Proposed EIC FST and Associated Physics}

\author[1]{Cheuk-Ping~Wong}
\author[2]{Xuan~Li}
\author{Melynda~Brooks}
\author{Matthew J. ~Durham}
\author{Ming Xiong~Liu}
\author{Astrid~Morreale}
\author{Cesar~da~Silva}
\author{Walter E.~Sondheim}

\affiliation{Los Alamos National Laboratory,\\
  Los Alamos, NM, USA}
  
\hfill
\hskip+4.0cm\begin{tabular}[t]{l@{}}
  \textbf{LANL Report: } \\
  \textit{LA-UR-20-26642}
\end{tabular}

\emailAdd{cpwong@lanl.gov}
\emailAdd{xuanli@lanl.gov}

\note{Primary Author and contact person}
\note{Primary Author and contact person}

\abstract{The future Electron-Ion Collider (EIC) will explore several fundamental questions in a broad Bjorken-x ($x_{BJ}$) and $Q^{2}$ phase space. Heavy flavor and jet products are ideal probes to precisely study the tomography of nucleon/nuclei structure, help solve the proton spin puzzle and understand the hadronizaton processes in vacuum or in the QCD medium. Due to the asymmetric collisions at the EIC, most of the final state hadrons are produced in the nucleon/nuclei beam going (forward) direction. A silicon vertex/tracking is critical to precisely measure these forward hadrons at the EIC. Details of different conceptual designs of the proposed Forward Silicon Tracker (FST) and the relevant detector performance are presented in this technical note. The associated heavy flavor and jet studies with the evaluated FST performance are discussed as well. }

%--------------------------------------------------------------------%

\maketitle

%--------------------------------------------------------------------%
%\linenumber
\section{Introduction}
\label{sec: intro}
\FloatBarrier
The future Electron-Ion Collider (EIC) \cite{eic_WP} at Brookhaven National Laboratory will open a new QCD frontier to explore fundamental questions in nuclear physics. Heavy flavor and jet measurements will help us 

1) precisely determine the initial nucleon/nuclei parton distribution functions, especially in the high Bjorken-x ($x_{BJ}$) region.

2) explore the hadronization processes in vacuum and medium.

3) improve the understanding of the proton spin structure including study of the gluon Sivers function.

4) characterize new exotic states, which may be pentaquarks, tetraquark or glueballs.

To realize the heavy flavor and jet measurements, a silicon vertex/tracking detector is essential for the future EIC. The LANL LDRD 20200022DR project focuses on the proposed forward silicon tracker design, R$\&$D, and simulation studies together with studies of how heavy flavor and jet probes can be used for the future EIC physics program. In this techinical note, we discuss details of the kinematic distributions of final particle production, the Forward Silicon Tracker (FST) design and the tracking performance, and heavy flavor and jet physics performance projections.
\section{Kinematics distributions}
\label{sec: kin}
\begin{figure}[ht]
    \centering
    \includegraphics[width=\textwidth]{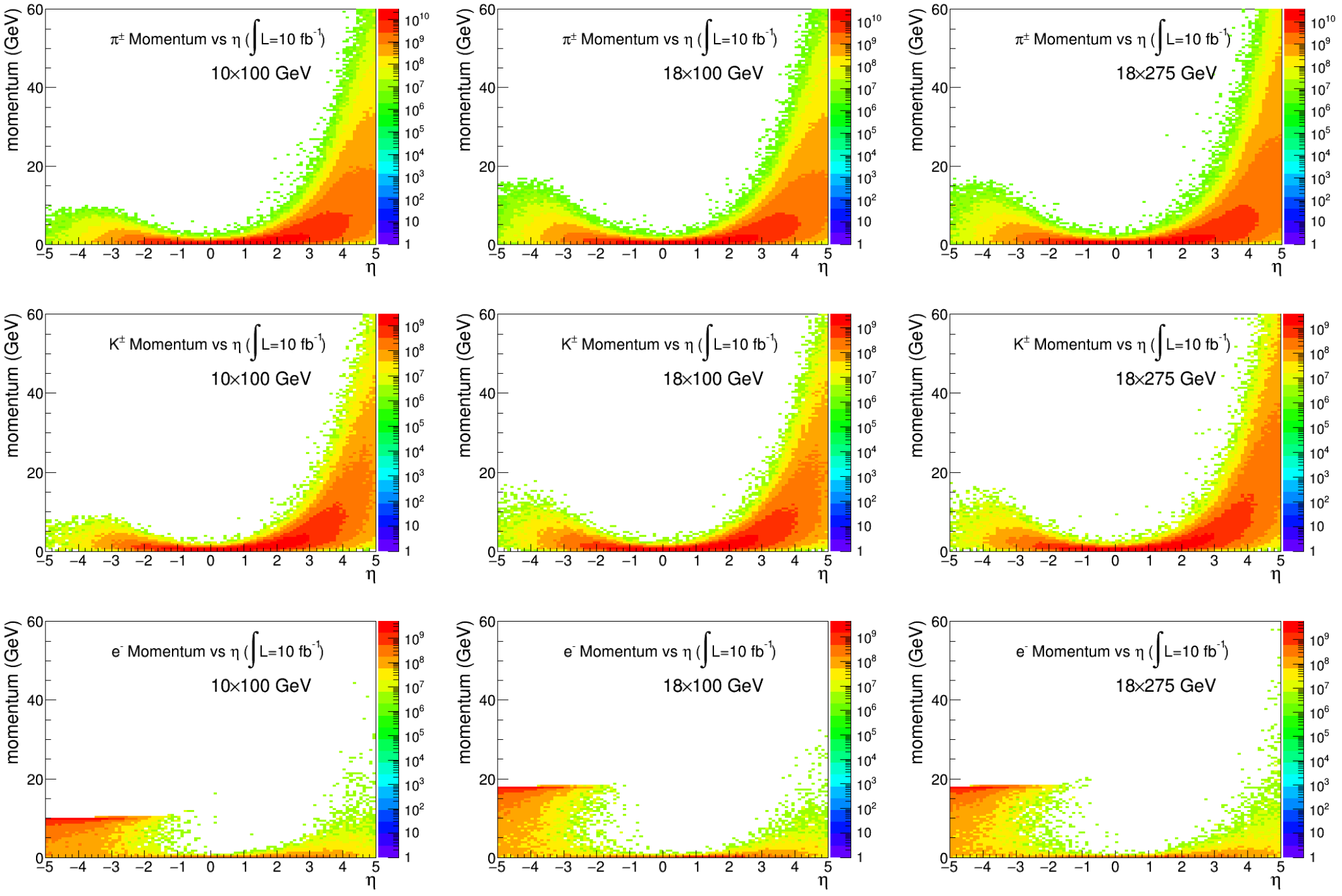}
    \caption{Inclusive decay $\pi^{\pm}$ (top row), $K^{\pm}$ (second row) and $e^{-}$ (bottom row) momenta as a function of pseudorapidity from electron-proton collisions at $10\times100$~GeV (left column), $18\times100$~GeV (center column) and $18\times275$~GeV (right column) using Pythia6 simulations.}
    \label{fig:kinematics_pythia6_eta}
\end{figure}

\begin{figure}[ht]
    \centering
    \includegraphics[width=\textwidth]{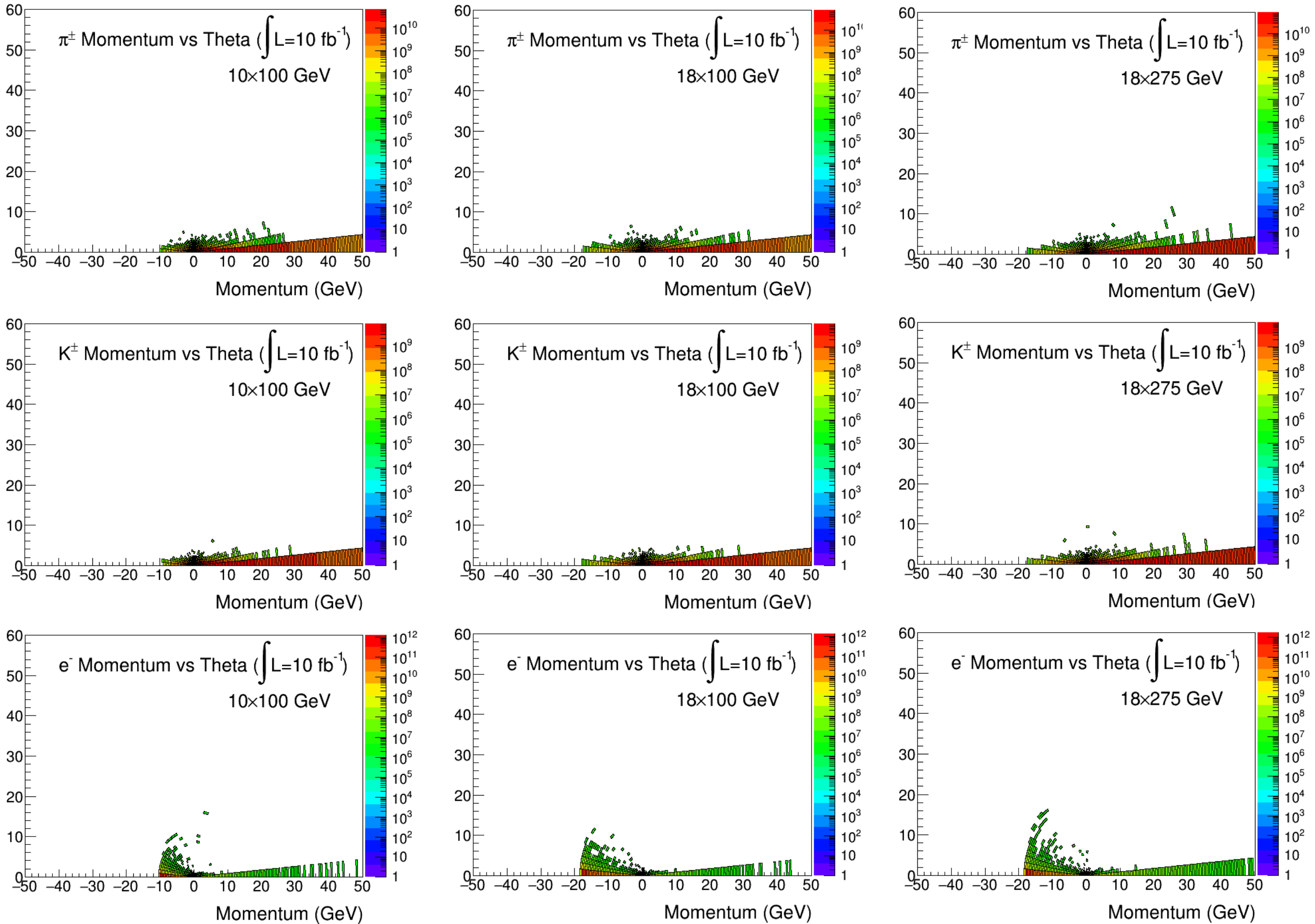}
    \caption{Inclusive decay $\pi^{\pm}$ (top row), $K^{\pm}$ (second row) and $e^{-}$ (bottom row) momentum distributions in polar coordinates from electron-proton collisions at $10\times100$~GeV (left column), $18\times100$~GeV (center column) and $18\times275$~GeV (right column) using Pythia6 simulations.}
    \label{fig:kinematics_pythia6_polar}
\end{figure}

%----------------------------------------------------------%
% Inclusive mesons
%----------------------------------------------------------%
The kinematic distributions of inclusive $K^\pm$, $\pi^\pm$ and $e^-$ obtained from Pythia6 simulation with radiative corrections~\cite{bib:pythia6,bib:pythiaeRHIC} as shown in Figure~\ref{fig:kinematics_pythia6_eta} and Figure~\ref{fig:kinematics_pythia6_polar}. Momenta of these inclusive products are plotted as a function of pseudorapidity in Figure~\ref{fig:kinematics_pythia6_eta}, while momenta plotted versus polar angle, $\theta$, is shown in Figure~\ref{fig:kinematics_pythia6_polar}. Figure~\ref{fig:kinematics_pythia6_eta} and Figure~\ref{fig:kinematics_pythia6_polar} show that the low momentum ($p\leq10$~GeV) inclusive charged meson products dominate in $0<\eta<4$ in the forward direction. The inclusive electrons, however, dominate in the backward direction ($\eta<0$), that is, the electron beam direction, as expected.
%
%\begin{landscape}
%\end{landscape}
%
%----------------------------------------------------------%
% B meson decay products
%----------------------------------------------------------%
\begin{figure}[ht]
    \centering
    \includegraphics[width=\textwidth]{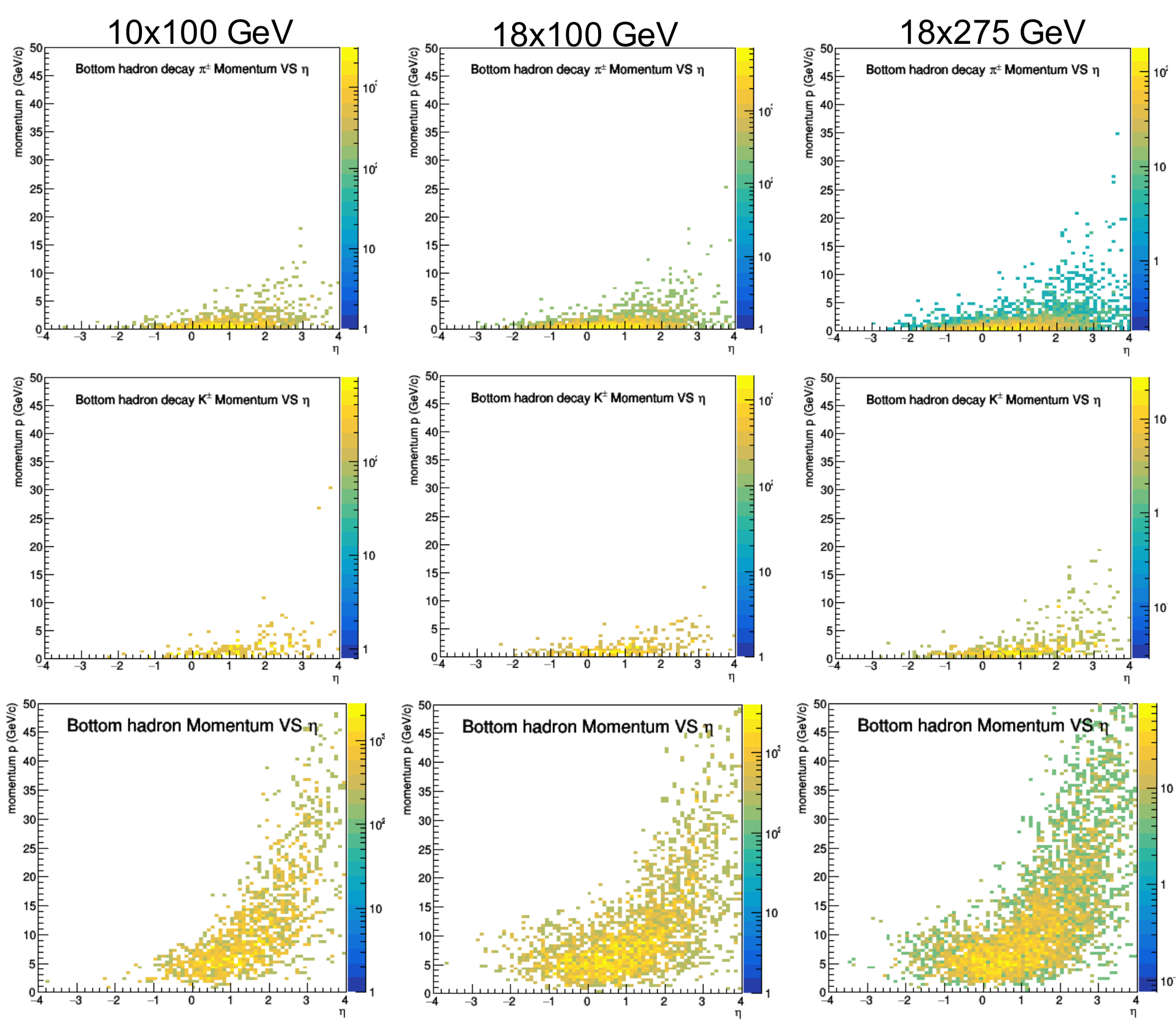}
    \caption{$B$ meson decay $\pi^{\pm}$ (top row), $K^{\pm}$ (second row) and $B$ meson (bottom row) momentum as a function of pseudorapidity from electron-proton collisions at $10\times100$~GeV (left column), $18\times100$~GeV (center column) and $18\times275$~GeV (right column) using Pythia8 simulations.}
    \label{fig:kinematics_pythia8_B_decay_mom_eta}
\end{figure}
\begin{figure}[ht]
    \centering
    \includegraphics[width=\textwidth]{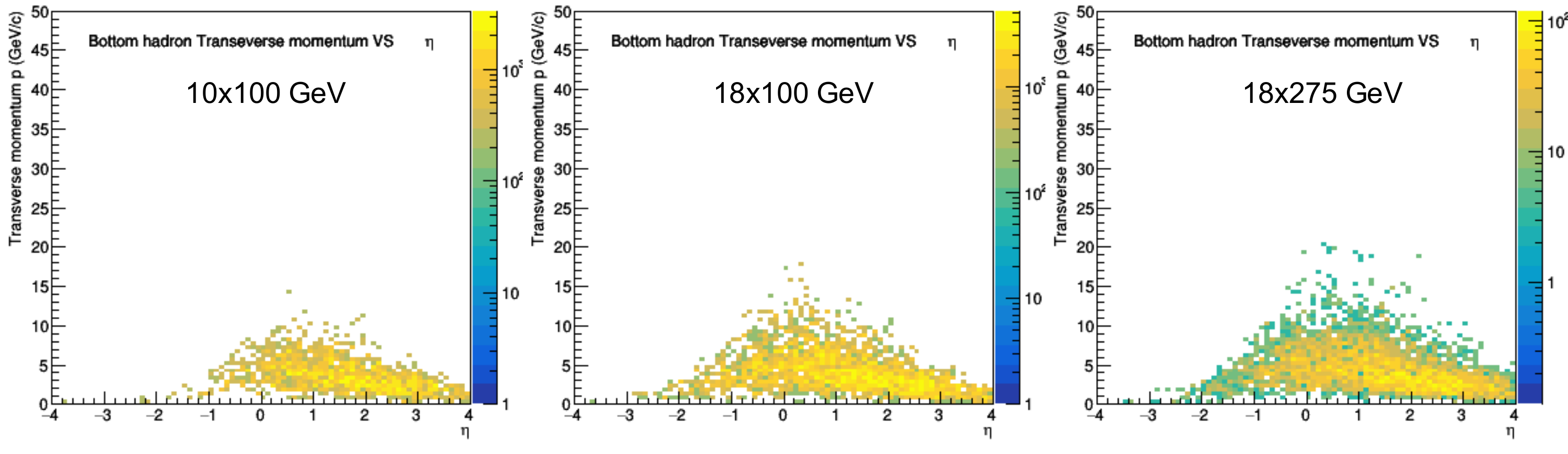}
    \caption{$B$ meson $p_T$ and $\eta$ distributions from electron-proton collisions at $10\times100$~GeV (left column), $18\times100$~GeV (center column) and $18\times275$~GeV (right column) using Pythia8 simulations.}
    \label{fig:kinematics_pythia8_B_pT_eta}
\end{figure}
\begin{figure}[ht]
    \centering
    \includegraphics[width=\textwidth]{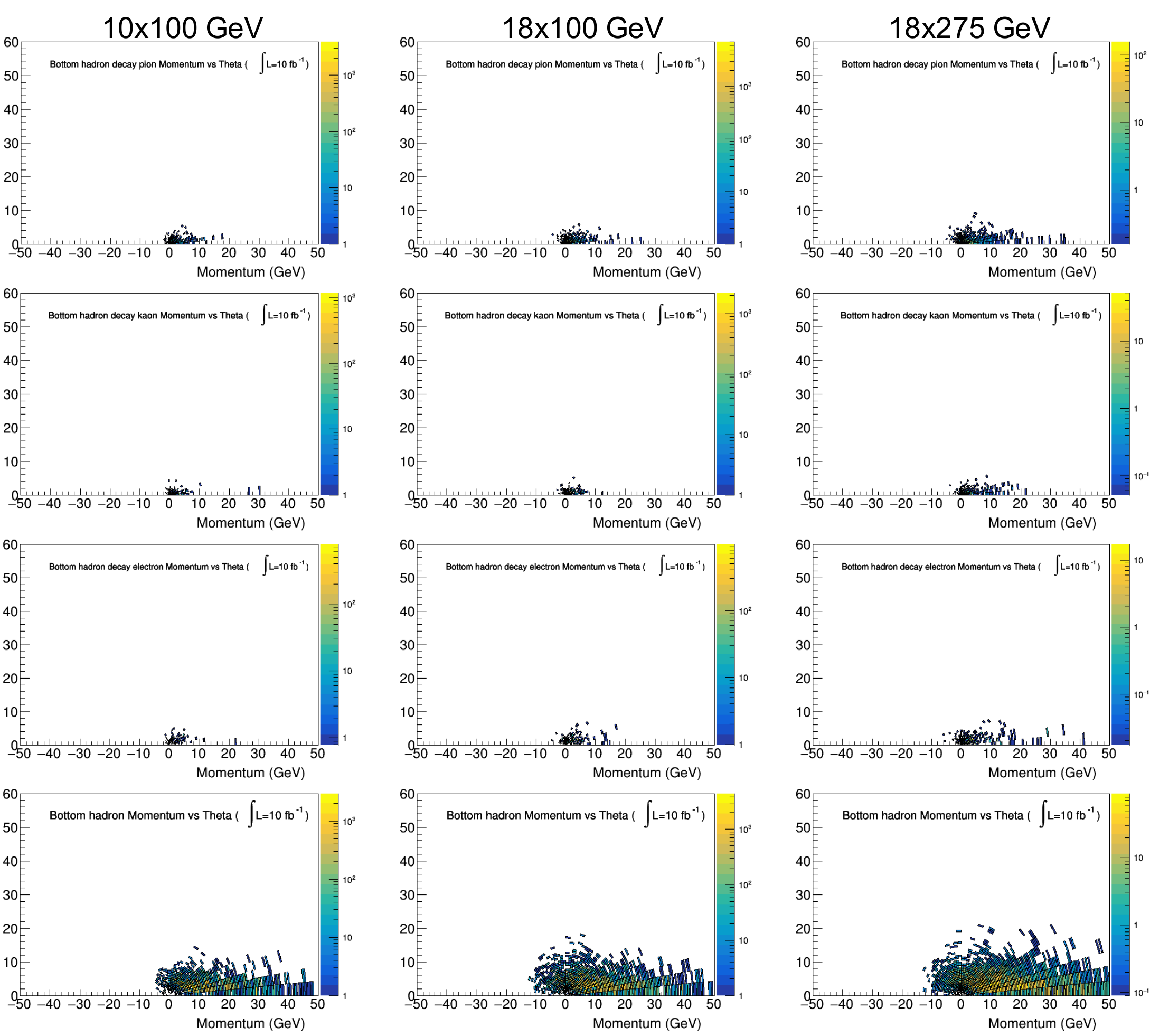}
    \caption{$B$ meson decay $\pi^{\pm}$ (top row), $K^{\pm}$ (second row), $e^-$ (third row) and $B$ meson (bottom row) momentum  distributions in polar coordinates from electron-proton collisions at $10\times100$~GeV (left column), $18\times100$~GeV (center column) and $18\times275$~GeV (right column) using Pythia8 simulations.}
    \label{fig:kinematics_pythia8_B_decay_mom_polar}
\end{figure}
%----------------------------------------------------------%
% D meson decay products
%----------------------------------------------------------%
\begin{figure}[ht]
    \centering
    \includegraphics[width=\textwidth]{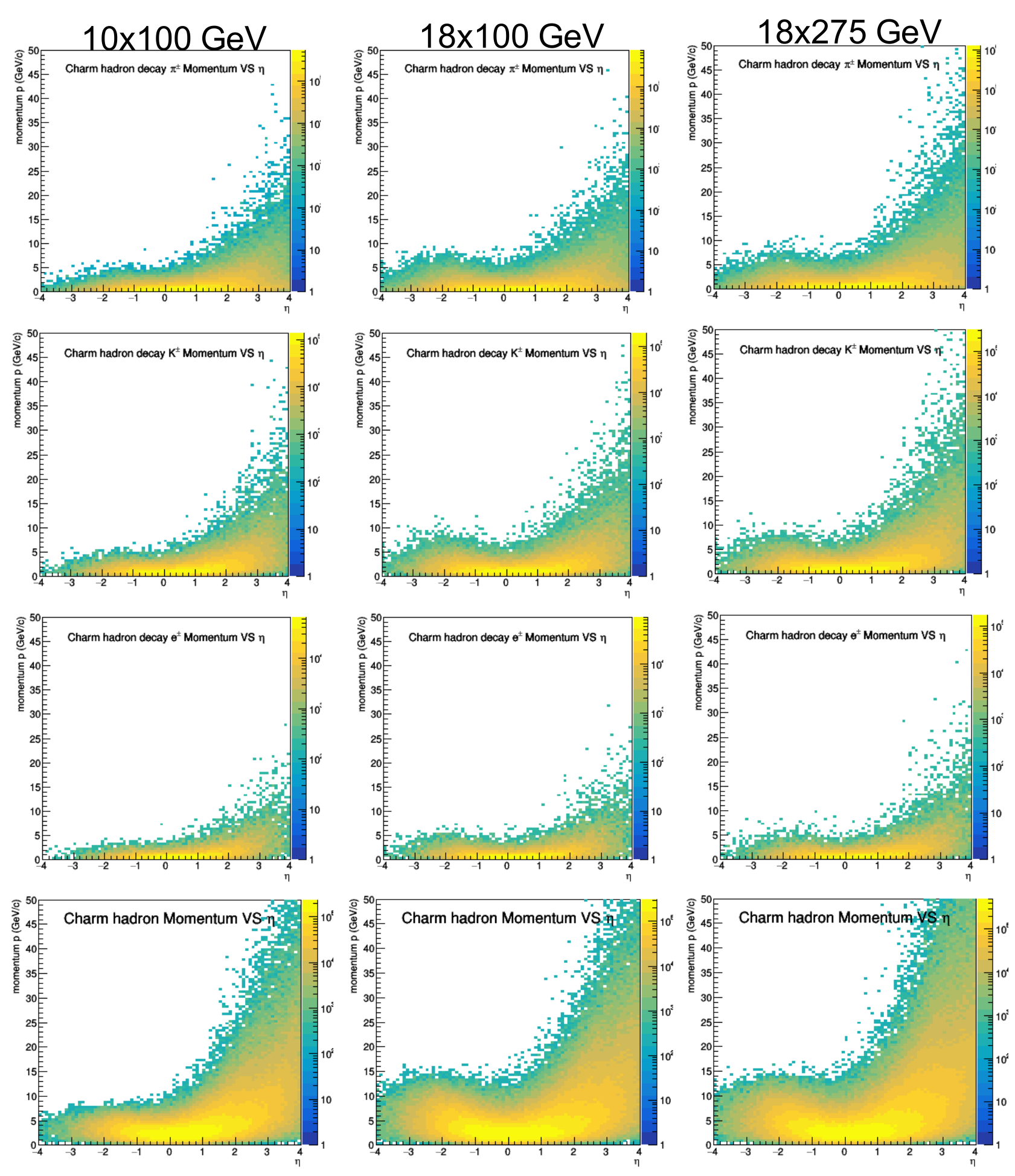}
    \caption{$D$ meson decay $\pi^{\pm}$ (top row), $K^{\pm}$ (second row) and $D$ meson (bottom row) momentum as a function of pseudorapidity from electron-proton collisions at $10\times100$~GeV (left column), $18\times100$~GeV (center column) and $18\times275$~GeV (right column) using Pythia8 simulations.}
    \label{fig:kinematics_pythia8_D_decay_mom_eta}
\end{figure}
\begin{figure}[ht]
    \centering
    \includegraphics[width=\textwidth]{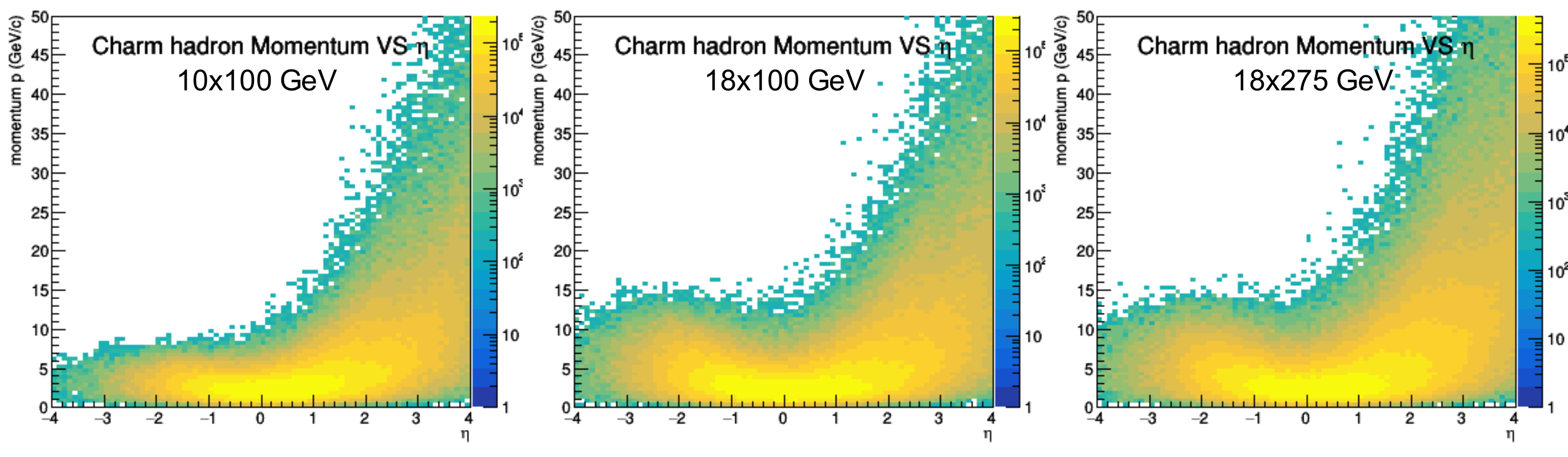}
    \caption{$D$ meson $p_T$ and $\eta$ distributions from electron-proton collisions at $10\times100$~GeV (left column), $18\times100$~GeV (center column) and $18\times275$~GeV (right column) using Pythia8 simulations.}
    \label{fig:kinematics_pythia8_D_pT_eta}
\end{figure}
\begin{figure}[ht]
    \centering
    \includegraphics[width=\textwidth]{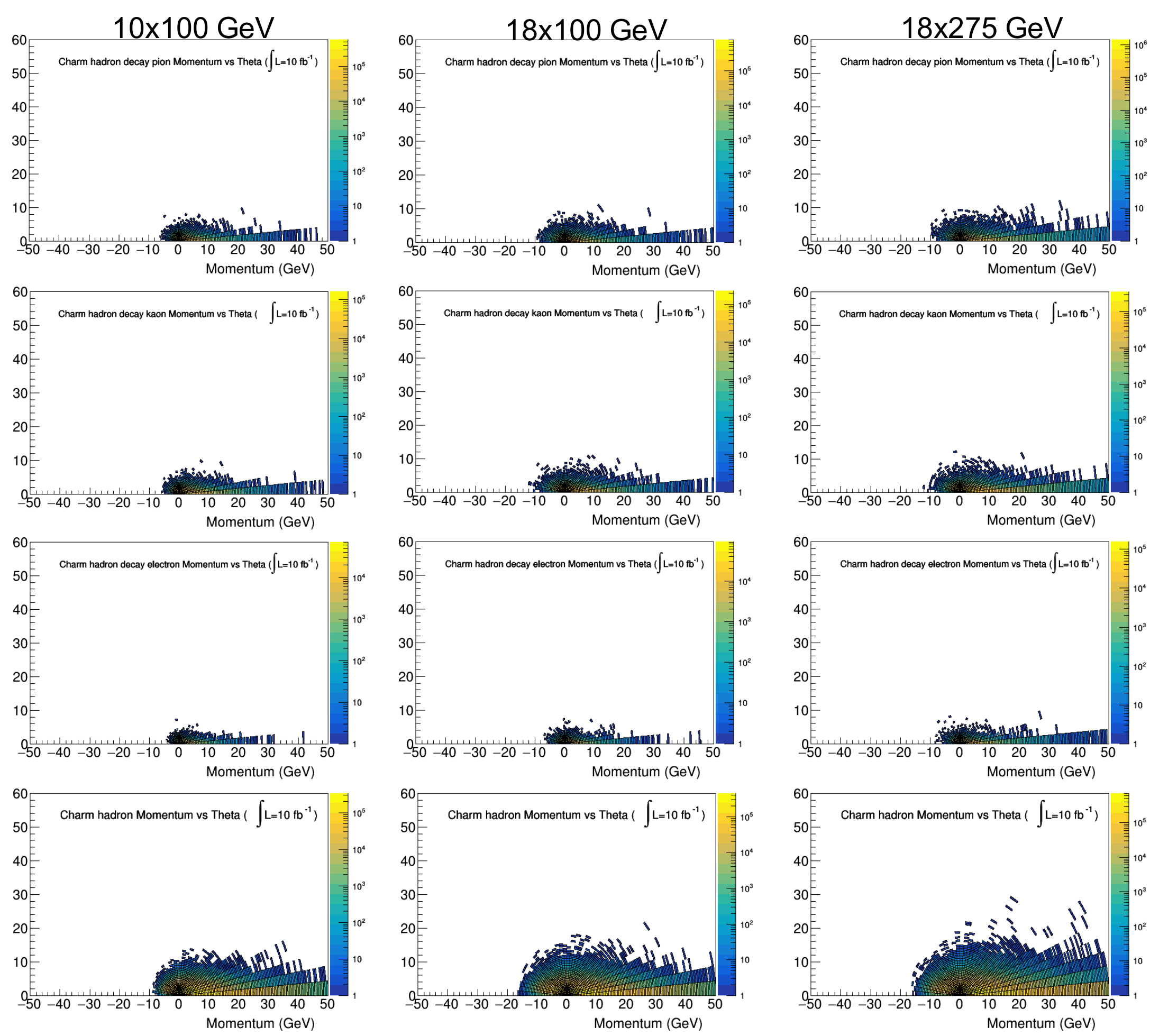}
    \caption{$D$ meson decay $\pi^{\pm}$ (top row), $K^{\pm}$ (second row), $e^-$ (third row) and $D$ meson (bottom row) momentum  distributions in polar coordinates from electron-proton collisions at $10\times100$~GeV (left column), $18\times100$~GeV (center column) and $18\times275$~GeV (right column) using Pythia8 simulations.}
    \label{fig:kinematics_pythia8_D_decay_mom_polar}
\end{figure}
\FloatBarrier
\section{Detector Design}
\label{sec: design}
Due to the asymmetric collisions that will occur at the future EIC, the majority of the final hadrons are produced in the pseudorapity region of $-2$ to $4$ (see the kinematic plots in section \ref{sec: kin}). The initial design is studied in the fast simulation with the LDT package, see the initial design in \cite{Li:2020sru}.

%coverage\\ geometry\\ material\\ sensor options\\
\FloatBarrier
\section{Fun4all Simulation}
\label{sec: fun4all}
The EIC Fun4all simulation is a modified version of the sPHENIX simulation. Both the Babar and BeAST magnets are tested in the simulation. The Babar magnetic field peaks at $1.4$T and the BeAST magnetic field peaks at $3$T. A $95$\% detection hit efficiency is used in both track and vertex reconstructions. In track reconstruction, the Kalman Filter algorithm is used and a $20$~\um vertex Gaussian smearing is applied to both $x$ and $y$ directions. In vertex reconstruction, no vertex smearing is applied and $10$~charged pions are launched per event. The number of degrees of freedom is set to $1$ in vertex reconstruction for the initial studies.

This section will show the the tracking performance from simulations with the preliminary detector designs detailed in Section~\ref{sec:geo}. These simulation results, including momentum resolution and distance of closest approach resolution, are applied to heavy meson reconstruction in physics simulation discussed in Section~\ref{sec: phy}. Both detector and physics simulation results will help evaluate and improve different designs.
%--------------------------------------------------------------------%
\subsection{Detector Setup}\label{sec:geo}
\begin{figure}[h]
\centering
\includegraphics[width=0.49\textwidth]{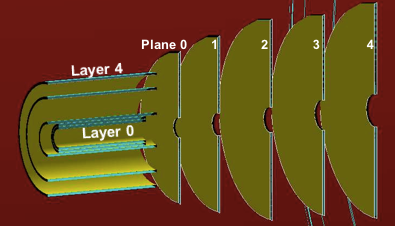}
\includegraphics[width=0.49\textwidth]{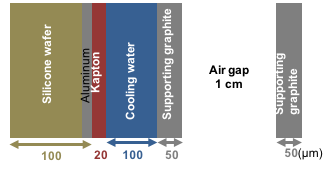}
\caption{Left: initial detector implementation in Fun4all simulation includes the five-plane FST system (right) along with the five-layer barrel system (left). Right: illustration of each layer of barrel detector and FST.}
\label{fig:ver0}
\end{figure}

\begin{figure}[h]
\centering
\includegraphics[width=0.49\textwidth]{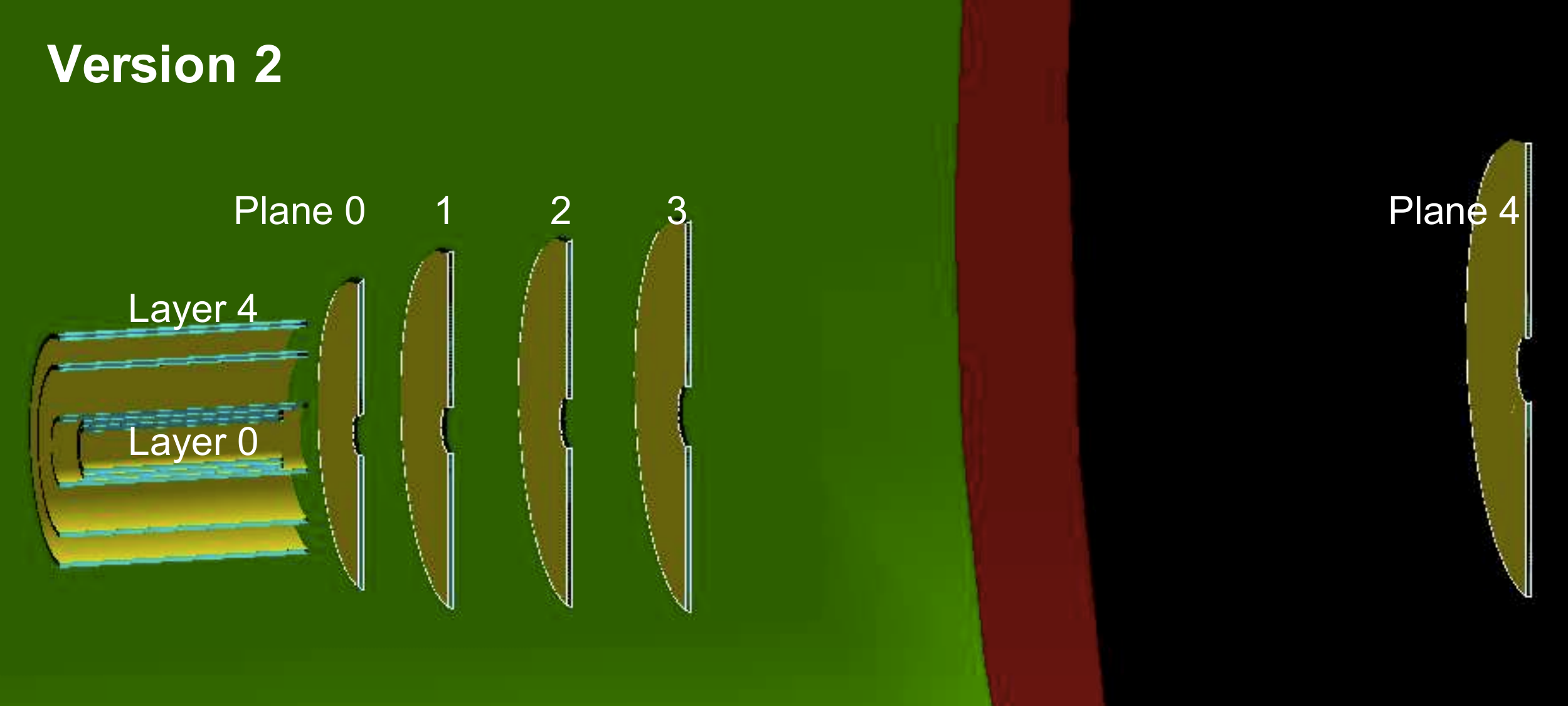}
\includegraphics[width=0.49\textwidth]{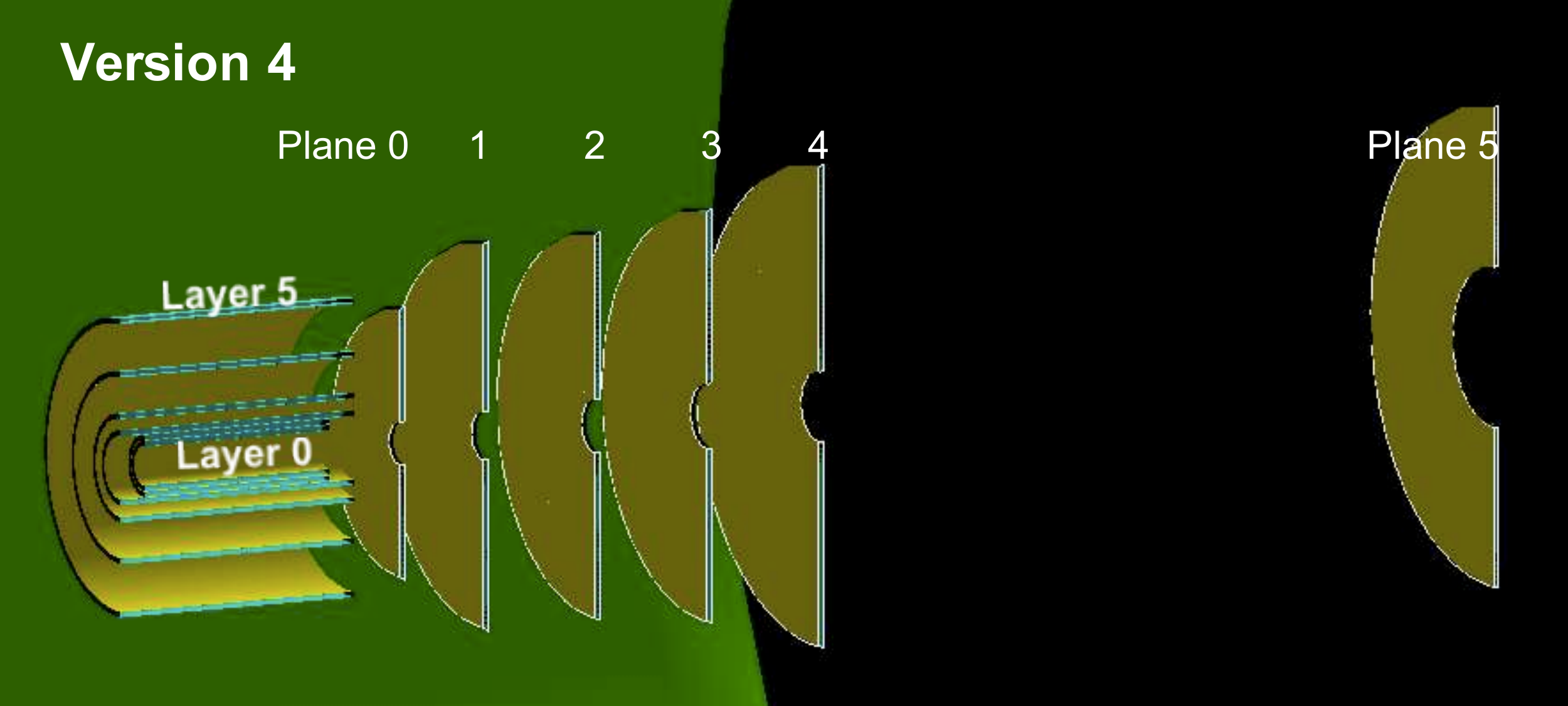}
\caption{Left: version~2 of FST design. Right: version~4 of FST design which is the same for version~4.1 and~4.2.}
\label{fig:ver2and4}
\end{figure}

The initial implementation of the FST simulation in Fun4all includes both the barrel detector system and the FST as shown in Figure~\ref{fig:ver0}. This section will detail three different FST designs, including geometry and material. The parameters of detector geometries are listed in Table~\ref{tab:simBarrel_Geo} and Table~\ref{tab:simFST_Geo}.
\begin{itemize}
    \item \textbf{Version~0} is the initial FST design. There are five-layer barrel detector and five-plane FST detector implemented in the Fun4all simulation as shown on the left of Figure~\ref{fig:ver0}. The inner radius of the FST planes increases along the $z$ positions to accommodate the ion beam pipe which gets larger at large $z$. The structure of each plane of FST and layer of barrel detector, which is illustrated on the right of Figure~\ref{fig:ver0}, includes a silicon wafer with a sheet of aluminum support, a thin layer of kapton followed by the cooling wafer and the supporting structure of graphite. The silicon wafer (including the aluminum base) implemented in both the barrel detector and the FST has a $100$~\um thickness in this initial design.
    \item \textbf{Version~1} design is derived from the version~0 design. Two changes are made in version~1 compared to version~0. First, the outer radii of plane~0 and plane~1 of the FST were changed as listed in Table~\ref{tab:simFST_Geo}. Plane~0 is smaller to reduce material budget at mid pseudorapidity while plane~1 is slightly larger by $1$~cm to improve the forward acceptance. Secondly, the thickness of the silicon wafer is different than in version~0. In the outer two layers (layer~3 and layer~4) of the barrel detector, the silicon wafer has a $100$~\um thickness. In the rest of the barrel detector and all five planes of the FST, the silicon wafer has a $50$~\um thickness.
    \item \textbf{Version~2}, which is shown on the left of Figure~\ref{fig:ver2and4}, is also derived from the version~0 design. This version moved the last plane of the FST, that is plane~4, from $z=1.25$~m to $z=2.7$~m to improve the detector performance at large pseudorapidity ($\eta=3$). As in version~1, in the outer two layers (layer~3 and layer~4) of the barrel detector the silicon wafer has a $100$~\um thickness. In the rest of the barrel detector and all five planes of the FST, the silicon wafer has a $50$~\um thickness.
    \item \textbf{Version~3} is derived from the version~1 design. The silicon thickness of the barrel detector is set to $35$~\um in version3. The last two planes, plane~3 and~4, of the FST have a $36.4$~\um pixel pitch and a $100$~\um thick silicon wafer.
    \item \textbf{Version~4}, which is shown on the right of Figure~\ref{fig:ver2and4}, is derived from version~2 design. Instead of moving plane~4 to the far-z location ($z=2.7$~m) as in version~2, an additional plane is placed at $z=2.7$~m. Furthermore, the barrel detector is also updated in version~4. Plane~3 and plane~4 have larger radii in version~4 than in version~4. Version~4 also has an additional layer, layer~6, with a $27$~cm radius at the barrel detector. There are two variations of version~4, namely version~4.1 and version~4.2, with larger pixel pitch and thicker silicon wafer at the last few planes as listed in Table~\ref{tab:simFST_Geo}.
\end{itemize}
\begin{table}[h]
    \centering
    \caption{\label{tab:simBarrel_Geo}Barrel detector geometry parameters}
    \begin{tabular}{c c c c c}
    \hline
    %-------------------------------%
    \multicolumn{5}{c}{Version 0}\\
    \hline
    Layer & half length (cm) & radius (cm) & pixel Pitch (\um) & silicon thickness (\um)\\ 
    \hline
    0 & 20  & 3.64 & 20 & 100\\
    1 & 20  & 4.81 & 20 & 100\\
    2 & 25  & 5.98 & 20 & 100\\
    3 & 25  & 16   & 20 & 100\\
    4 & 25  & 22   & 20 & 100\\
    \hline
    %-------------------------------%
    \multicolumn{5}{c}{Version 1}\\
    \hline
    Layer & half length (cm) & radius (cm) & pixel Pitch (\um) & silicon thickness (\um)\\ 
    \hline
    0 & 20  & 3.64 & 20 & 50\\
    1 & 20  & 4.81 & 20 & 50\\
    2 & 25  & 5.98 & 20 & 50\\
    3 & 25  & 16   & 20 & 100\\
    4 & 25  & 22   & 20 & 100\\
    \hline
    %-------------------------------%
    \multicolumn{5}{c}{Version 2}\\
    \hline
    Layer & half length (cm) & radius (cm) & pixel Pitch (\um) & silicon thickness (\um)\\ 
    \hline
    0 & 20  & 3.64 & 20 & 50\\
    1 & 20  & 4.81 & 20 & 50\\
    2 & 25  & 5.98 & 20 & 50\\
    3 & 25  & 16   & 20 & 100\\
    4 & 25  & 22   & 20 & 100\\
    \hline
    %-------------------------------%
    \multicolumn{5}{c}{Version 3}\\
    \hline
    Layer & half length (cm) & radius (cm) & pixel Pitch (\um) & silicon thickness (\um)\\ 
    \hline
    0 & 20  & 3.64 & 20 & 35\\
    1 & 20  & 4.81 & 20 & 35\\
    2 & 25  & 5.98 & 20 & 35\\
    3 & 25  & 16   & 20 & 35\\
    4 & 25  & 22   & 20 & 35\\
    \hline
    %-------------------------------%
    \multicolumn{5}{c}{Version 4, 4.1 and 4.2}\\
    \hline
    Layer & half length (cm) & radius (cm) & pixel Pitch (\um) & silicon thickness (\um)\\ 
    \hline
    0 & 20  & 3.64 & 20 & 50\\
    1 & 20  & 4.81 & 20 & 50\\
    2 & 25  & 5.98 & 20 & 50\\
    3 & 25  & 9.2  & 20 & 100\\
    4 & 25  & 17   & 20 & 100\\
    5 & 25  & 27   & 20 & 100\\
    \hline
    \end{tabular}
\end{table}
\begin{table}[h]
    \centering
    \caption{\label{tab:simFST_Geo}Forward plane detector geometry parameters}
    \small
    \begin{tabular}{c c c c c c}
    \hline
    %-------------------------------%
    \multicolumn{6}{c}{Version 0}\\
    \hline
    Plane & z (cm) & inner radius (cm) & outer radius (cm) & pixel Pitch (\um) & silicon thickness (\um)\\ 
    \hline
    0 & 35  & 4   & 30   & 20 & 100\\
    1 & 53  & 4.5 & 35   & 20 & 100\\
    2 & 77  & 5   & 36   & 20 & 100\\
    3 & 101 & 6   & 38.5 & 20 & 100\\
    4 & 125 & 6.5 & 45   & 20 & 100\\
    \hline
    %-------------------------------%
    \multicolumn{6}{c}{Version 1}\\
    \hline
    Plane & z (cm) & inner radius (cm) & outer radius (cm) & pixel Pitch (\um) & silicon thickness (\um)\\ 
    \hline
    0 & 35  & 4   & 25   & 20 & 50\\
    1 & 53  & 4.5 & 36   & 20 & 50\\
    2 & 77  & 5   & 36   & 20 & 50\\
    3 & 101 & 6   & 38.5 & 20 & 50\\
    4 & 125 & 6.5 & 45   & 20 & 50\\
    \hline
    %-------------------------------%
    \multicolumn{6}{c}{Version 2}\\
    \hline
    Plane & z (cm) & inner radius (cm) & outer radius (cm) & pixel Pitch (\um) & silicon thickness (\um)\\ 
    \hline
    0 & 35  & 4   & 30   & 20 & 50\\
    1 & 53  & 4.5 & 35   & 20 & 50\\
    2 & 77  & 5   & 36   & 20 & 50\\
    3 & 101 & 6   & 38.5 & 20 & 50\\
    4 & 270 & 6.5 & 45   & 20 & 50\\
    \hline
    %-------------------------------%
    \multicolumn{6}{c}{Version 3}\\
    \hline
    Plane & z (cm) & inner radius (cm) & outer radius (cm) & pixel Pitch (\um) & silicon thickness (\um)\\ 
    \hline
    0 & 35  & 4   & 25   & 20   & 50\\
    1 & 53  & 4.5 & 36   & 20   & 50\\
    2 & 77  & 5   & 36   & 20   & 50\\
    3 & 101 & 6   & 38.5 & 36.4 & 100\\
    4 & 125 & 6.5 & 45   & 36.4 & 100\\
    \hline
    %-------------------------------%
    \multicolumn{6}{c}{Version 4}\\
    \hline
    Plane & z (cm) & inner radius (cm) & outer radius (cm) & pixel Pitch (\um) & silicon thickness (\um)\\ 
    \hline
    0 & 35  & 4   & 25   & 20   & 50\\
    1 & 53  & 4.5 & 36   & 20   & 50\\
    2 & 77  & 5   & 36   & 20   & 50\\
    3 & 101 & 6   & 38.5 & 20   & 50\\
    4 & 125 & 6.5 & 45   & 20   & 50\\
    5 & 270 & 15  & 45   & 20   & 50\\
    \hline
    %-------------------------------%
    \multicolumn{6}{c}{Version 4.1}\\
    \hline
    Plane & z (cm) & inner radius (cm) & outer radius (cm) & pixel Pitch (\um) & silicon thickness (\um)\\ 
    \hline
    0 & 35  & 4   & 25   & 20   & 50\\
    1 & 53  & 4.5 & 36   & 20   & 50\\
    2 & 77  & 5   & 36   & 20   & 50\\
    3 & 101 & 6   & 38.5 & 36.4 & 100\\
    4 & 125 & 6.5 & 45   & 36.4 & 100\\
    5 & 270 & 15  & 45   & 36.4 & 100\\
    \hline
    %-------------------------------%
    \multicolumn{6}{c}{Version 4.2}\\
    \hline
    Plane & z (cm) & inner radius (cm) & outer radius (cm) & pixel Pitch (\um) & silicon thickness (\um)\\ 
    \hline
    0 & 35  & 4   & 25   & 20   & 50\\
    1 & 53  & 4.5 & 36   & 20   & 50\\
    2 & 77  & 5   & 36   & 20   & 50\\
    3 & 101 & 6   & 38.5 & 20   & 50\\
    4 & 125 & 6.5 & 45   & 36.4 & 100\\
    5 & 270 & 15  & 45   & 36.4 & 100\\
    \hline
    \end{tabular}
\end{table}
\FloatBarrier
\FloatBarrier
\subsection{Material Budget}
Different thicknesses of the silicon wafer are considered. The material scan with no silicon wafer shows the material budget of the supporting and cooling structures of the detector. The scan with the silicon wafer of a conservative $100$~\um thickness gives the upper bound estimation for the material budget. The mixed use of thickness of silicon wafers shows the realistic estimation. In the material scan of mixed silicon thicknesses, the silicon wafers in the outer layers (layer~$3$ and~$4$) of the barrel detector are set to $100$~\um thick, while the rest of the barrel detector layers as well as the FST have silicon wafers with a $50$~\um thickness.

\begin{figure}[htbp]
    \centering 
    \includegraphics[width=\textwidth]{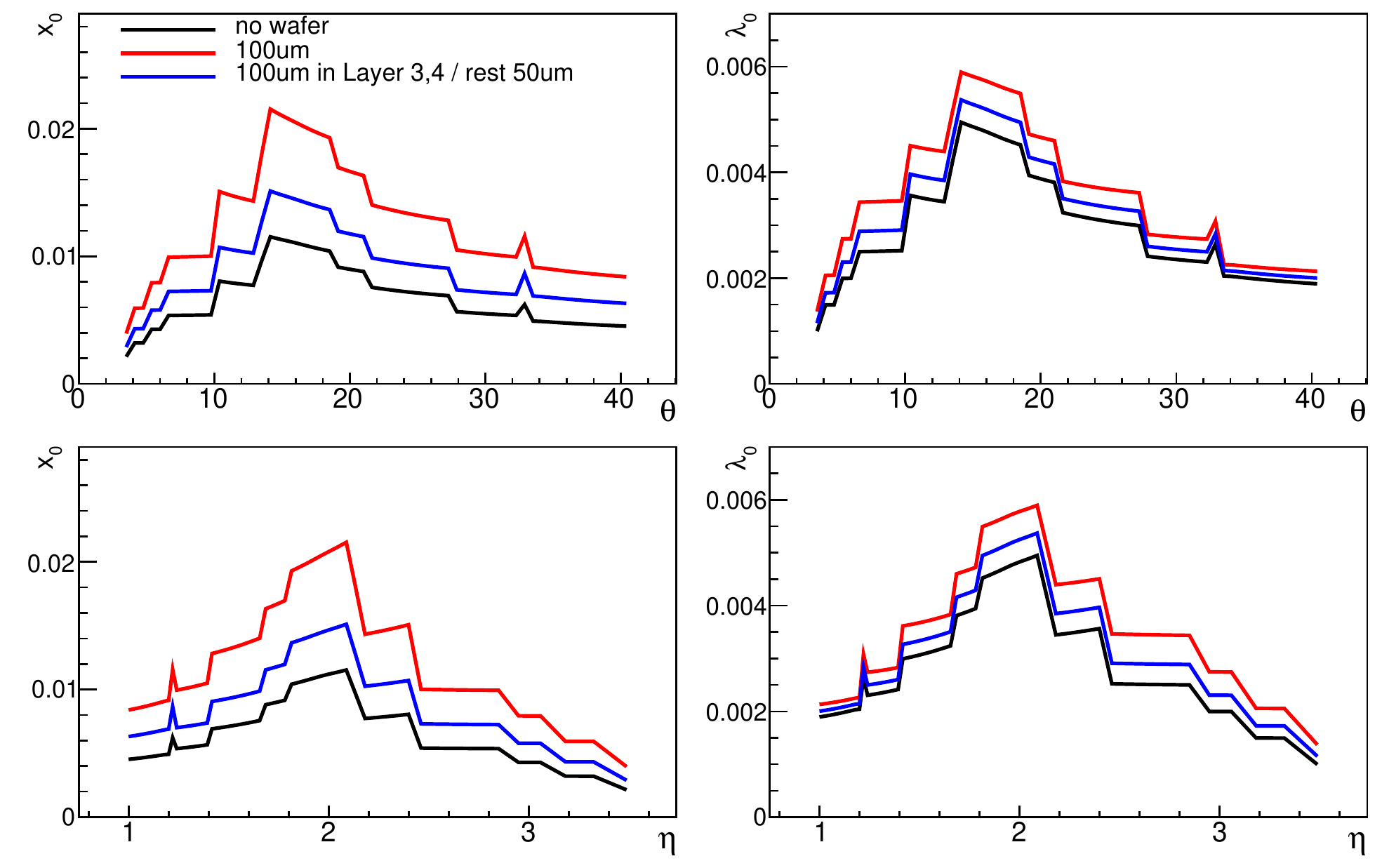}
    \vspace{-8mm}
    \caption{Material budget of the version~0 FST design.}
    \label{fig:matScan_ver0}
\end{figure}

Figure~\ref{fig:matScan_ver0}, which shows the material budget of the version~0 FST design, shows that the highest material budget occurs at $\theta\approx14^\circ$ ($\eta\approx2.1$) with about $2.2\%x_0$, $1.4\%x_0$ and $1.1\%x_0$ when using the $100$~\um silicon wafers in both barrel and FST systems, different thicknesses of silicon wafer in both barrel and FST systems and no silicon wafer in the detector systems, respectively.

\begin{figure}[htbp]
    \centering
    \includegraphics[width=\textwidth]{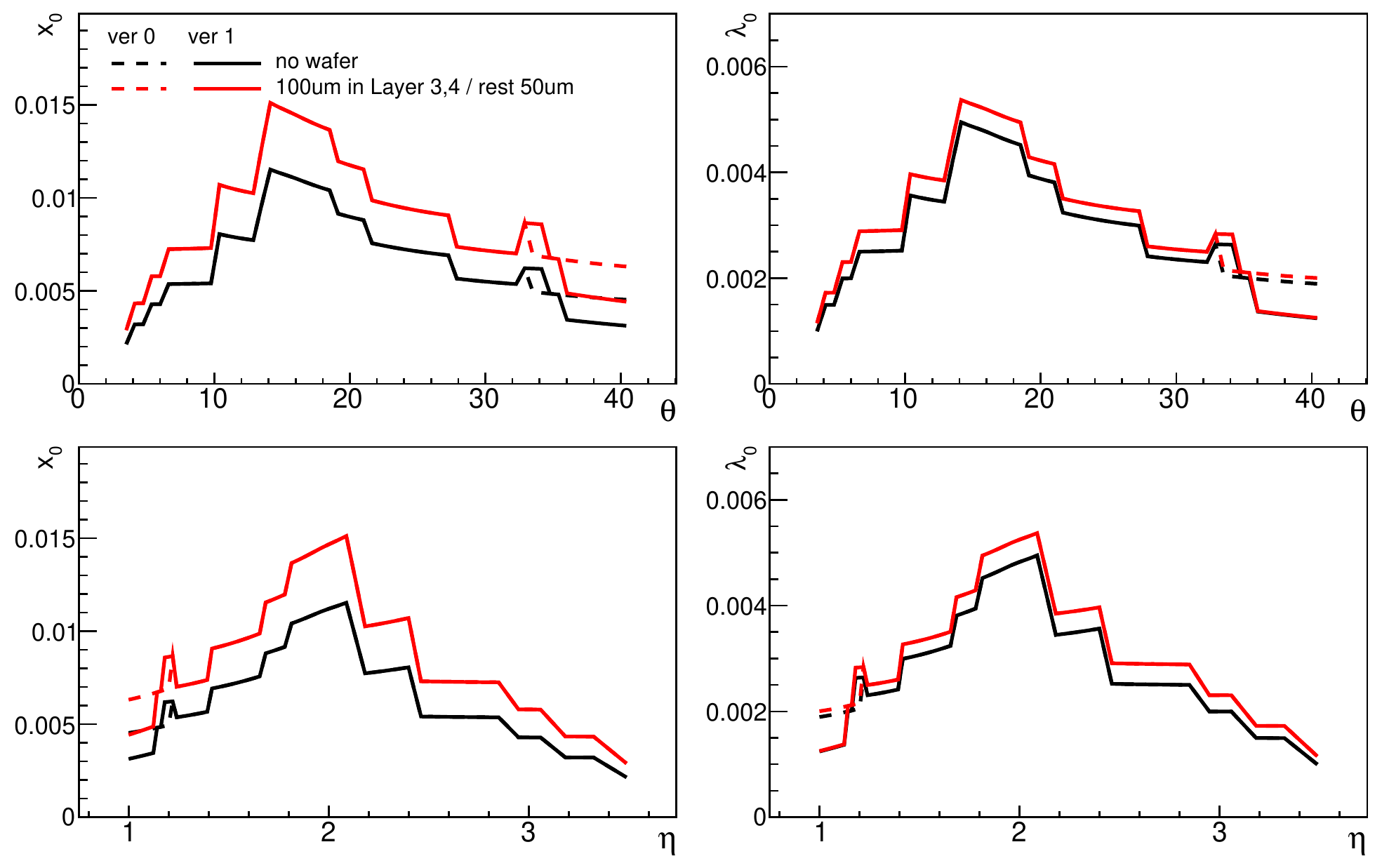}
    \vspace{-8mm}
    \caption{Material budget of the version~1 FST design compared to version~0 design.}
    \label{fig:matScan_ver0_ver1}
\end{figure}

\begin{figure}[htbp]
    \centering
    \includegraphics[width=\textwidth]{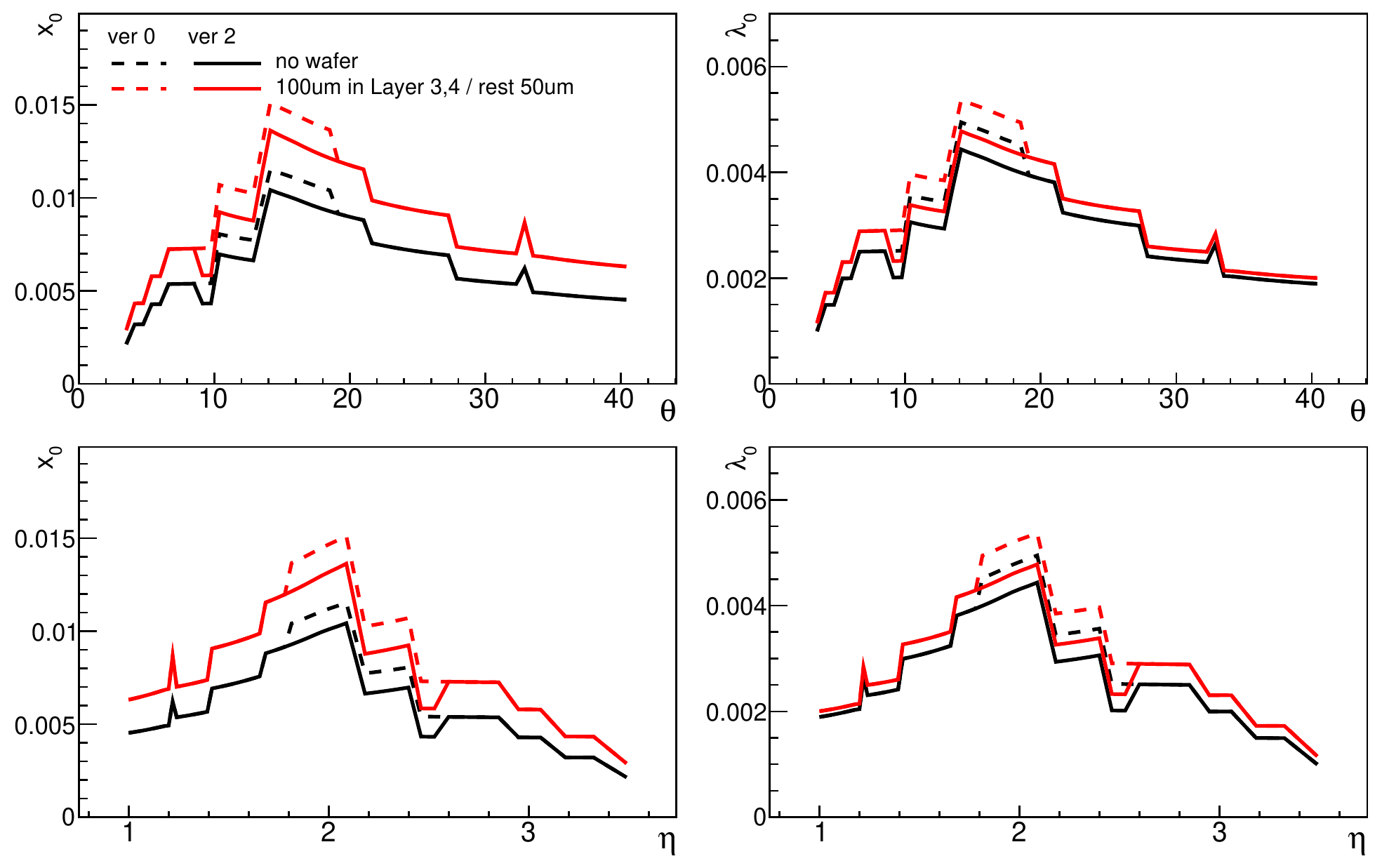}
    \vspace{-8mm}
    \caption{Material budget of the version~2 FST design compared to version~0 design.}
    \label{fig:matScan_ver0_ver2}
\end{figure}

Version~1 and version~2 designs give slightly different material budgets at different polar angles. The material budget of version~1 at $\theta\approx34^\circ$ ($\eta\approx1.2$) is about $0.2\%$ higher than version~0, as shown in Figure~\ref{fig:matScan_ver0_ver1}, due to the larger size of plane~1 in version~1. However, at $\theta>36^\circ$ ($\eta<1.2$), the material budget is about $0.2\%$ lower in version~1 than in version~0 because of the smaller plane~0 in version~1. The material budget of version~2 design is about $0.2\%$ lower than version~0 design between $10^\circ<\theta<20^\circ$ ($1.8<\eta<2.6$), as shown in Figure~\ref{fig:matScan_ver0_ver2}, as the last plane (plane~4) is moved from $z=1.25$~m to $z=2.7$~m.
\FloatBarrier
\subsection{Momentum Resolution}
The momentum resolution is defined as the width of the distribution of the relative difference between the reconstructed and the input (true) momenta, that is 
\begin{align}
    \frac{\Delta p}{p_{input}}=\frac{p_{reco}-p_{input}}{p_{input}}\text{ ,}
\end{align}
where $p_{reco}$ and $p_{input}$ are the reconstructed momentum and input momentum, respectively. This section will discuss the effects on the momentum resolution of different magnetic fields, detector geometry and pixel pitch.

%--------------------------------------------------------------------%
\subsubsection{Magnetic Field Study}
\begin{figure}[h]
    \centering
    \includegraphics[width=\textwidth]{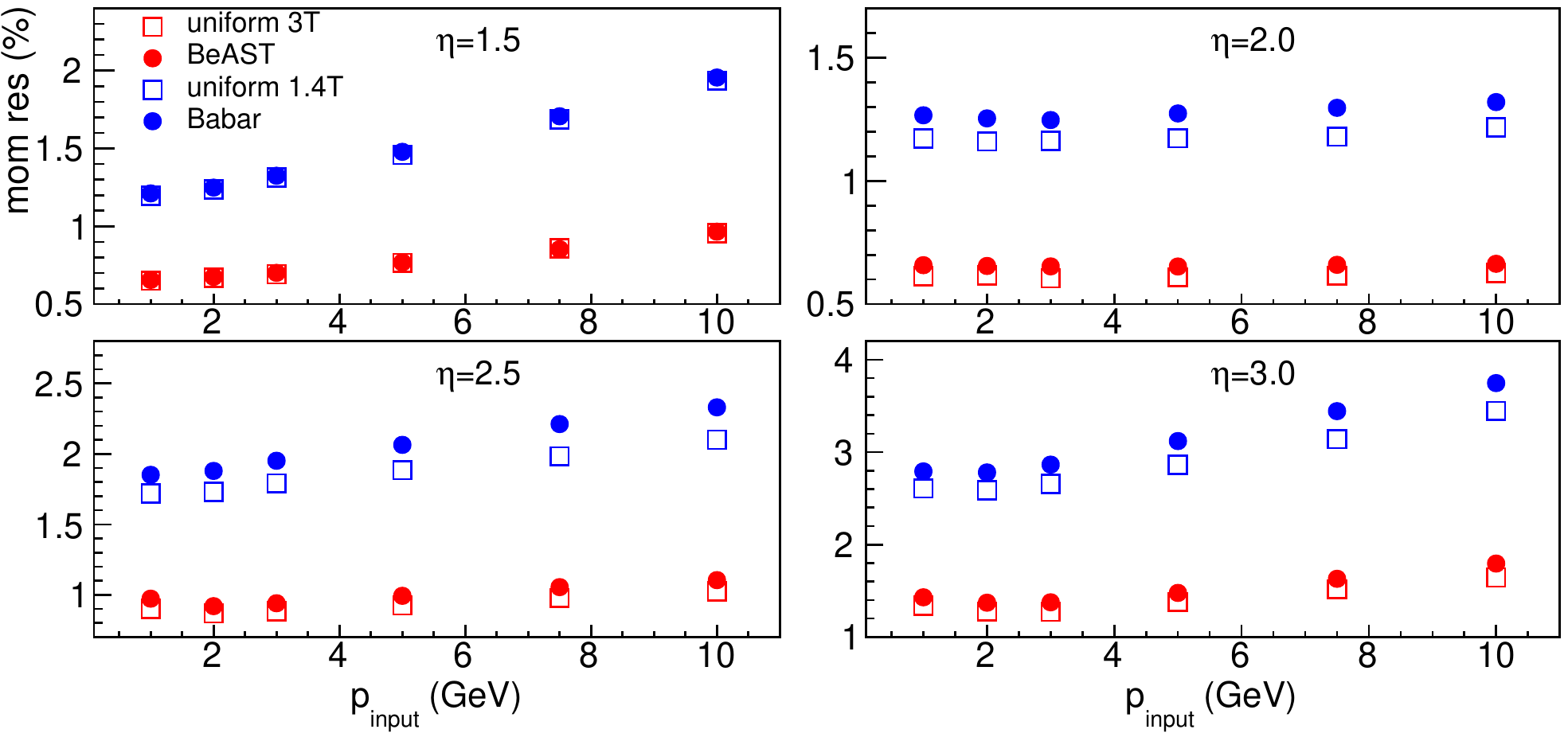}
    \caption{Comparison of Momentum resolution of the tracking system in different magnetic field in Fun4all simulation. Version~0 of FST design with $100$~\um thickness of silicon wafers are used in the simulation.}
    \label{fig:momRes_Bfield_0_10GeV}
\end{figure}

With uniform $1.4$~T and $3$~T magnetic fields, the Babar and the BeAST magnet are implemented in the Fun4all simulations. Figure~\ref{fig:momRes_Bfield_0_10GeV} shows the momentum resolutions of the version~0 design from simulations with different magnetic fields. The version~0 design can achieve lower momentum resolution with the use of the BeAST magnet, which has a magnetic field peaked at $3$~T, compared to the use of the Babar magnet with a magnetic field peaked at $1.4$~T. Furthermore, Figure~\ref{fig:momRes_Bfield_0_10GeV} also shows that a uniform magnetic field gives slightly better momentum resolution compared to the nominal Babar or BeAST magnetic field.

%--------------------------------------------------------------------%
\subsubsection{Detector Geometry Study}
\begin{figure}[h]
    \centering
    \includegraphics[width=\textwidth]{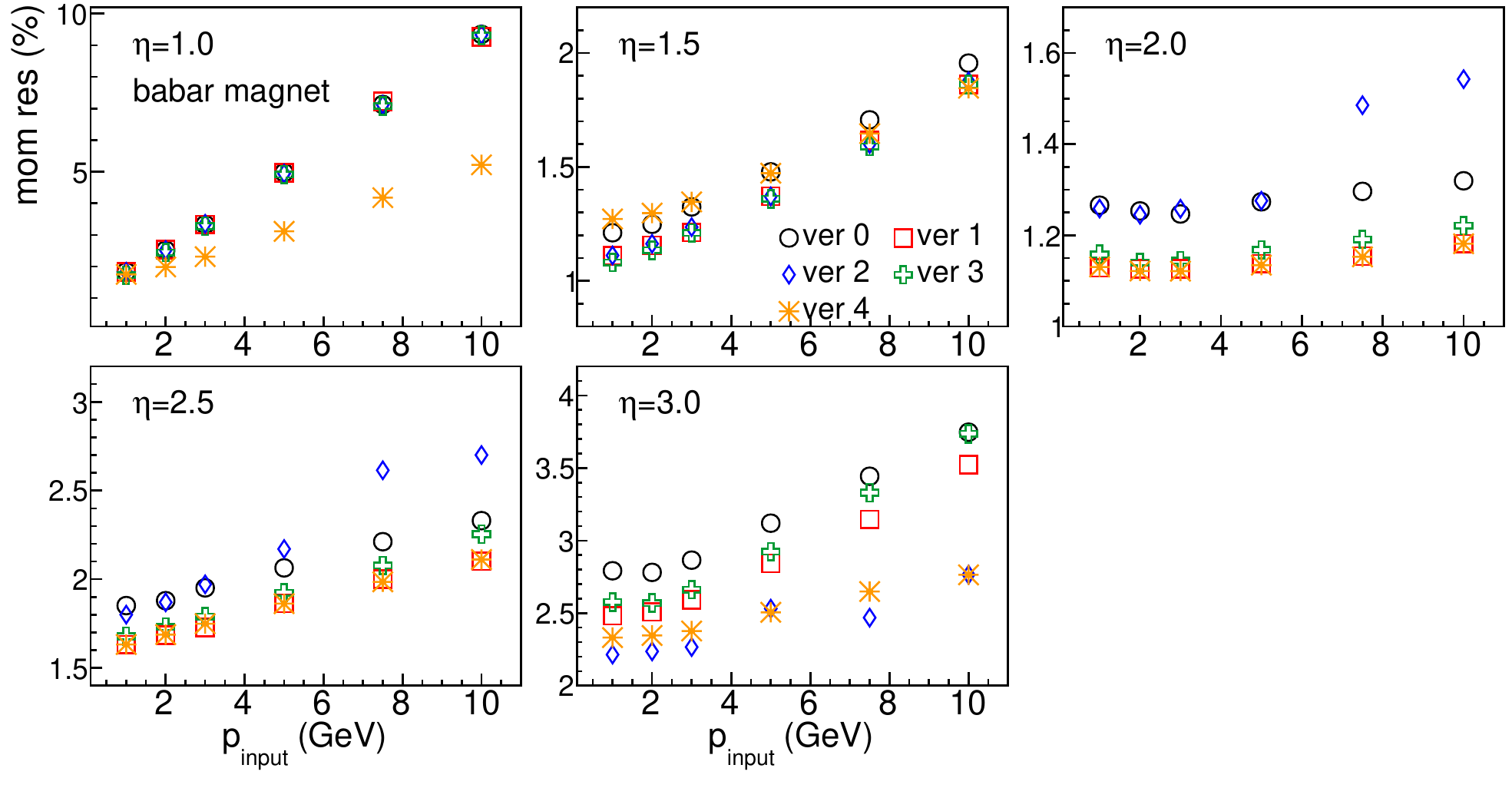}
    \vspace{-8mm}
    \caption{\label{fig:momRes_babar_0_10GeV}Comparison of Momentum resolution of the different FST design from Fun4All simulation with the Babar magnet.}
\end{figure}

\begin{figure}[h]
    \centering
    \includegraphics[width=\textwidth]{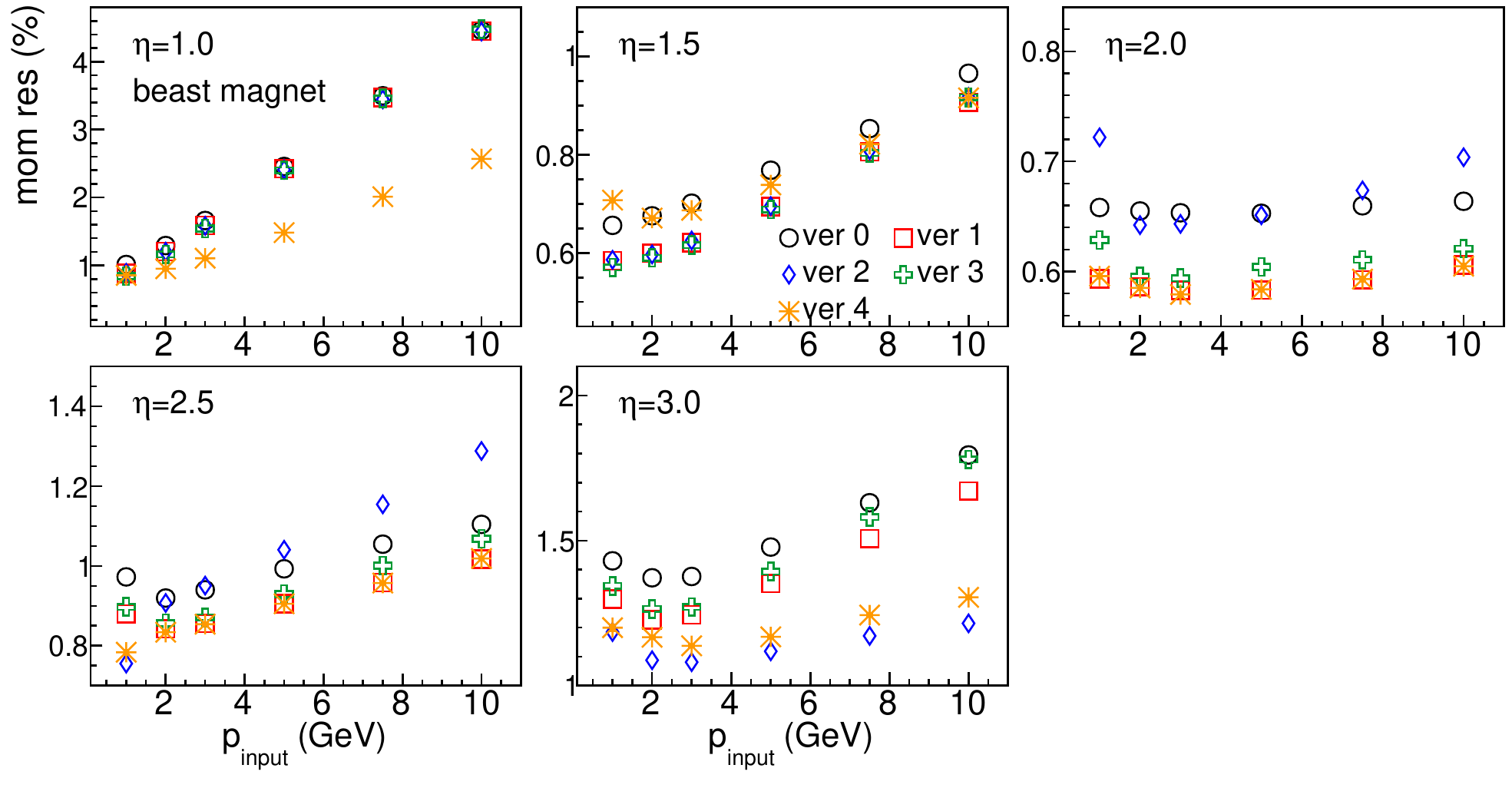}
    \vspace{-8mm}
    \caption{\label{fig:momRes_beAST_0_10GeV}Comparison of Momentum resolution of the different FST design from Fun4All simulation with the BeAST magnet.}
\end{figure}

Figure~\ref{fig:momRes_babar_0_10GeV} and~\ref{fig:momRes_beAST_0_10GeV} show the momentum resolution comparisons of different FST designs with the use of Babar and BeAST magnets, respectively. The comparison of version~0 to version~1 shows the momentum resolution decreases as the material budget is reduced in version~1. Version~1 and version~3 share the same geometry, but version~3 uses thinner ($35$~\um) silicon wafers in the barrel detector than version~1. However, there are no significant differences in momentum resolution at $\eta=1$. Version~3 also use thicker silicon wafers ($100$~\um) and larger pixel pitches ($36.4$~\um) for the plane~3 and~4 of the FST than version~1 which causes a small increase ($<0.3$\%) in the momentum resolution at $\eta\geq2$ compared to version~1.

The comparison between version~0 and version~2 shown in Figure~\ref{fig:momRes_babar_0_10GeV} and~\ref{fig:momRes_beAST_0_10GeV} demonstrates that with a plane at a far-z location ($z=2.7$~m), the momentum resolution at $10$~GeV and $\eta=3$ reduces from about $3.8$\% ($1.8$\%) to about $2.8$\% ($1.3$\%) when the Babar (BeAST) magnet is used. However, moving plane~4 of the FST to the far-z location reduces the detector performance at $2\leq\eta\leq2.5$. Therefore, version~4 with an additional FST plane at a far-z location is introduced to maintain the performance at $2\leq\eta\leq2.5$, while improving the performance at large pseudorapidity ($\eta=3$). The barrel detector geometry in version~4 is updated as well. Layer~3 and~4 of the barrel detector in version~4 has smaller radius, but a barrel layer with a $27$~cm radius is added in version4. This updated design of the barrel in version~4 reduces the momentum resolution at $10$~GeV and $\eta=1$ from about $10$\% ($4.5$\%) to $5$\% ($2.5$\%) when the Babar (BeAST) magnet is used.

To summarize, the version~4 design with an additional barrel layer and FST plane gives better momentum resolutions. However, the additional layer and plane could be costly. Therefore, a FST-GEM combined tracking system is under consideration to maintain cost effectiveness. Study of a combined silicon tracker and GEM system is done in Eicroot as shown in Section~\ref{sec:EICroot_momRes}. The same study will also be carried out in a Fun4All simulation.
%--------------------------------------------------------------------%
\subsubsection{Pixel Pitch Study}
\begin{figure}[h]
    \centering
    \includegraphics[width=0.9\textwidth]{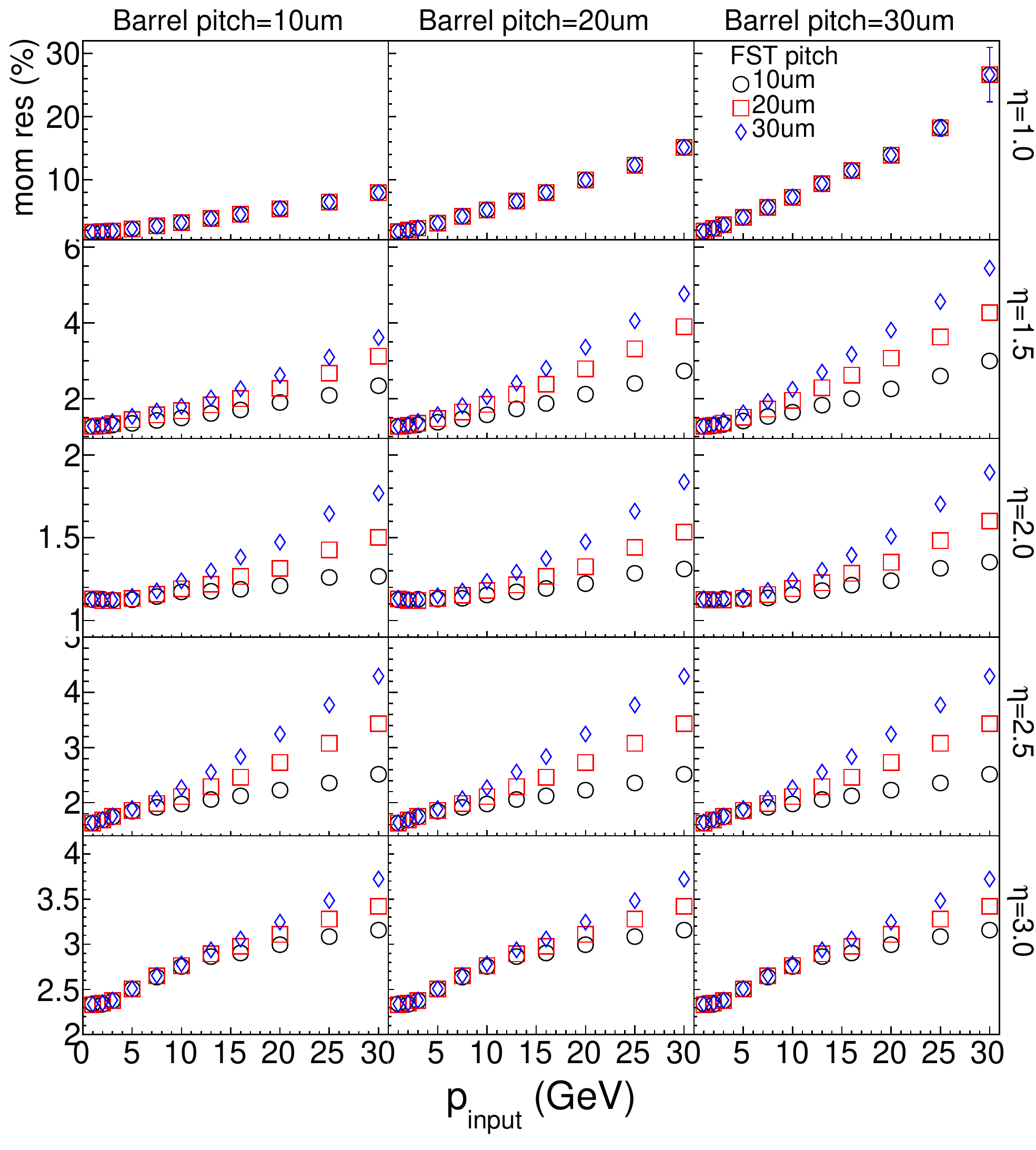}
    \caption{Momentum resolution as a function input (true) momentum from Fun4all simulation using version~4 of FST design and Babar magnet.}
    \label{fig:momRes_ver4_0_30GeV}
\end{figure}

\begin{figure}[h]
    \centering
    \includegraphics[width=0.9\textwidth]{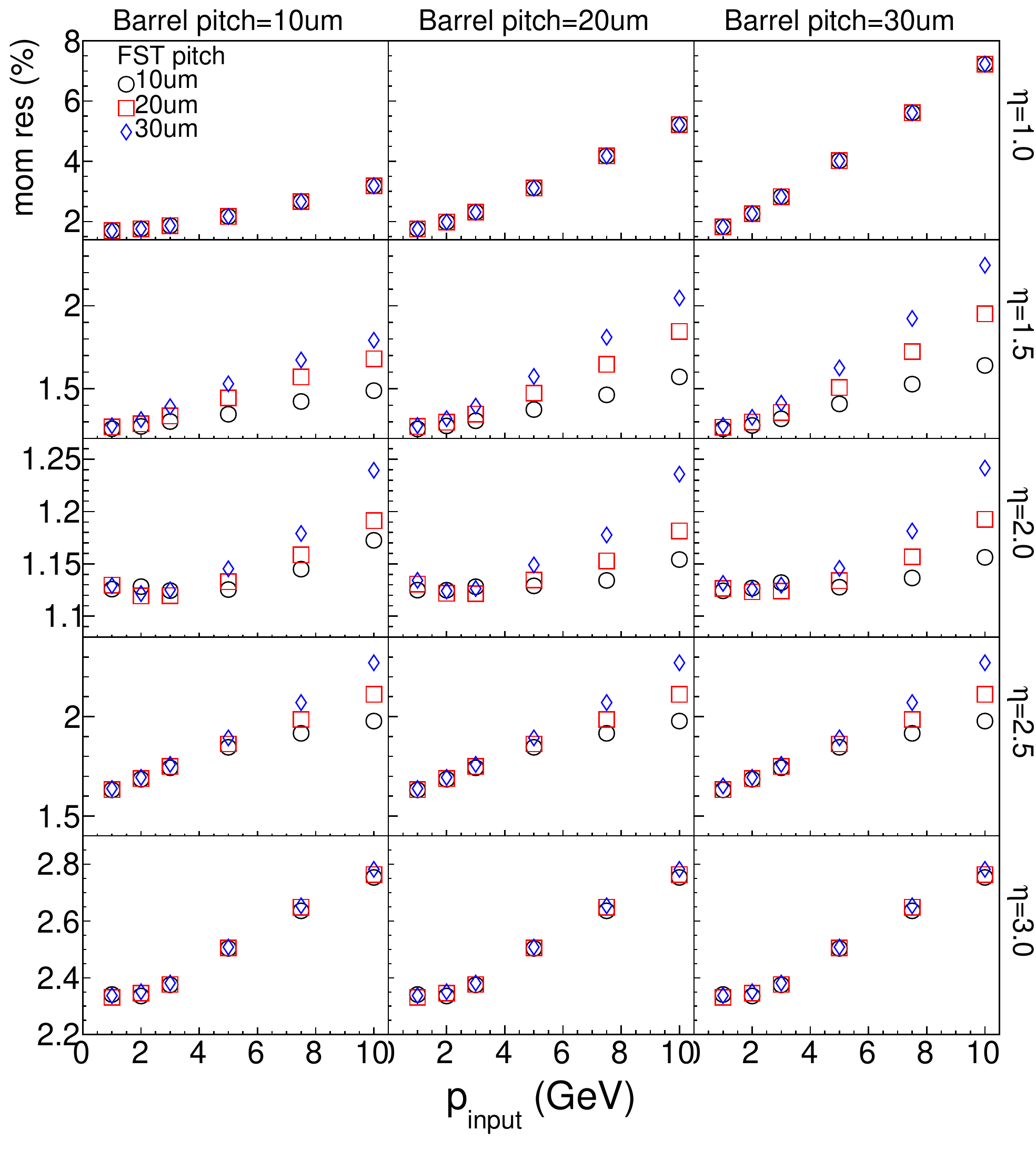}
    \caption{Momentum resolution as a function input (true) momentum from Fun4all simulation using version~4 of FST design and Babar magnet.}
    \label{fig:momRes_ver4_0_10GeV}
\end{figure}

Version~4 detector design is used in the simulation to study pixel pitch size effect on momentum resolution. Three pixel pitch sizes, $10$~\um, $20$~\um and $30$~\um, are used in the simulation with the implementation of the Babar magnet. Figure~\ref{fig:momRes_ver4_0_30GeV} shows the pixel pitch dependence of momentum resolution in different pseudorapidity. At $\eta=1$, the momentum resolution is highly dependent on the pixel pitch size at the barrel detector as tracks with a $\eta=1$ passes through multiple layers of the barrel detector. At $1.5\leq\eta\leq2$, the momentum resolution also relies on the pixel pitch size at the FST detector. At $\eta\geq2$, the momentum resolution becomes dependent on the pixel pitch size of the FST alone as the track with large pseudorapidity only passes through the FST. Figure~\ref{fig:momRes_ver4_0_10GeV} shows the same results as figure~\ref{fig:momRes_ver4_0_30GeV}, but zoom in to the $0$--$10$~GeV region. The same pixel pitch dependency shown at high momentum ($>10$~GeV) is also found at low momentum ($<10$~GeV). However, at $\eta=3$, there is no noticeable pixel pitch dependence shown at low momentum. 

\begin{figure}[H]
    \centering
    \includegraphics[width=0.95\textwidth]{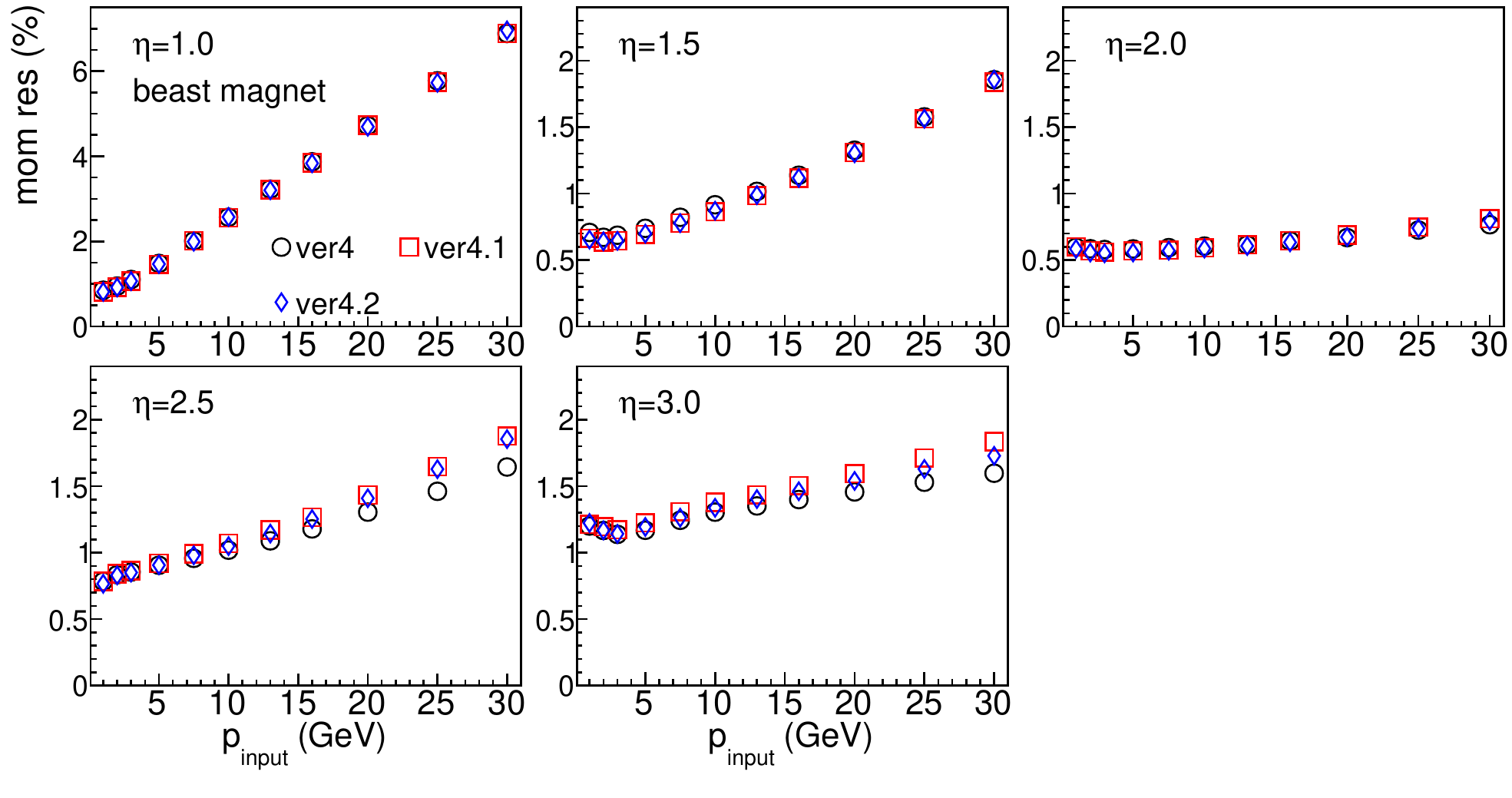}
    \vspace{-4mm}
    \caption{Momentum resolution as a function input (true) momentum from Fun4all simulation using version~4, 4.1 and 4.2 of FST design and BeAST magnet.}
    \label{fig:momRes_ver4_4dot1_4dot2_0_30GeV}
\end{figure}

Figure~\ref{fig:momRes_ver4_4dot1_4dot2_0_30GeV} shows the comparison of momentum resolutions between version~4, 4.1 and 4.2 from simulation with the implementation of the BeAST magnet. The latter two versions of FST design use larger pixel pitch and thicker silicon wafer in the last few planes as listed in Table~\ref{tab:simFST_Geo}. The momentum resolutions of version~4, 4.1 and 4.2 are consistent at $\eta\leq2$ as tracks with small pseudorapidity do not pass through the last few planes of FST. However, the differences between these three designs are noticeable at $\eta\geq2.5$, but limited within $0.3$\%.
\FloatBarrier
\FloatBarrier
\subsection{DCA Resolution}
\begin{figure}[htbp]
    \centering
    \includegraphics[width=0.95\textwidth]{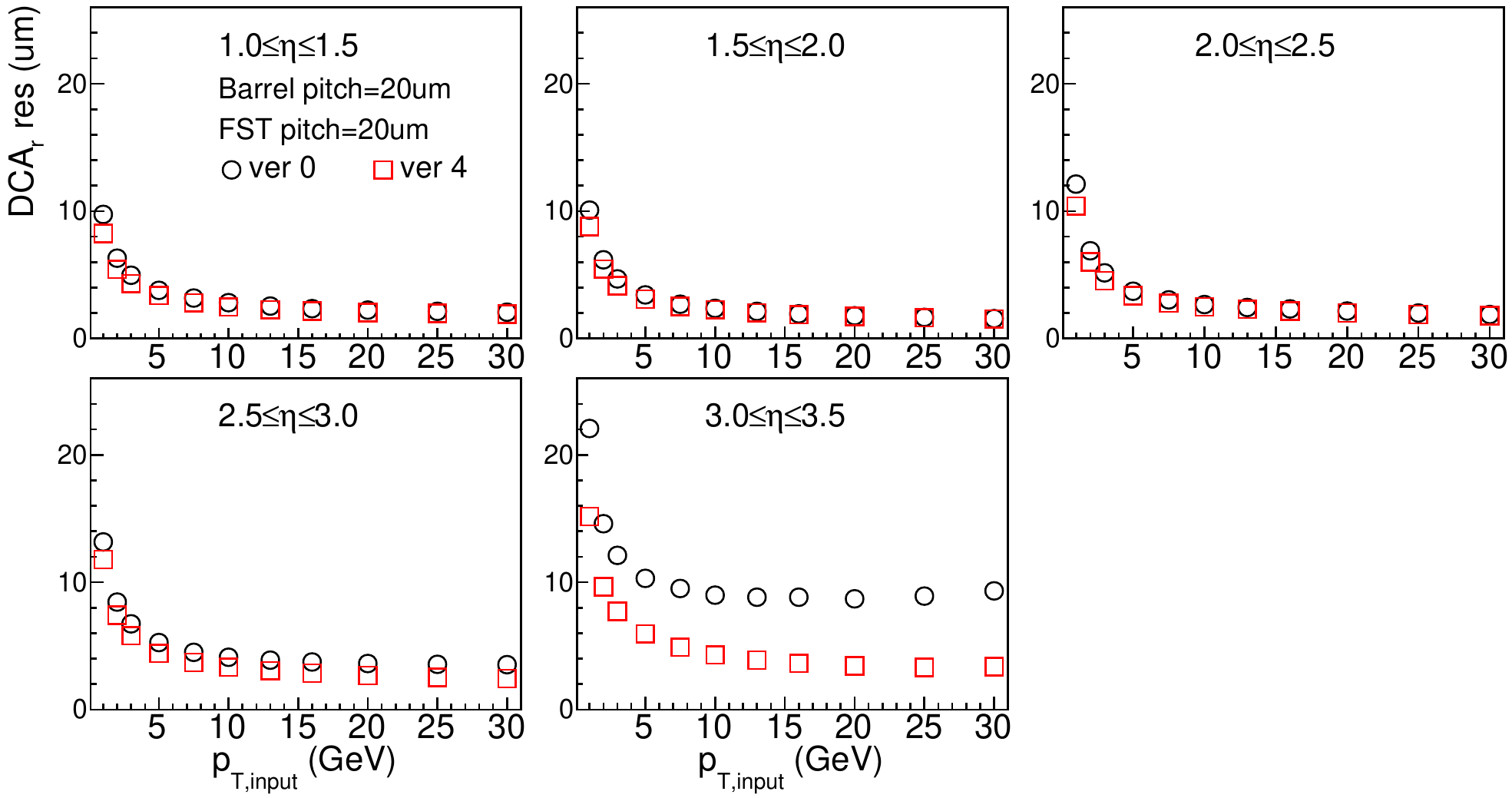}
    %\vspace{-4mm}
    \caption{$DCA_r$ resolution as a function input (true) transverse momentum from Fun4all simulation using version~0 and version~4 FST design with BeAST magnetic field.}
    \label{fig:DCArRes_ver0_ver4}
\end{figure}

\begin{figure}[htbp]
    \vspace{-5mm}
    \centering
    \includegraphics[width=0.95\textwidth]{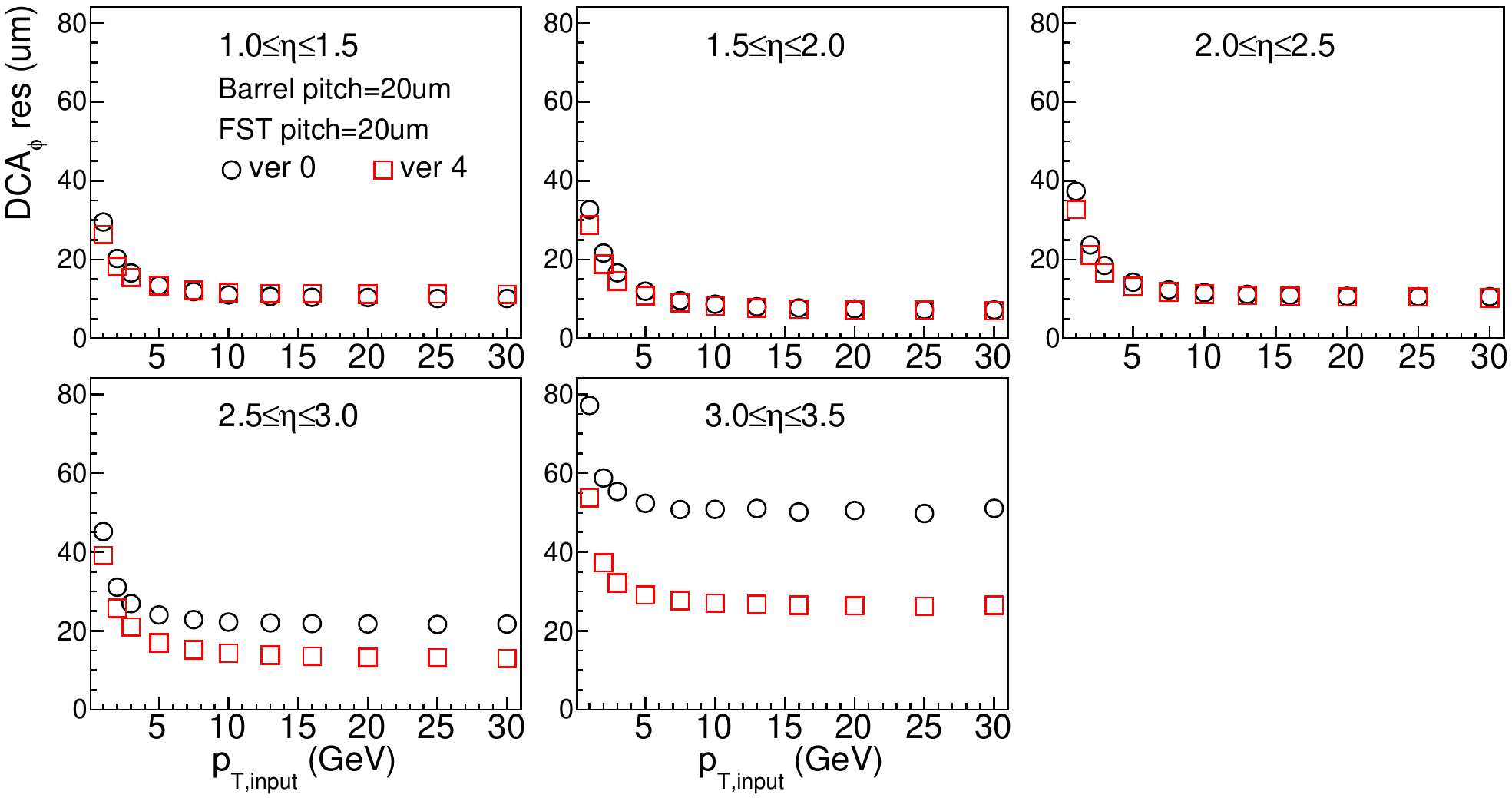}
    %\vspace{-4mm}
    \caption{$DCA_\phi$ resolution as a function input (true) transverse momentum from Fun4all simulation using version~0 and version~4 FST design with BeAST magnetic field.}
    \label{fig:DCAphiRes_ver0_ver4}
    \vspace{-5mm}
\end{figure}

\begin{figure}[htbp]
    %\vspace{-8mm}
    \centering
    \includegraphics[width=0.95\textwidth]{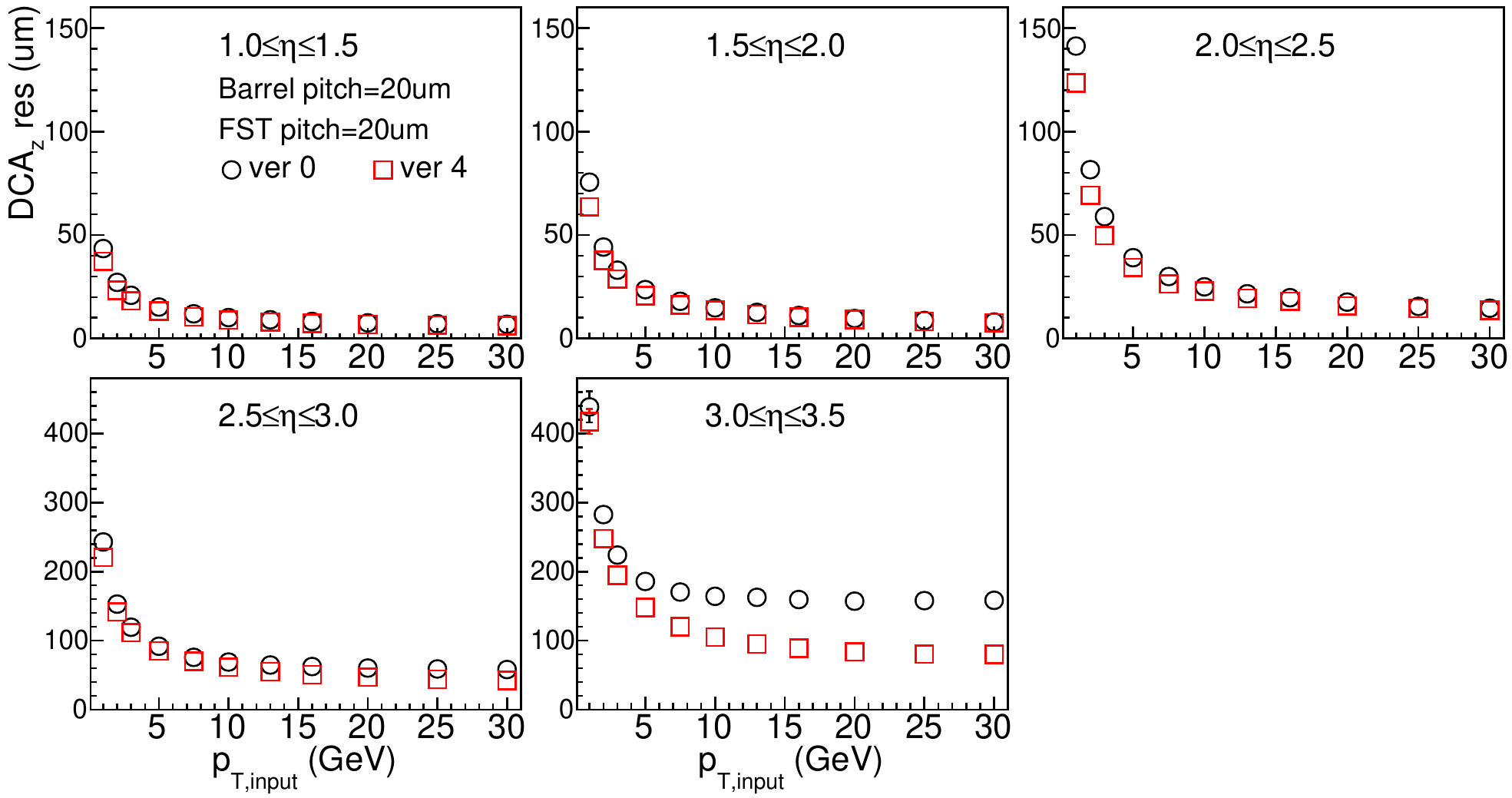}
    %\vspace{-4mm}
    \caption{$DCA_z$ resolution as a function input (true) transverse momentum from Fun4all simulation using version~0 and version~4 FST design with BeAST magnetic field.}
    \label{fig:DCAzRes_ver0_ver4}
\end{figure}

\begin{figure}[htbp]
    \centering
    \includegraphics[width=0.95\textwidth]{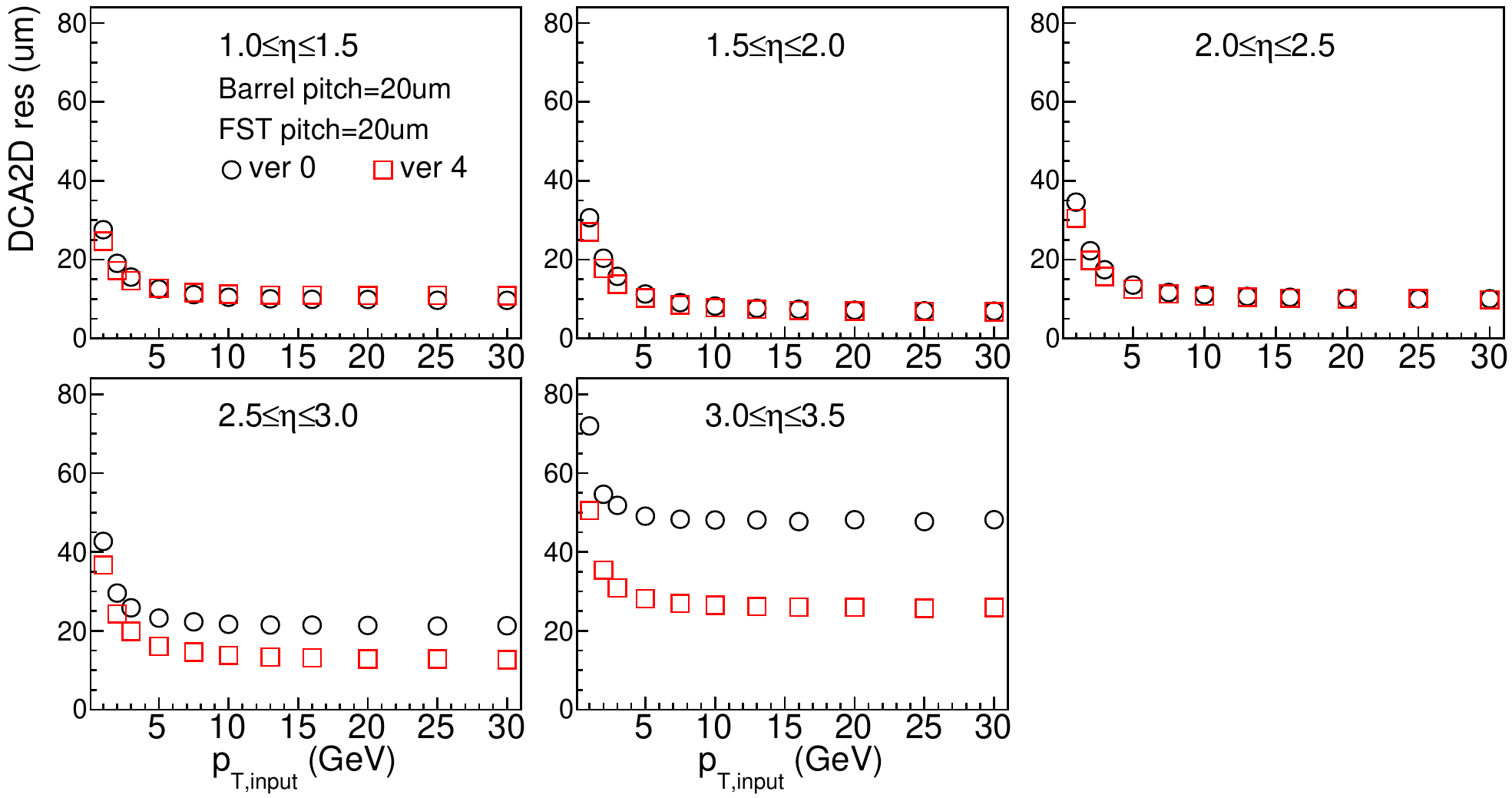}
    %\vspace{-4mm}
    \caption{$DCA2D$ resolution as a function input (true) transverse momentum from Fun4all simulation using version~0 and version~4 FST design with BeAST magnetic field.}
    \label{fig:DCA2DRes_ver0_ver4}
\end{figure}

Four distance of the closest approach (DCA) values, $DCA_r$, $DCA_\phi$, $DCA_z$ and $DCA2D$ are measured in the simulations. These DCA values are defined as
\begin{align}
    DCA_r&=(pca-vtx)\cdot p_T\text{ ,}\\
    DCA_\phi&=(pca-vtx)\times p_T\text{ ,}\\
    DCA_z&=(pca_z-vtx_z)\cdot p_z\text{ ,}\\
    DCA2D&=(pca-vtx)\cdot(p_T\times\hat{z})\text{ ,}
\end{align}
where $pca$ is the point of closest approach of the tracks to the primary vertex and $vtx$ is the reconstructed vertex. Different $DCA$ resolutions from Fun4All simulations using version~0, version~4 FST designs and BeAST are shown in Figure~\ref{fig:DCArRes_ver0_ver4} to Figure~\ref{fig:DCA2DRes_ver0_ver4}. With the additional layer of barrel detector and plane detector at far-z location, the DCA resolutions of version~4 are lower than version~0 especially at $3\leq\eta\leq3.5$.
%\FloatBarrier
%--------------------------------------------------------------------%
\subsubsection{Pixel Pitch and Silicon thickness study}\label{sec:DCA_ver4.1_4.2}
\begin{figure}[htbp]
    \centering
    \includegraphics[width=0.95\textwidth]{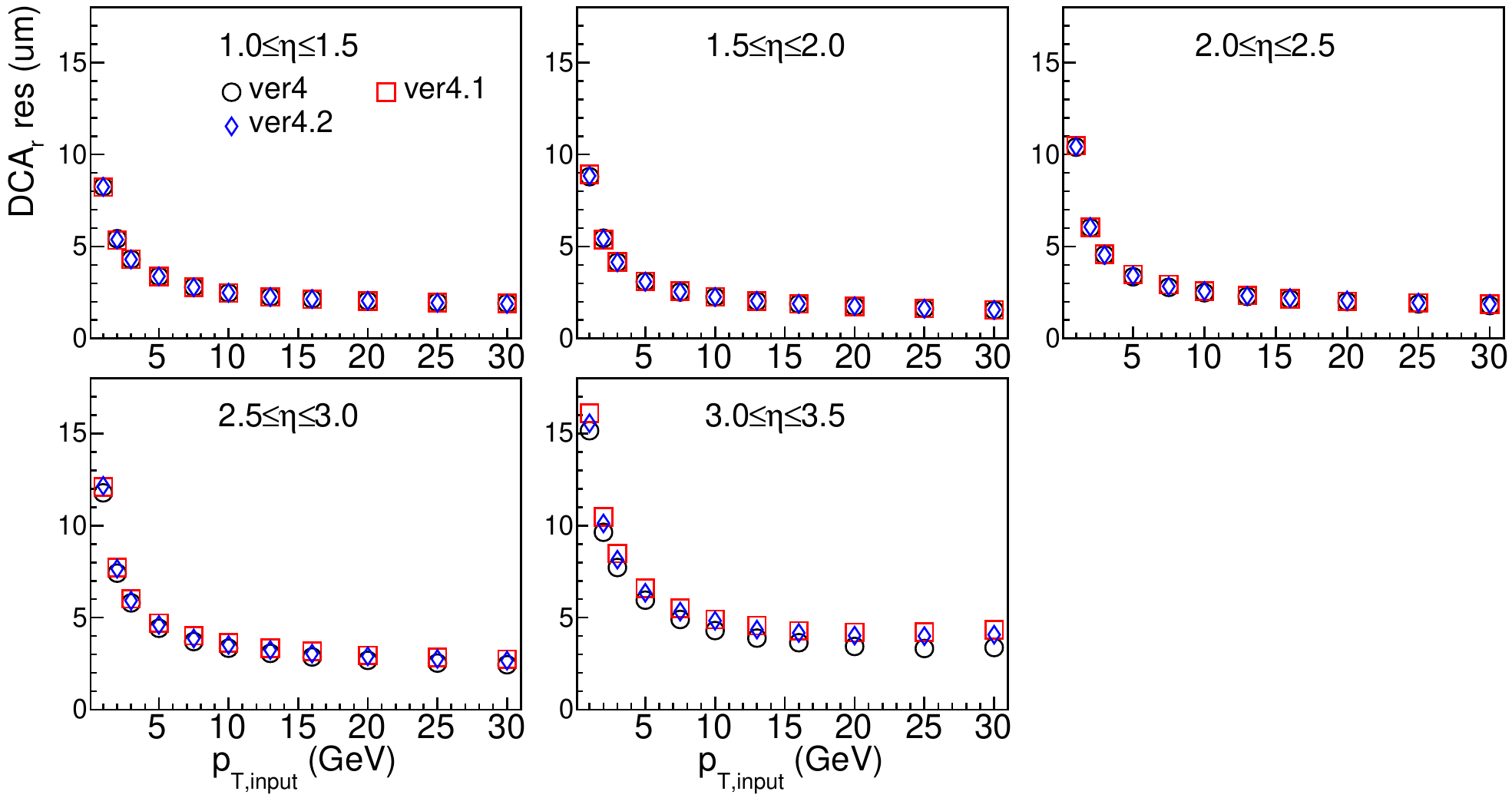}
    \caption{$DCA_r$ resolution as a function input (true) transverse momentum from Fun4all simulation using version~4, 4.1 and 4.2 FST design.}
    \label{fig:DCArRes_ver4dot1_4dot2_30GeV}
\end{figure}

\begin{figure}[htbp]
    \centering
    \includegraphics[width=0.95\textwidth]{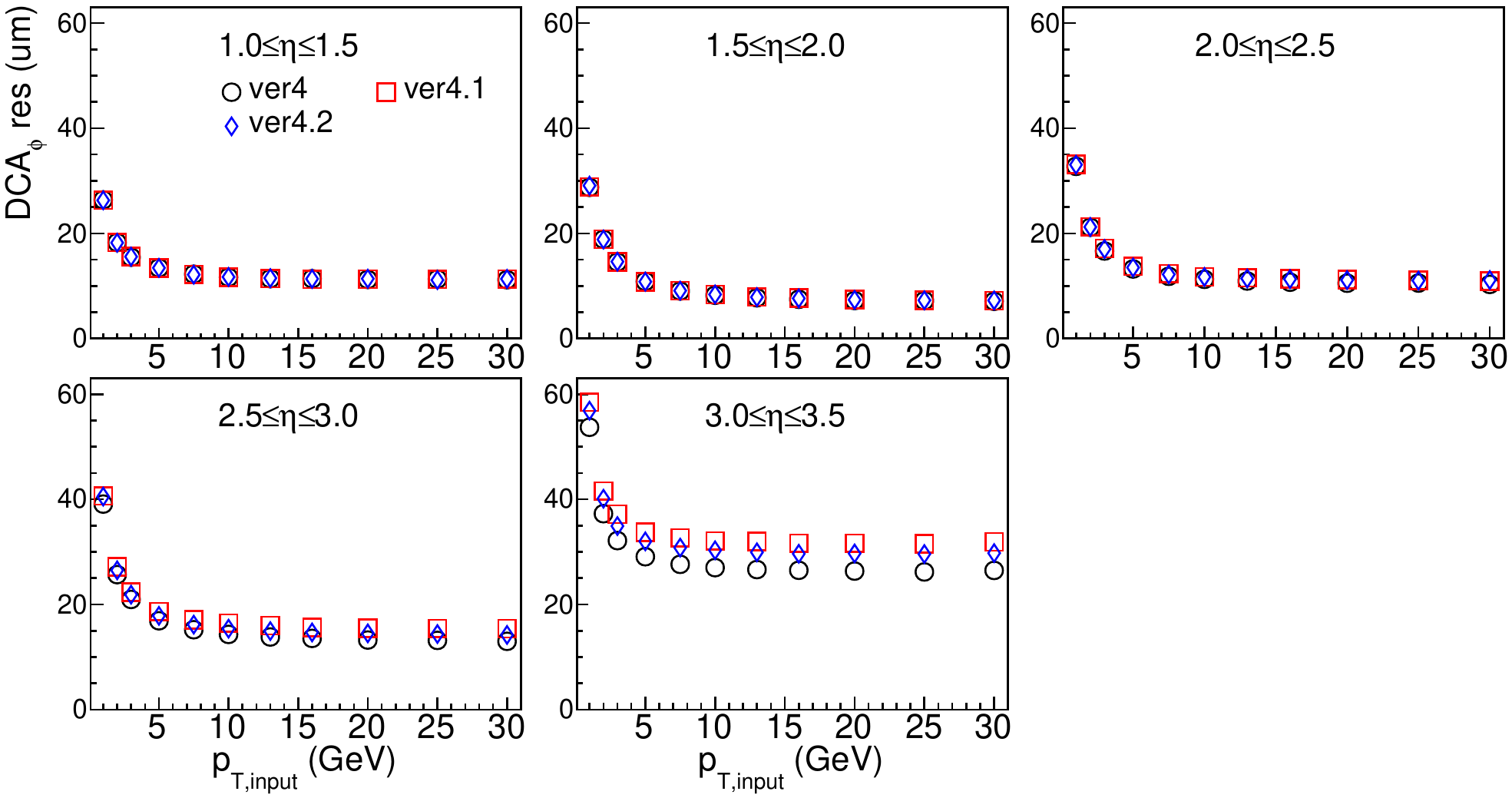}
    \caption{$DCA_\phi$ resolution as a function input (true) transverse momentum from Fun4all simulation using version~4, 4.1 and 4.2 FST design.}
    \label{fig:DCAphiRes_ver4dot1_4dot2_30GeV}
\end{figure}

\begin{figure}[htbp]
    \centering
    \includegraphics[width=0.95\textwidth]{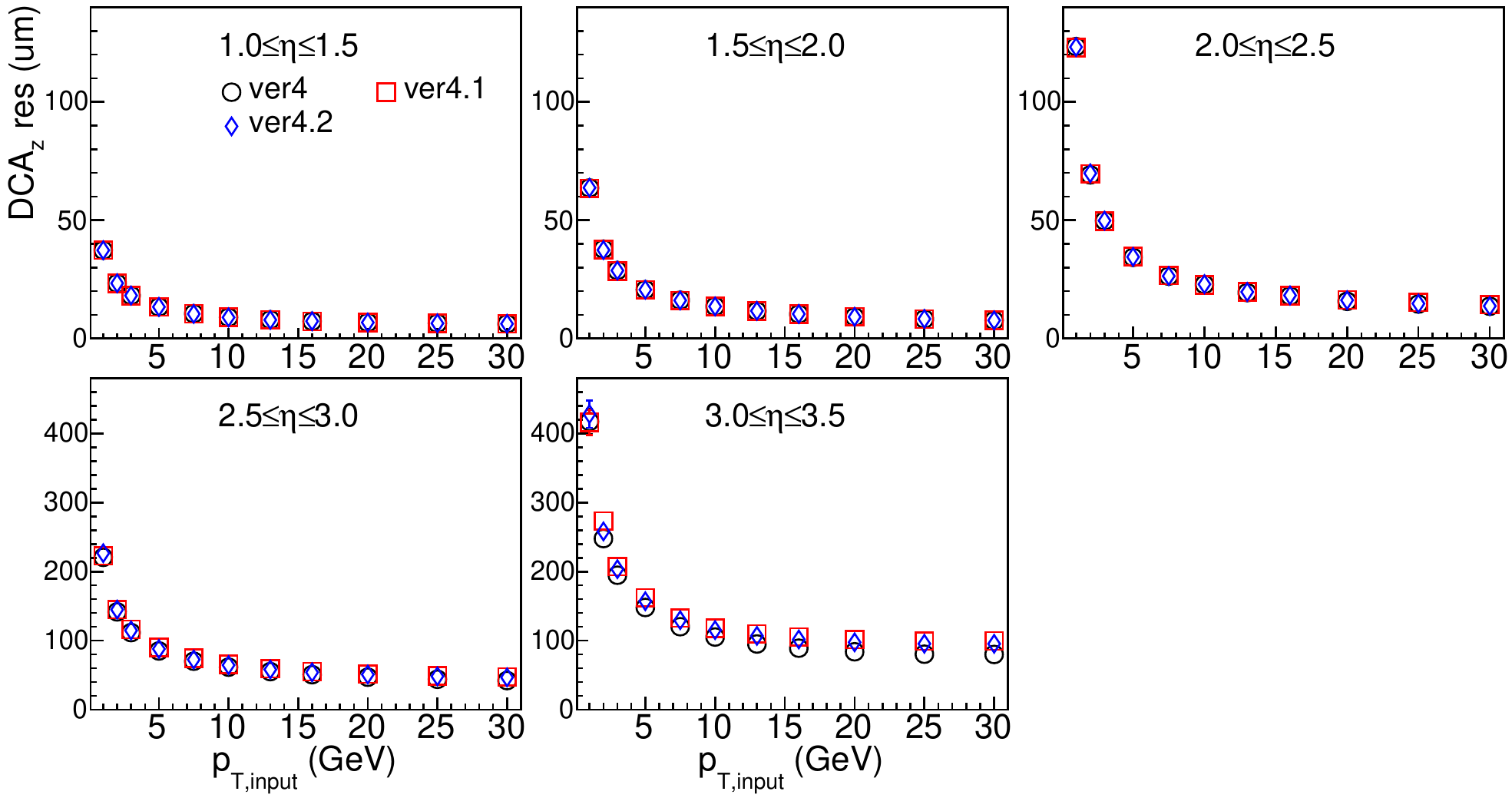}
    \caption{$DCA_z$ resolution as a function input (true) transverse momentum from Fun4all simulation using version~4, 4.1 and 4.2 FST design.}
    \label{fig:DCAzRes_ver4dot1_4dot2_30GeV}
\end{figure}

\begin{figure}[htbp]
    \centering
    \includegraphics[width=0.95\textwidth]{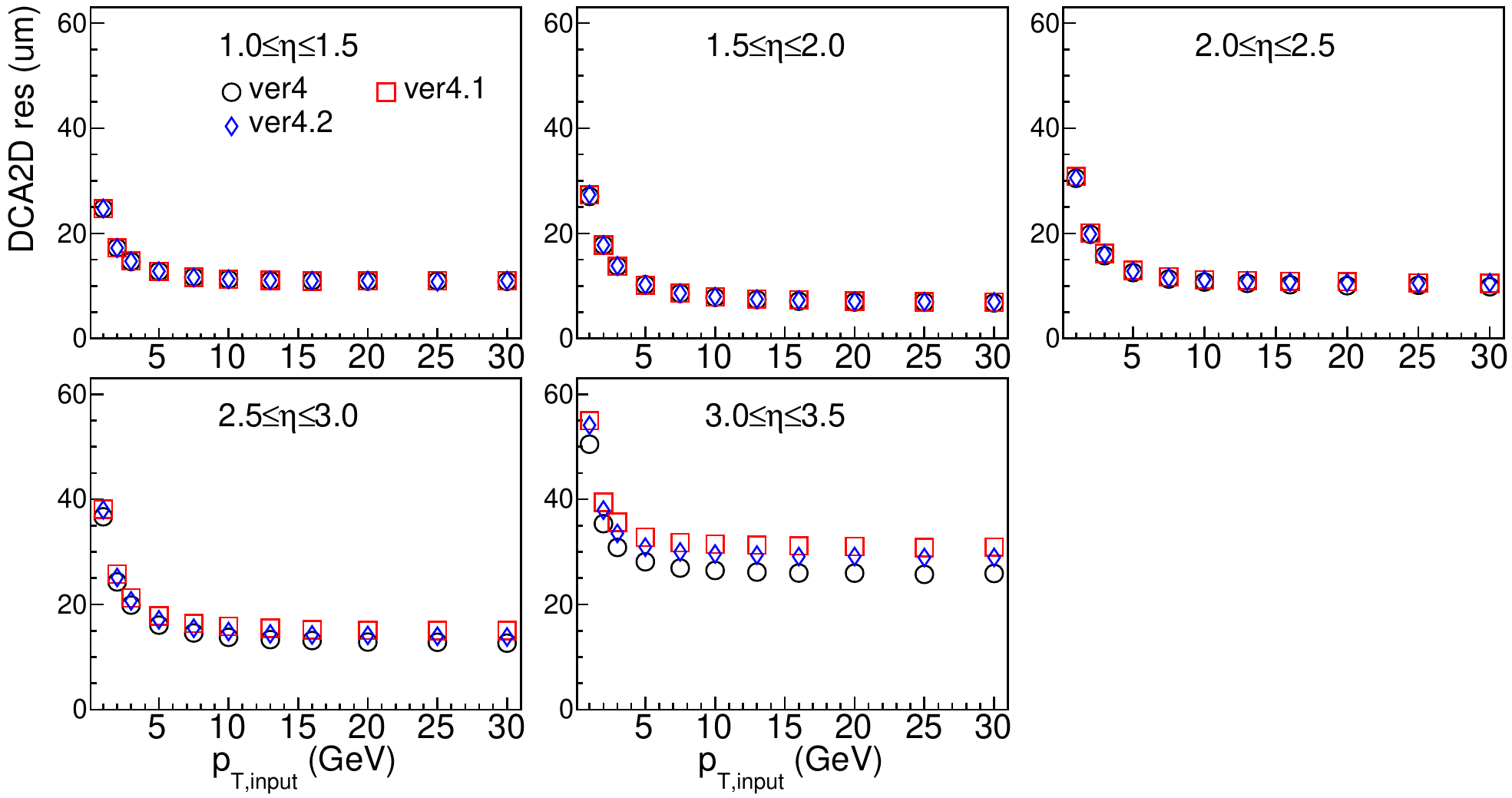}
    \caption{$DCA2D$ resolution as a function input (true) transverse momentum from Fun4all simulation using version~4, 4.1 and 4.2 FST design.}
    \label{fig:DCA2DRes_ver4dot1_4dot2_30GeV}
\end{figure}

To study effect from different silicon sensor technologies, version~4.1 and version~4.2 FST designs are introduced. Version~4.1 (Version~4.4) and version~4 use the same detector geometry as version~4, but larger pixel pitch ($36.4$~\um) and thicker silicon wafer ($100$~\um) are implemented in the last three (two) planes as shown in Table~\ref{tab:simFST_Geo}. The comparisons of DCA resolutions of version~4, version~4.1 and version~4.2 are shown in Figure~\ref{fig:DCArRes_ver4dot1_4dot2_30GeV} to~\ref{fig:DCA2DRes_ver4dot1_4dot2_30GeV}. These figures show that the changes of pixel pitch and silicon wafer thickness at the last two or three planes do not affect the resolutions at $\eta\leq2.5$ as the tracks does not pass through the last few planes. At $\eta>2.5$, the effect of pixel pitch and silicon wafer thickness become. noticeable. Version~4.1 with three planes that have a pixel pitch ($36.4$~\um) and $100$~\um thick silicon wafer fives the highest DCA resolutions, while version~$4$ gives the lowest DCA resolutions amount these three designs. The differences between Version~4 and Version~4.1 are about $2$~\um in $DCA_r$, $7$~\um in $DCA_\phi$, $20$~\um in $DCA_z$ and $5$~\um in $DCA2D$.
\FloatBarrier
\subsection{Vertex Resolution}
\begin{figure}[h]
    \centering
    \includegraphics[width=0.88\textwidth]{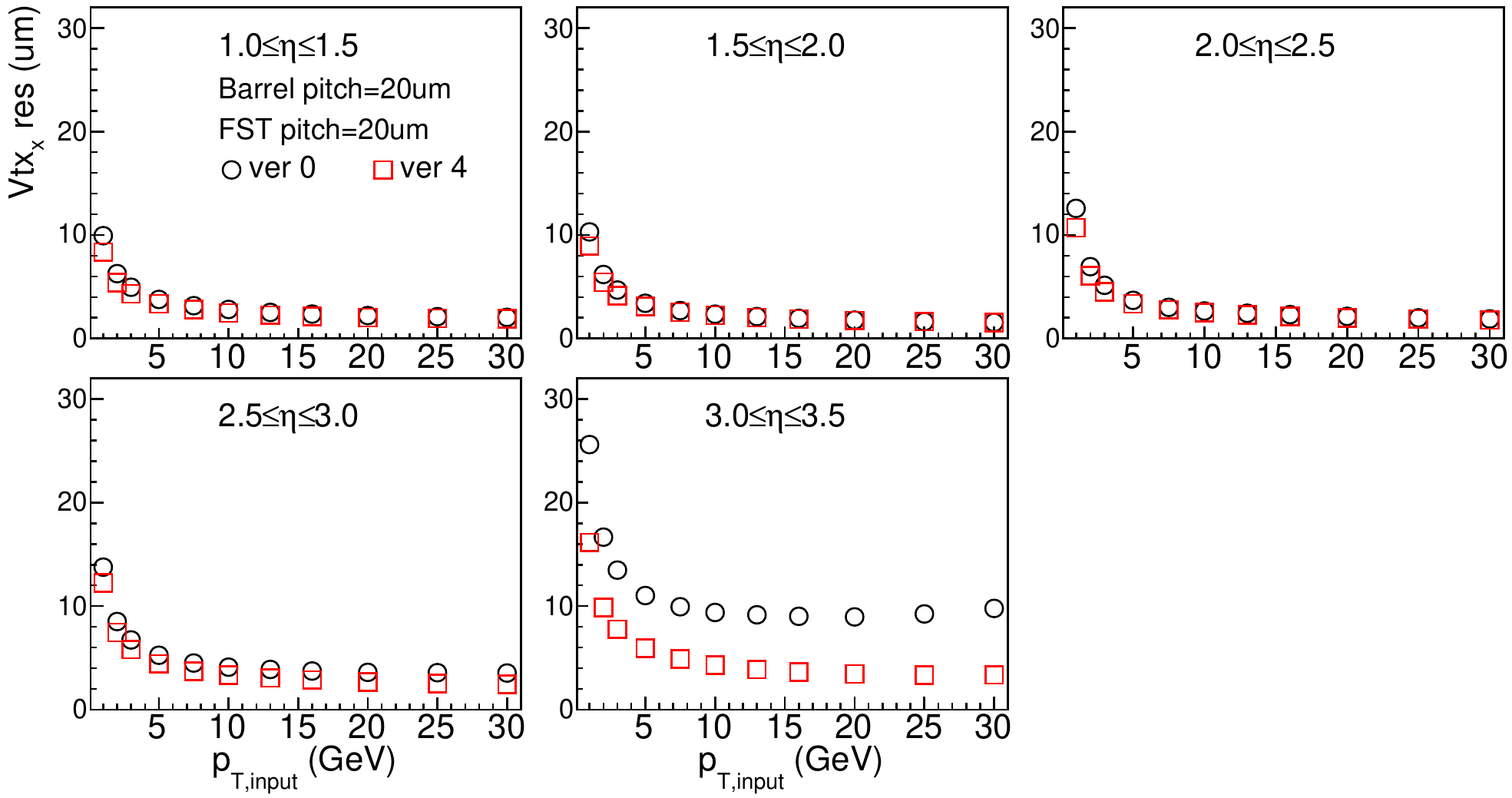}
    \vspace{-4mm}
    \caption{\label{fig:vtx_x_Res_ver0_ver4}x vertex resolution as a function input (true) transverse momentum from Fun4all simulation using version~0 and version~4 FST design with BeAST magnetic field.}
\end{figure}

\begin{figure}[h]
    \vspace{-6mm}
    \centering
    \includegraphics[width=0.88\textwidth]{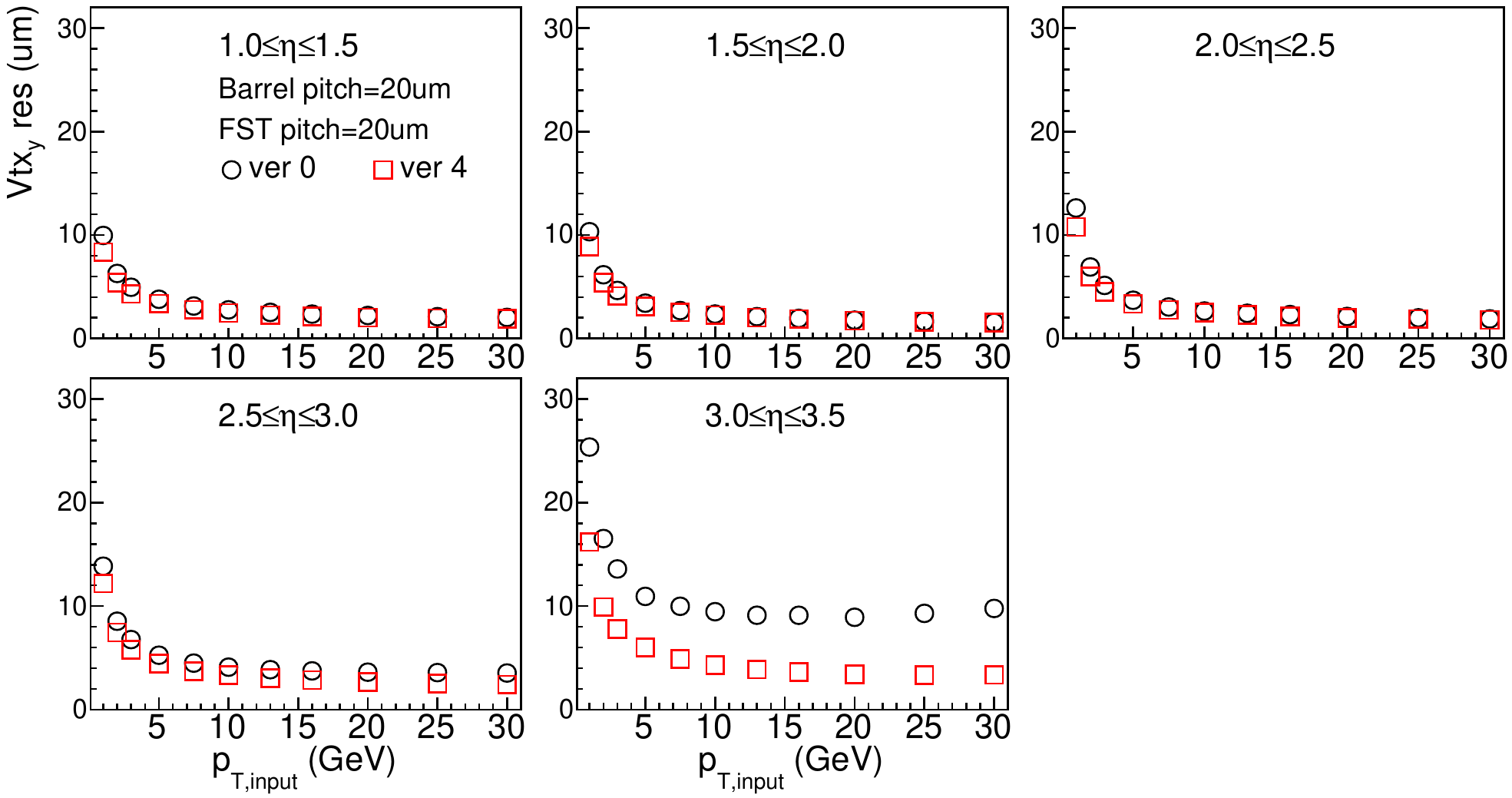}
    \vspace{-4mm}
    \caption{\label{fig:vtx_y_Res_ver0_ver4}y vertex resolution as a function input (true) transverse momentum from Fun4all simulation using version~0 and version~4 FST design with BeAST magnetic field.}
    \vspace{-8mm}
\end{figure}

\begin{figure}[h]
    \centering
    \includegraphics[width=0.88\textwidth]{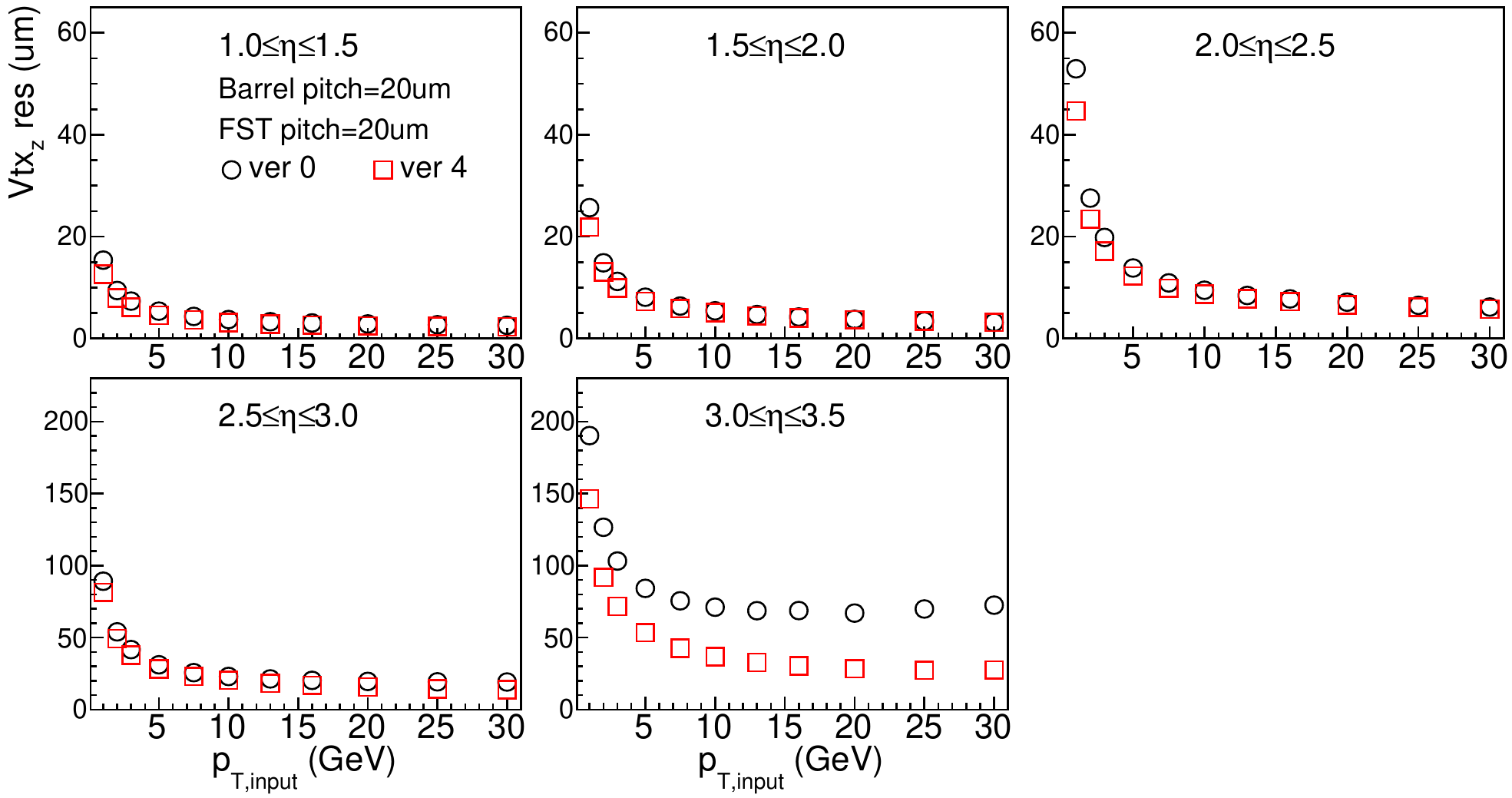}
    \vspace{-4mm}
    \caption{\label{fig:vtx_z_Res_ver0_ver4}z vertex resolution as a function input (true) transverse momentum from Fun4all simulation using version~0 and version~4 FST design with BeAST magnetic field.}
\end{figure}

Vertex resolutions in x, y and z directions are defined as the Gaussian width of the difference of the reconstructed vertex ($vtx_{x,reco}$, $vtx_{y,reco}$ and $vtx_{z,reco}$) and input (true) vertex  ($vtx_{x,input}$, $vtx_{y,input}$ and $vtx_{z,input}$), that is written as
\begin{align}
    \Delta vtx_x&=vtx_{x,reco}-vtx_{x,input}\text{ ,}\\
    \Delta vtx_y&=vtx_{y,reco}-vtx_{y,input}\text{ ,}\\
    \Delta vtx_z&=vtx_{z,reco}-vtx_{z,input}\text{ .}
\end{align}
The vertex resolutions of version~0 and version~4 FST designs from Fun4All simulation with BeAST magnetic field are shown in Figure~\ref{fig:vtx_x_Res_ver0_ver4} to Figure~\ref{fig:vtx_z_Res_ver0_ver4}. The x vertex and y vertex resolutions shown in Figure~\ref{fig:vtx_x_Res_ver0_ver4} and Figure~\ref{fig:vtx_y_Res_ver0_ver4} are consistent because of the symmetric detector geometry. The x and y vertex resolutions are below $14$~\um at $\eta<=2.5$, they increase to about $26$~\um ($16$~\um) at $1$~GeV when version~0 (version~4) design is implemented. The z vertex resolution shown in Figure~\ref{fig:vtx_z_Res_ver0_ver4} increases from about $18$~\um to about $200$~\um for version~0 design and $150$~\um for version~4 design at $1$~GeV as pseudorapidity increases. 
\FloatBarrier
%--------------------------------------------------------------------%
\subsubsection{Pixel Pitch and Silicon thickness study}
\begin{figure}[h]
    \centering
    \includegraphics[width=0.95\textwidth]{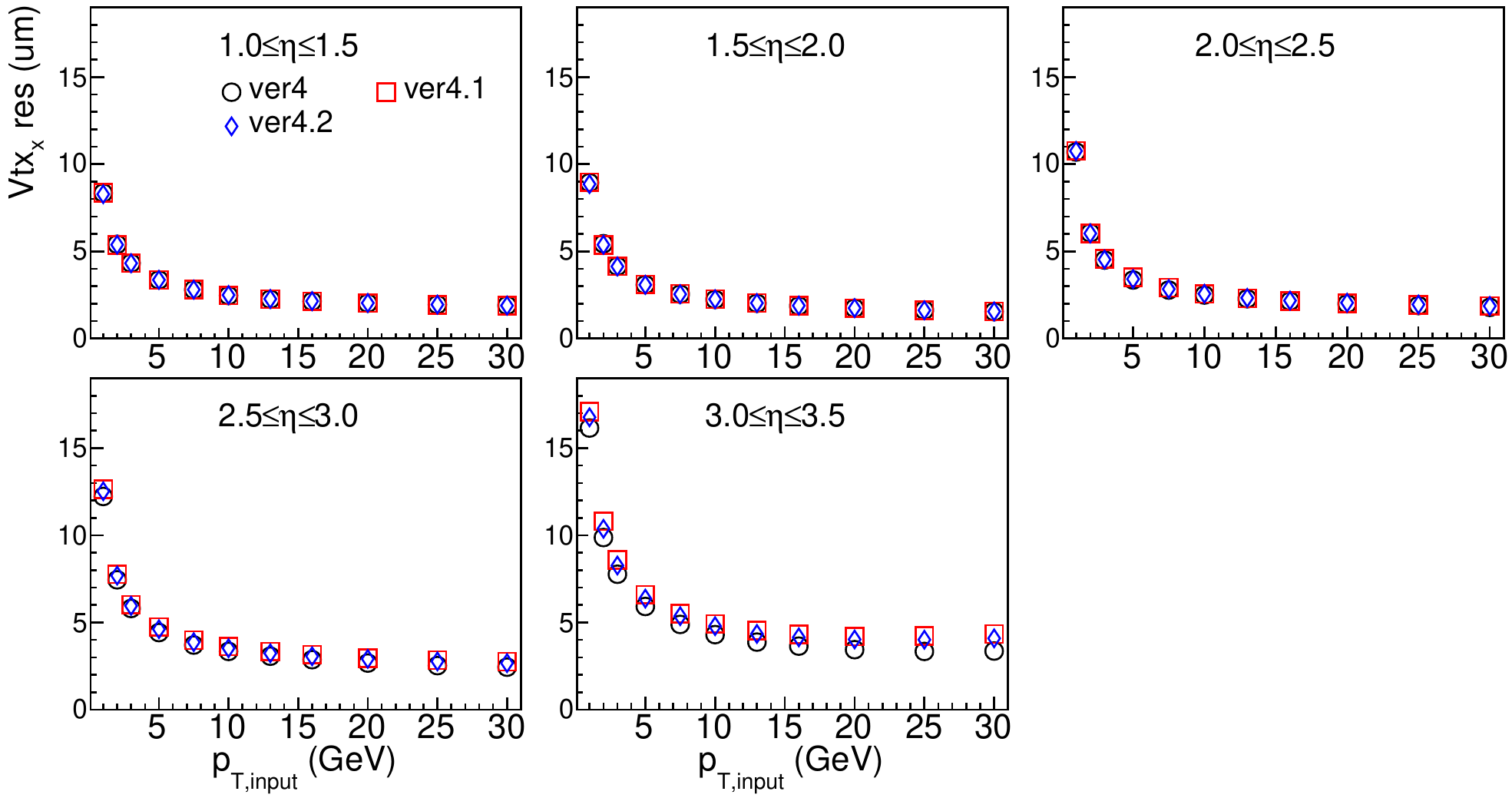}
    \caption{x vertex resolution as a function input (true) transverse momentum from Fun4all simulation using version~4, 4.1 and 4.2 FST design.}
    \label{fig:vtx_x_Res_ver4dot1_4dot2_30GeV}
\end{figure}

\begin{figure}[h]
    \centering
    \includegraphics[width=0.95\textwidth]{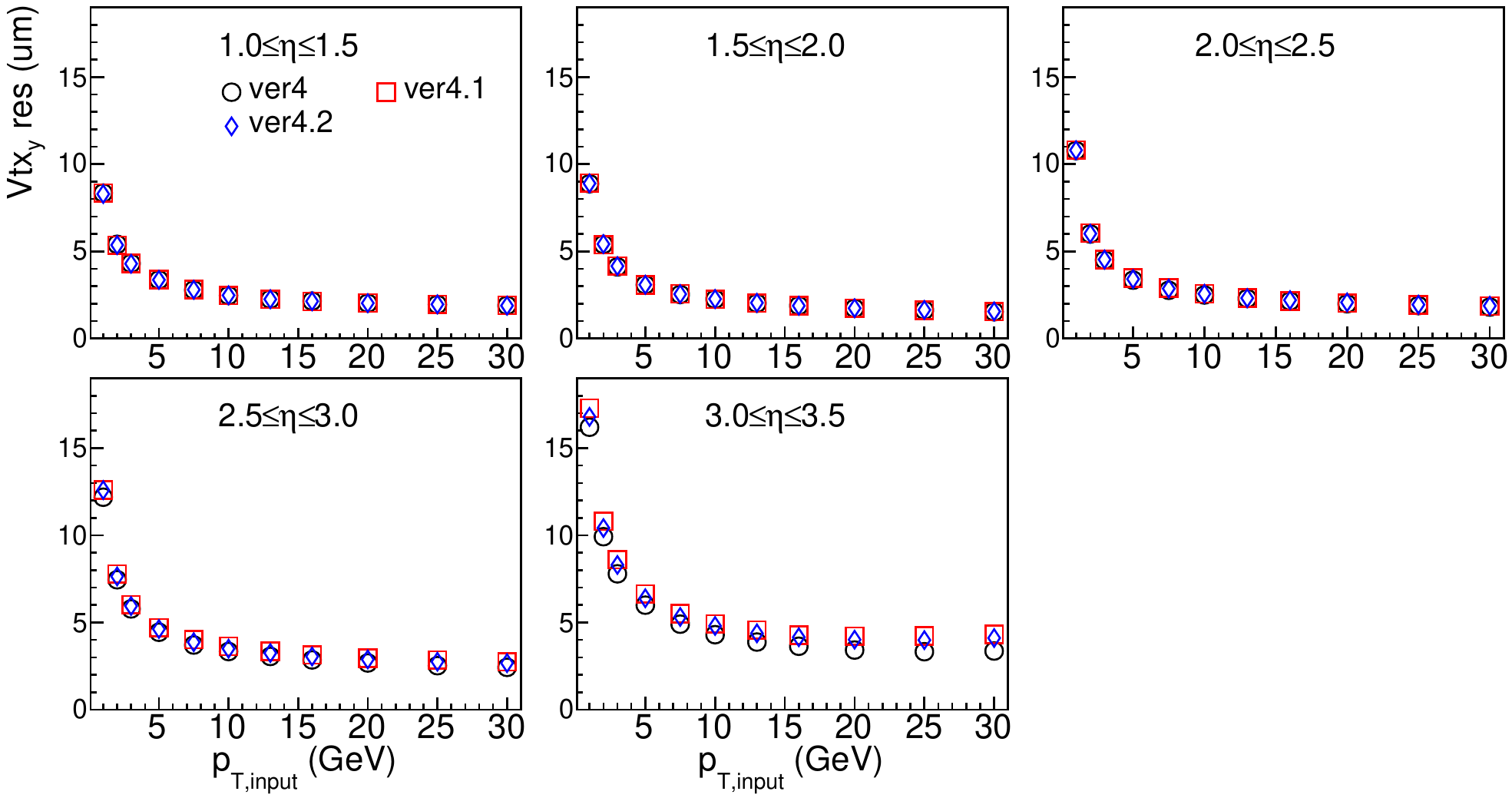}
    \caption{y vertex resolution as a function input (true) transverse momentum from Fun4all simulation using version~4, 4.1 and 4.2 FST design.}
    \label{fig:vtx_y_Res_ver4dot1_4dot2_30GeV}
\end{figure}

\begin{figure}[h]
    \centering
    \includegraphics[width=0.95\textwidth]{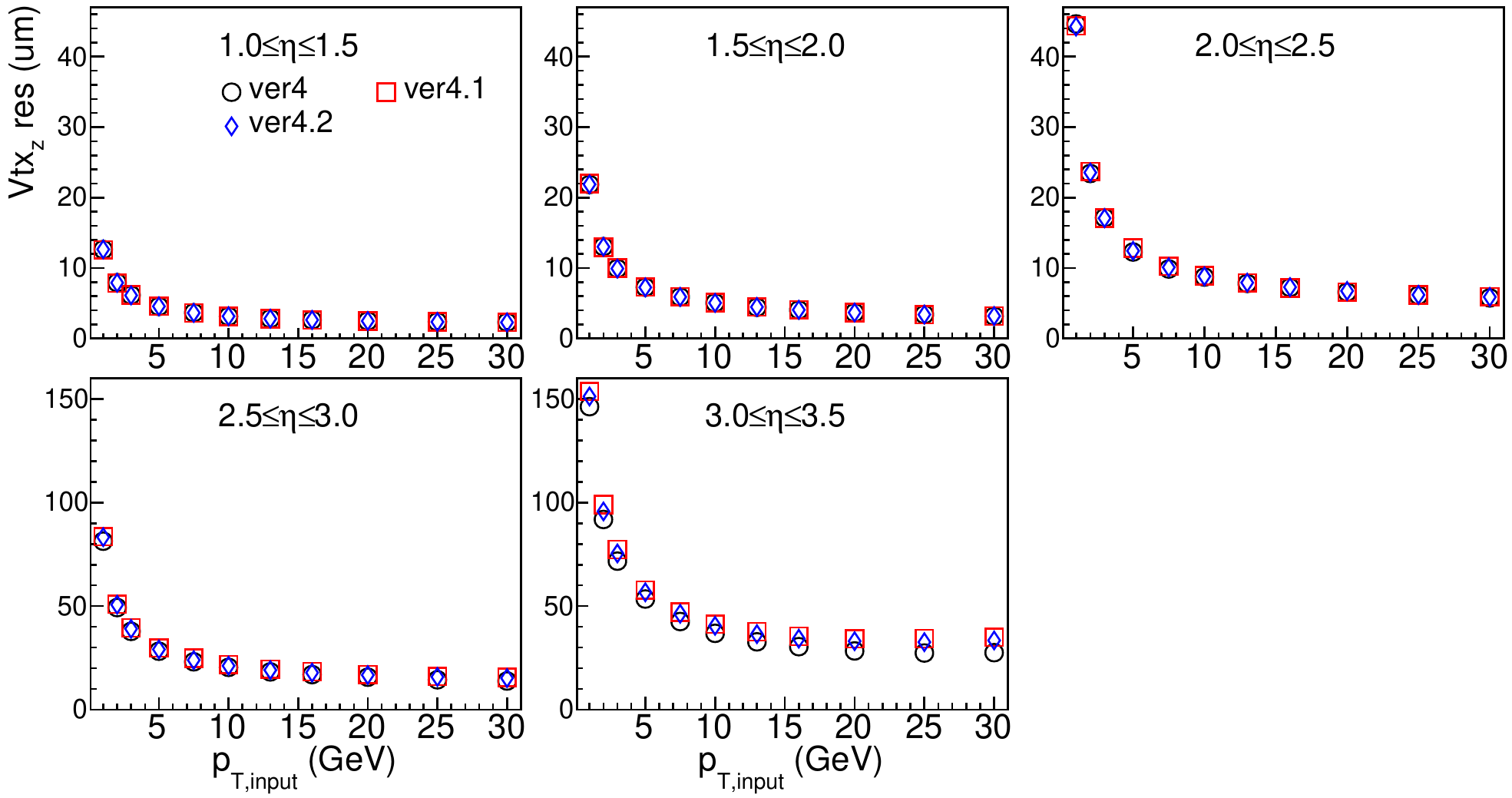}
    \caption{z vertex resolution as a function input (true) transverse momentum from Fun4all simulation using version~4, 4.1 and 4.2 FST design.}
    \label{fig:vtx_z_Res_ver4dot1_4dot2_30GeV}
\end{figure}

Version~4.1 and version~4.2, which are described in Section~\ref{sec:DCA_ver4.1_4.2} and Table~\ref{tab:simFST_Geo}, are also tested for vertex resolutions as shown in Figure~\ref{fig:vtx_x_Res_ver4dot1_4dot2_30GeV} to~\ref{fig:vtx_z_Res_ver4dot1_4dot2_30GeV}. Similar to DCA resolutions shown in Section~\ref{sec:DCA_ver4.1_4.2}, the vertex resolutions are the same between version~4, 4.1 and 4.2 at $\eta\leq3$. At $\eta>3$, the differences are about $1.5$~\um in x and y vertex resolutions, and about $10$~\um in z vertex resolution.
\FloatBarrier

\newpage
\section{Eicroot Simulation}
Tracking performance with additional GEM detectors studies were done using Eicroot simulation with a five-plane silicon detector setup. 
\subsection{Detector Setup}
The position and radii of the silicon disks are summarized in Table~\ref{tab:SiMay11}. The addition of large gaseous detectors (GEMS~\cite{gems:SAULI20162}) at large $z$ ($>1$~m) in addition to the five silicon planes was also studied and added to the tracking reconstruction in Geant4. The GEMS' geometrical constraints are summarized in Table~\ref{tab:GEMSMay11}.
\begin{table}[H]
    \centering
    \caption{\label{tab:SiMay11}Silicon detector planes geometrical configurations.}
    \begin{tabular}{|c|c|c|c|c|c|c|c|}
    \hline
     &disk 1& disk 2& disk 3& disk 4 & disk 5\\
    \hline
    r$_{1}$& 40 mm& 40 mm & 50 mm&60 cm& 65mm\\ 
    r$_{2}$& 300 mm & 350 mm& 400mm& 400mm& 440 mm\\
    \hline
    &z$_{1}$&z$_{2}$&z$_{3}$&z$_{4}$&z$_{5}$\\
    \hline
    Config 1 (default) & 48cm& 63cm& 78cm& 93cm& 113cm\\\
    Config 2& 35cm& 63cm& 78cm& 93cm& 113cm\\
    Config 3& 25cm& 63cm& 78cm& 93cm& 113cm\\
    \hline
    \end{tabular}
\end{table}
\begin{table}[h]
    \centering
    \caption{\label{tab:GEMSMay11}GEM detector planes geometrical configuration.}
    \begin{tabular}{|c|c|c|}
    \hline
    z    & r$_{1}$ &r$_{2}$ \\
    \hline
    120cm &23.4mm& 775mm \\
    264cm &23.4mm& 775mm \\
    273cm &23.4mm& 775mm \\
    280cm &23.4mm& 775mm \\
    \hline
    \end{tabular}   
\end{table}
\subsection{Momentum Resolution}\label{sec:EICroot_momRes}
\begin{figure}[h]
    \centering
    \includegraphics[width=0.9\textwidth]{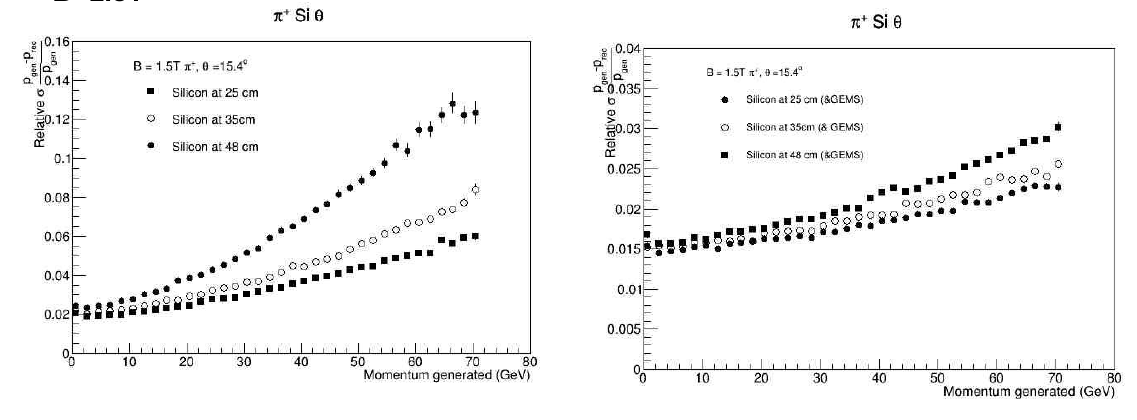}
    \caption{Relative resolution under 1.5T. The left panel shows the dispersion obtained when varying the first plane $z$ position. The right panel shows the equivalent result when adding the GEMS in addition to the 3 silicon plane detector.}
    \label{fig:firstdiskBabar}
\end{figure}

\begin{figure}[h]
    \centering
    \includegraphics[width=0.9\textwidth]{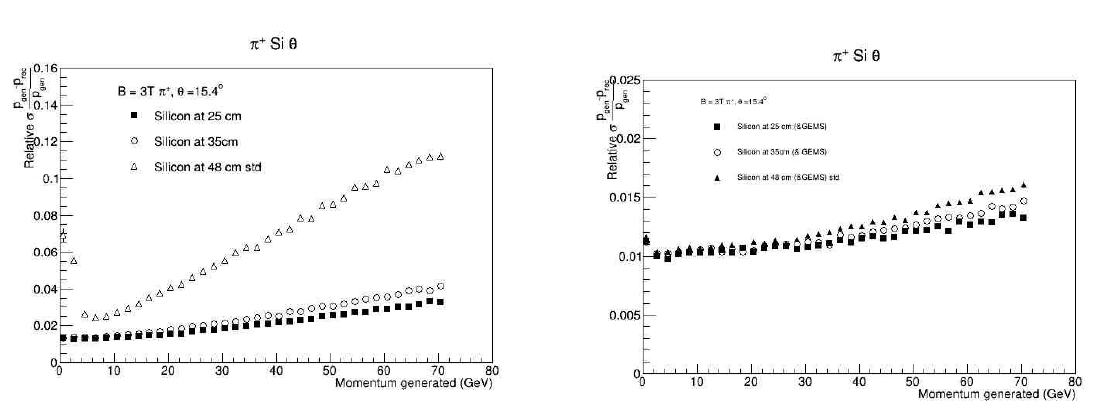}
    \caption{Relative resolution under 3T. The left panel shows the dispersion obtained when varying the first plane $z$ position. The right panel shows the equivalent result when adding the GEMs in addition to the 3 silicon plane detector.}
    \label{fig:firstdiskBeast}
\end{figure}

To evaluate the effect of the z position of the first plane (closest to the interaction point), three z positions were evaluated as described in Table~\ref{tab:SiMay11}. Figure~\ref{fig:firstdiskBabar} (left panel) shows that a first plane positioned at 25~cm from the interaction point gives the best overall relative resolution beyond momentum of 10~GeV. The right panel of the same figure on the other hand shows that the inclusion of the GEMS reduces the overall effect of this plane position. Using a magnetic field of 3T as it is shown in Figure~\ref{fig:firstdiskBeast} shows similar trends as before but with an improved resolution across all momentum. 
%
%%%%%%%%%%%%%%%%%%%
%keep
%
%%%%%%%%%%%%%%%%%%%
%keep
%--------------------------------------------------------------------%
%\subsubsection{Detector performance with electron-proton simulations}
%Electron proton collisions were simulated with 20GeV (250 GeV) electron (proton) beams. 400k events were generated at Q$^{2}>$0.8. No radiative corrections were applied. Charged pions were reconstructed using the Montecarlo generated PID information. Both the initial 3 plane forward silicon tracker and the Silicon plus GEM design was evaluated. The results are shown in Figure~\ref{fig:epcollisions}. The reconstruction versus momentum (left panel) struggles to find the pions despite the use of the montecarlo PID. Addition of the GEMS improves drastically the reconstruction. 
%As a function of $\eta$ the reconstruction is improved mostly at values above 2.

%\begin{figure}[h!]
%    \centering
%    \includegraphics[width=0.49\textwidth]{fig/EICRootSim/epvsMomenta.png}\includegraphics[width=0.49\textwidth]{fig/EICRootSim/epvseta.png}
%    \caption{Performance of track reconstruction of pions in electron pion simulations. Left panel shows as a function of the pion's momentum. Right panel is the performance versus pseudo-rapidity. Open circles assume a silicon only tracking detector while closed circles include the GEMS.}
%    \label{fig:epcollisions}
%\end{figure}

\FloatBarrier
\subsection{Sagitta measurements}
\begin{figure}[ht!]
    \centering
    \includegraphics[width=0.6\textwidth]{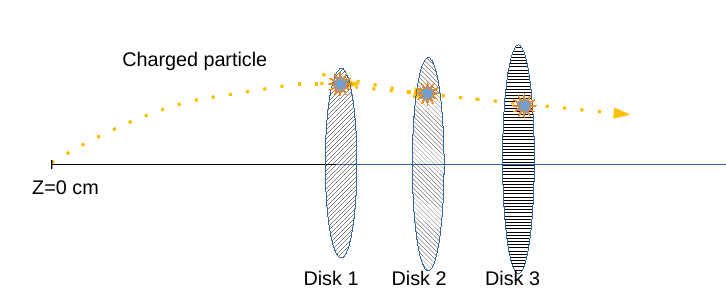}
    \caption{Illustration of the trajectory of a charged particle whose curvature depends on the magnetic field strength.}
    \label{fig:sagitta}
\end{figure}

\begin{figure}[ht!]
    \centering
    \includegraphics[width=0.45\textwidth]{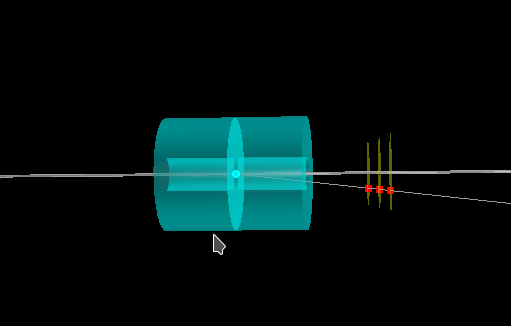}
    \includegraphics[width=0.45\textwidth]{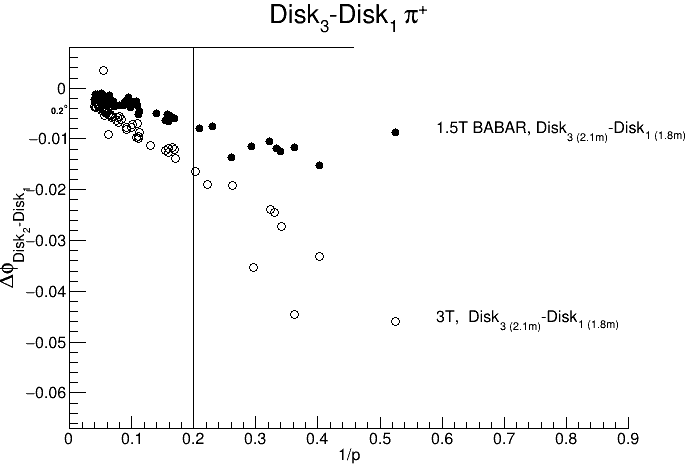}
    \caption{Left: illustration of a pion beam tracked by three-plane tracker. The first plane is positioned at 1.8~m from the vertex. Right: track curvature evaluations in two magnetic field strengths. Ordinate is the  hit azimuth position difference in radians. Abscissa is $1/p$ average reconstructed of the track. }
    \label{fig:sagitta3T-std-Z}
\end{figure}

Momentum of a charged particle is determined by the degree of curvature of a track. To evaluate the curvature one can look at the azimuth location of a single track hit in sequential or alternating plane locations as it is illustrated in Figure~\ref{fig:sagitta}. The larger the difference between the azimuth location in two planes the larger the curvature. In the following azimuth difference studies, three silicon planes were used at a starting distance of 1.8~m from the interaction region (Figure~\ref{fig:sagitta3T-std-Z} left panel). Pions were generated at a vertex position of (0,0,0) with a flat $p_{T}$ distribution ranging 2-25~GeV. The plane positions were shifted in $z$ and the azimuth difference was re-evaluated. Figure~\ref{fig:sagitta3T-std-Z} (right) shows the difference of azimuth position between the third and first plane as a function of $1/p$ . 
\begin{figure}[ht!]
    \centering
    \includegraphics[width=0.45\textwidth]{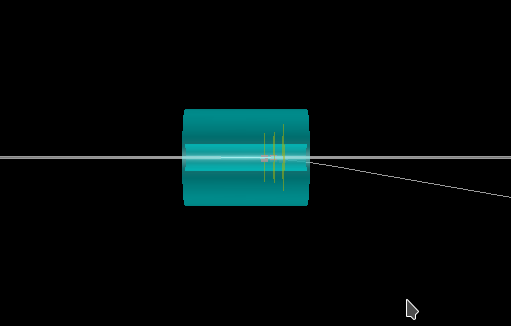}
     \includegraphics[width=0.45\textwidth]{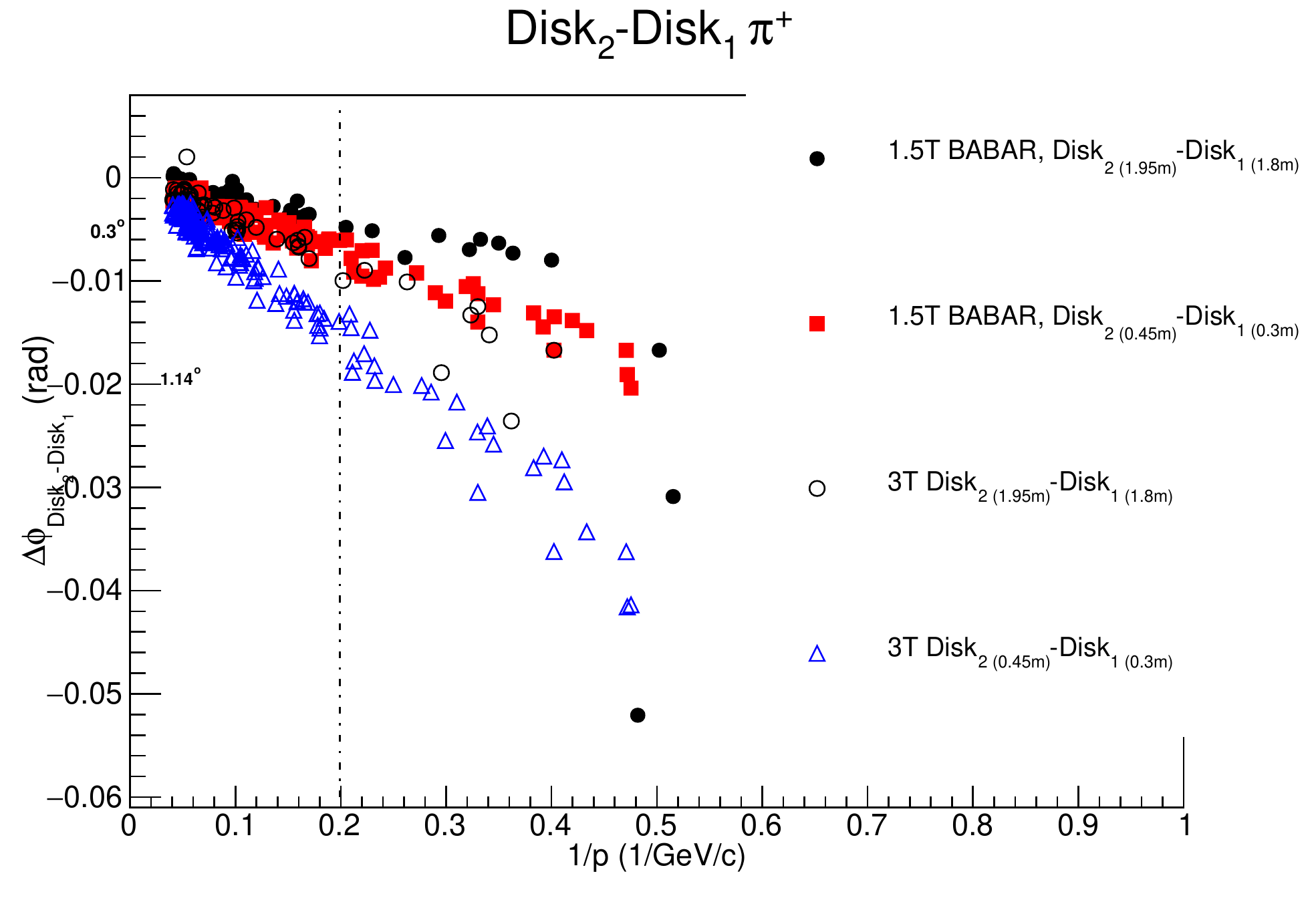}
    \caption{Azimuth difference between two adjacent planes sitting at 1.8~m and 1.95~m (black). Re-evaluation after a shifting closer to $z$ both planes. Two magnetic fields were considered with 3T (blue triangles and open black circles) giving a larger difference as expected.} 
    \label{fig:sagitta3T-300mm}
\end{figure}
\begin{figure}[ht!]
    \centering
    \includegraphics[width=0.45\textwidth]{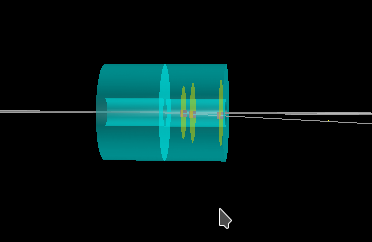}
    \includegraphics[width=0.45\textwidth]{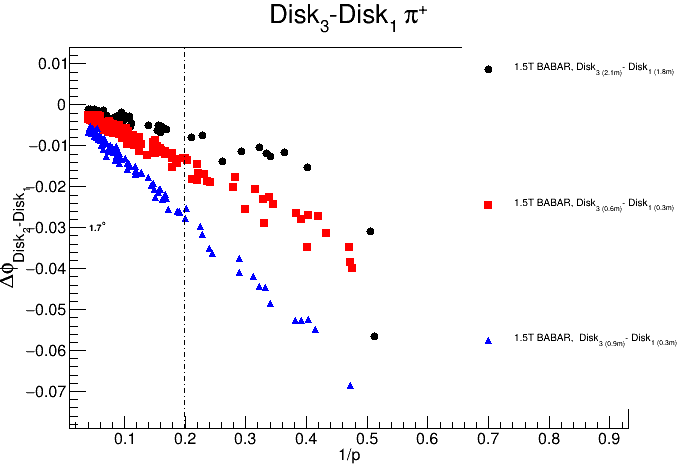}
    \caption{Comparison of azimuth differences between the third and first plane assuming equidistant planes as presented before (red and black markers) and with non equidistant planes (blue markers) 0.~3m and 0.9~m}
    \label{fig:AsymBabarField}
\end{figure}

\begin{figure}
    \centering
    \includegraphics[width=0.5\textwidth]{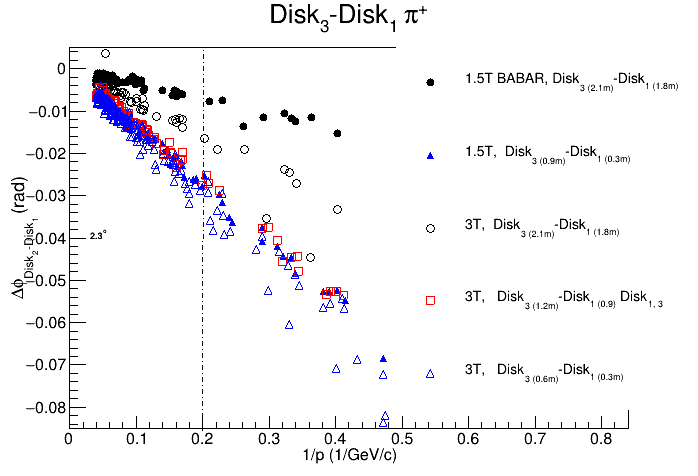}
    \caption{Summary of results}
    \label{fig:babarbeastSagitta}
\end{figure}

Figure~\ref{fig:sagitta3T-300mm} shows the results of the difference in azimuth of two adjacent planes while moving the planes closer to the vertex. Plane~1 was shifted from 1.8~m to 0.3~m while plane~2 was shifted from 1.95~m to 0.45~m. The two-plane configurations considered before assume equidistant silicon planes. An asymmetric plane spacing was also studied and the results are given in Figure~\ref{fig:AsymBabarField}. Finally, Figure~\ref{fig:babarbeastSagitta}, which summarizes all results indicates that equivalent results can be obtained with both field strengths if different $z$ positions are considered in each case.

%\subsubsection{Sagitta results for GEM detectors}
%Addition of GEM detectors at large z ($z>250$cm) while retaining silicon detectors below 115~cm. Four GEM detectors were added to the study of curvature as it was done in the previous section. The results of these studies are shown in figs.

\begin{figure}
    \centering
    \includegraphics[width=0.6\textwidth]{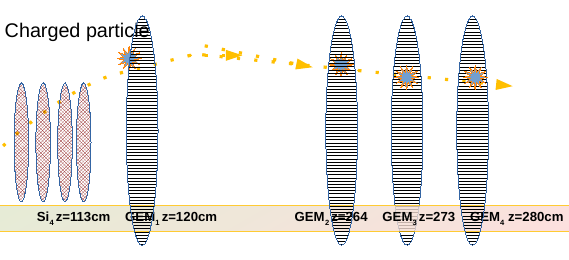}
    \caption{Cartoon illustrating the relative position of the GEM detectors with respect to the silicon detectors}
    \label{fig:siGemsDrawingSagitta}
\end{figure}
\subsection{Pointing resolution evaluation as a function of disk z position}
The distance from the interaction point to the vertex tracker has a significant effect on the pointing resolution from tracks back to the collision vertex.  In general, a tracking detector with angular resolution $d\theta$ that is a distance $r$ from the vertex will produce a pointing uncertainty $rd\theta$ when projecting tracks back to the vertex.  Therefore, minimizing the distance to the vertex as well as enhancing the angular resolution are both necessary for optimal vertexing performance.  Momentum resolution improves as particles are tracked over a longer lever arm.  However, the silicon area required to cover a given angular acceptance increases with the square of the distance from the vertex, so minimizing that distance also reduces the detector area and therefore cost and complexity.  A compromise between lever arm and detector size must therefore be made.

Simulations are used to study the track pointing resolution to the vertex as a function of the silicon plane positions. Five planes of silicon tracker are placed at 50, 63, 78, 93, and 113~cm from the vertex. Tracks are launched at 22.5$^{\circ}$ above the $z$ axis, and the hit positions at the planes are smeared with a 30~\um Gaussian resolution.  The smeared points are fit with a straight line, and the resulting tracks are projected back to the collision vertex $z$ position. The distance from the projected track to the actual origin are collected into histograms and fit with a Gaussian. The width of that Gaussian is taken as the vertex pointing resolution for each detector configuration.

\begin{figure}[H]
    \centering
    \hspace{\fill}
    \includegraphics[width=0.4\textwidth]{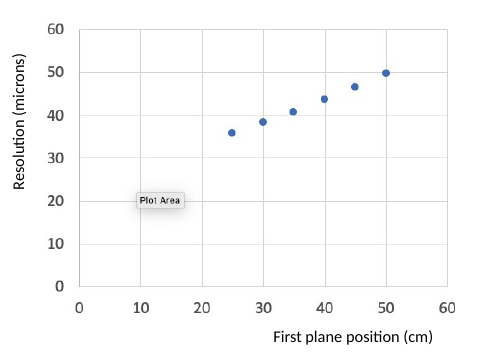}
    \hfill
    \includegraphics[width=0.42\textwidth]{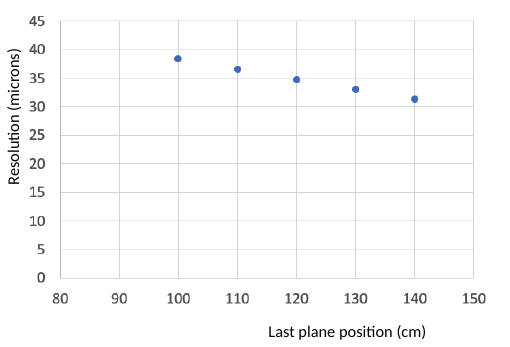}
    \hspace{\fill}
    \caption{Effects of moving the first (left) and last (right) planes in the tracker along the $z$ axis}
    \label{fig:DCA2}
\end{figure}

In the nominal case, where the first plane of the tracker is 50~cm from the vertex, a vertex pointing resolution of 50~\um is found. To examine the effect of distance on the pointing resolution, the position of the first plane is varied while the positions of the other planes are held constant. The resulting resolutions are shown on the right of Figure~\ref{fig:DCA2} which shows that a vertex pointing resolution of approximately 35~\um is obtained when the first plane is 25~cm from the vertex.

The effect of the position of the last plane was also evaluated, with results shown on the right of Figure~\ref{fig:DCA2}. Here positions of the first four planes were held constant at 25, 63, 78, and 93~cm from the vertex, and the last plane was moved. Plot on the right of Figure~\ref{fig:DCA2} shows that the resolution varies by 4~\um from approximately 38~\um to 32~\um when the last plane is moved away from $z$=100~cm to 140~cm. This would, however, necessitate an increase in size of the last detector by a factor of 2 to have the same angular acceptance. As the position of this last plane has relatively little effect on the vertex pointing resolution, detailed studies of the trade-off between plane position, area and momentum resolution are needed to fully evaluate this detector design.

\newpage
\section{Heavy Flavor and Jet Physics studies}
\label{sec: phy}
\begin{figure}[H]
\centering
\includegraphics[width=0.98\textwidth]{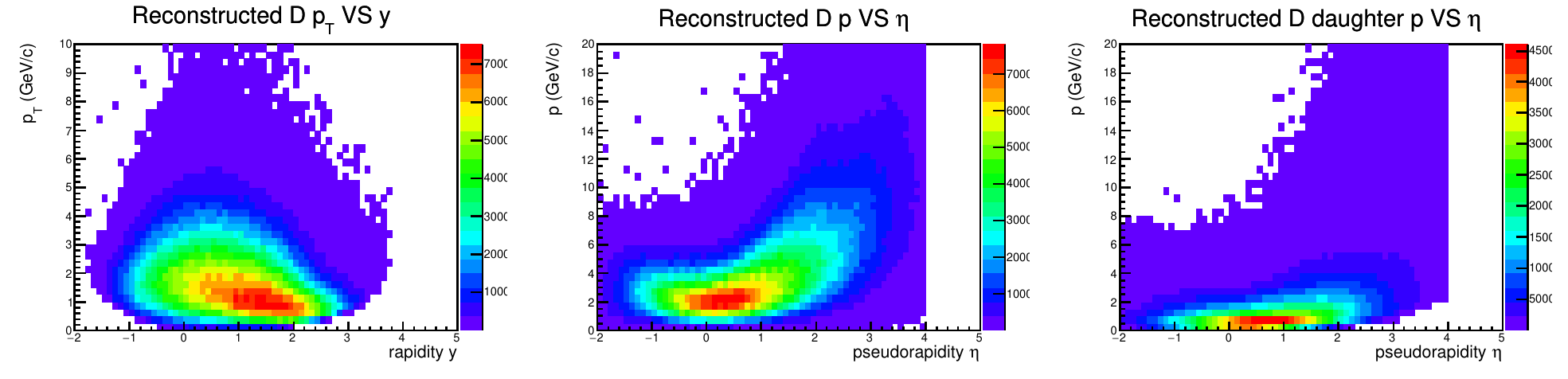}
\caption{\label{fig:recD_kin} Kinematic distributions for reconstructed D-meson and its daughters in 10 $fb^{-1}$ 10 GeV electron and 100 GeV protons collisions. Reconstructed D-meson $p_{T}$ VS rapidity y is shown in the left. Reconstructed D-meson $p_{T}$ VS pseudorapidity $\eta$ is shown in the middle. D-meson decayed daughter $p_{T}$ VS pseudorapidity $\eta$ is shown in the right.}
\end{figure}

As shown in the various kinematic distributions of heavy flavor hadrons and their decay daughters in section \ref{sec: kin}, majority of the final state particles produced at the future EIC are within pseudorapidity region from -2 to 4 (see example in Figure~\ref{fig:recD_kin} which use one EIC collision combination). The EIC recommend collision energy combinations are listed below:

\begin{itemize}
\item $E_{e}$ = 5 GeV, $E_{p/A}$ = 41 GeV 
\item $E_{e}$ = 5 GeV, $E_{p}$ = 100 GeV 
\item $E_{e}$ = 10 GeV, $E_{p/A}$ = 100/110 GeV 
\item $E_{e}$ = 18 GeV, $E_{p/A}$ = 100/110 GeV 
\item $E_{e}$ = 18 GeV, $E_{p}$ = 275 GeV 
\end{itemize}

\begin{figure}[H]
\centering
\includegraphics[width=0.9\textwidth]{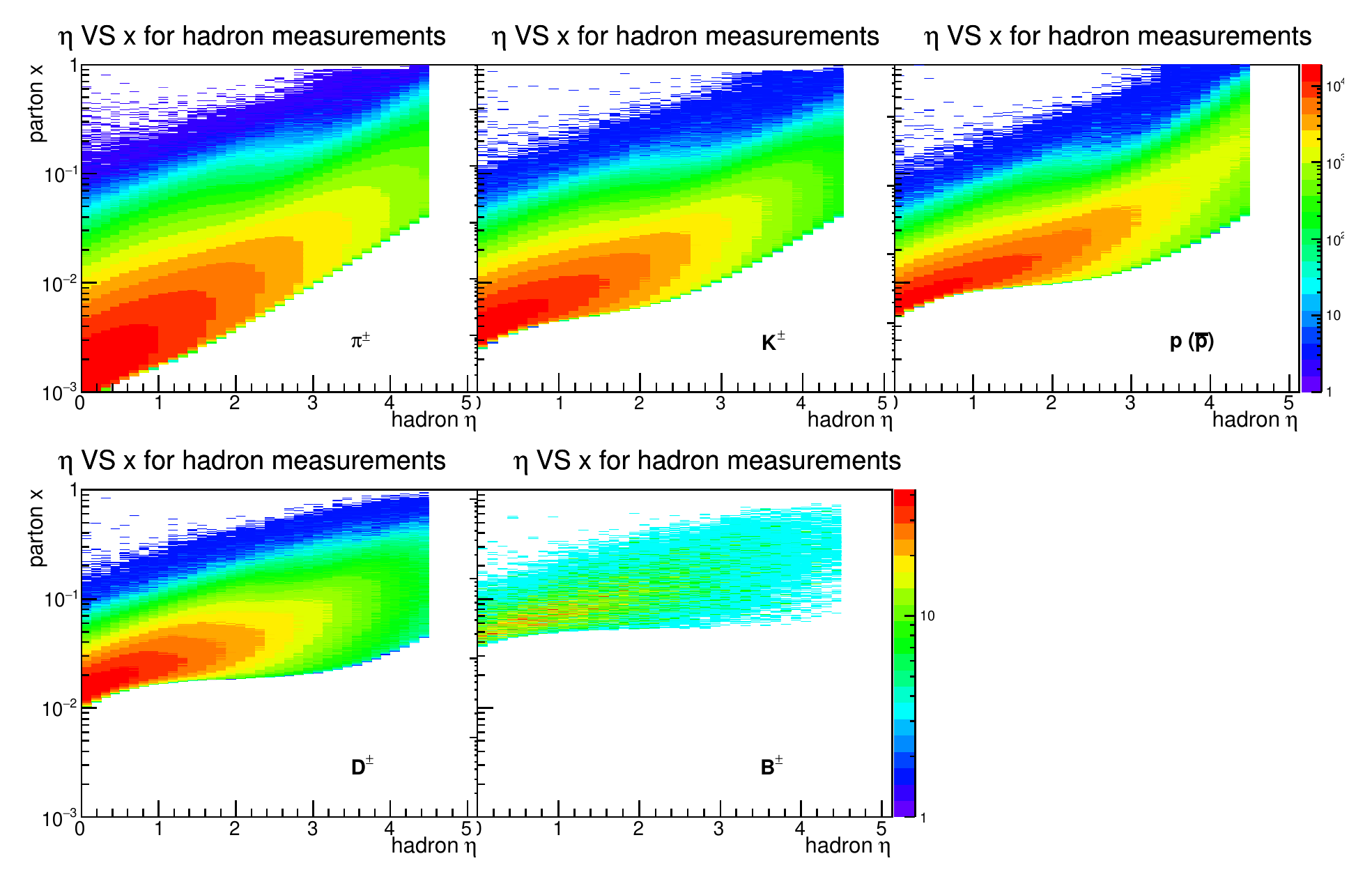}
\caption{\label{fig:gen_had_etax} Pseudorapidity versus Bjorken-x for different flavor hadrons at generation level in 10 $fb^{-1}$ 10 GeV electron and 100 GeV protons collisions.}
\end{figure}

We use one collision energy combination which is the 10 GeV electron and 100 GeV protons collisions as our reference system to study the heavy flavor and jet measurements. This energy can access a wide Bjorken-x region especially for the high-x region as shwon in Figure ~\ref{fig:gen_had_etax}. Initial heavy flavor studies have been carried out with detector performance in fast simulation \cite{Li:2020sru,Li:2020hp}. In the following sections, we will start with the open heavy flavor hadron studies (see section \ref{sec: OHF_had}), then followed by the heavy flavor jet tagging and jet angularity studies (in section \ref{sec: OHF_jet}). Beyond the normal heavy flavor and jet observable, new exotic states such as the X(3872) will be explored at the future EIC. Section \ref{sec: exotic} will introduce the ongoing work of the exotic states.

\begin{figure}[H]
\begin{center}
\includegraphics[width=0.33\textwidth]{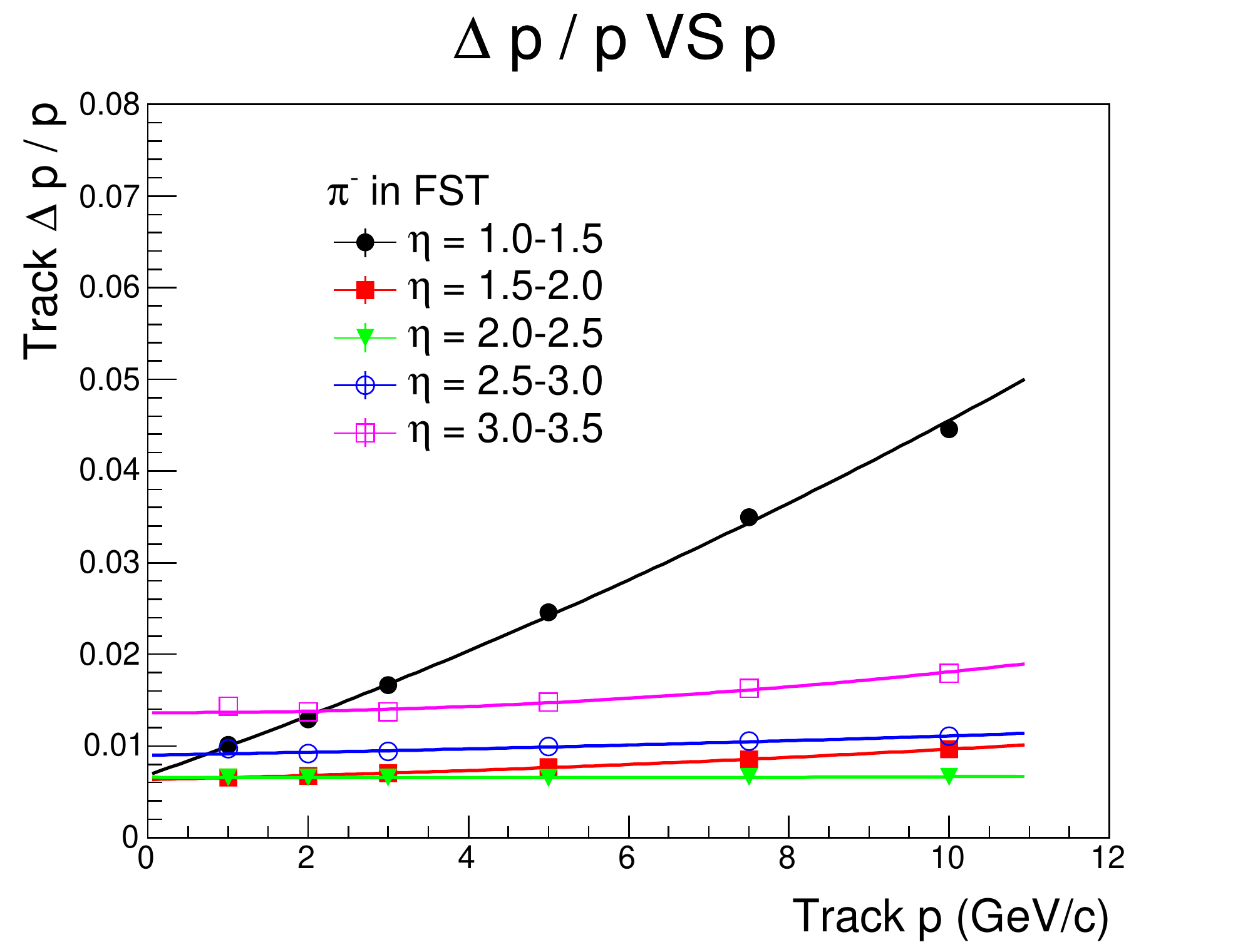}
\includegraphics[width=0.33\textwidth]{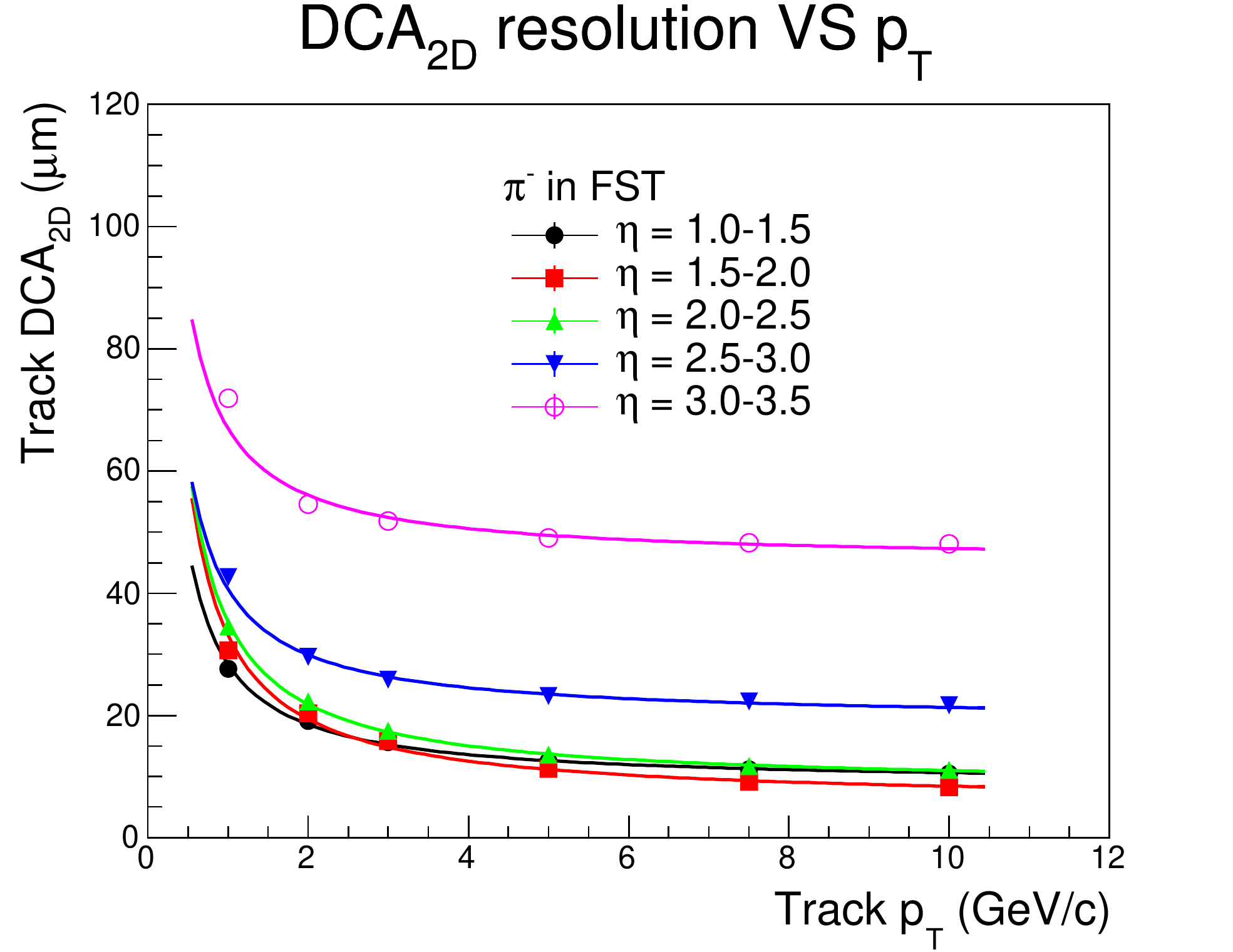}
\includegraphics[width=0.33\textwidth]{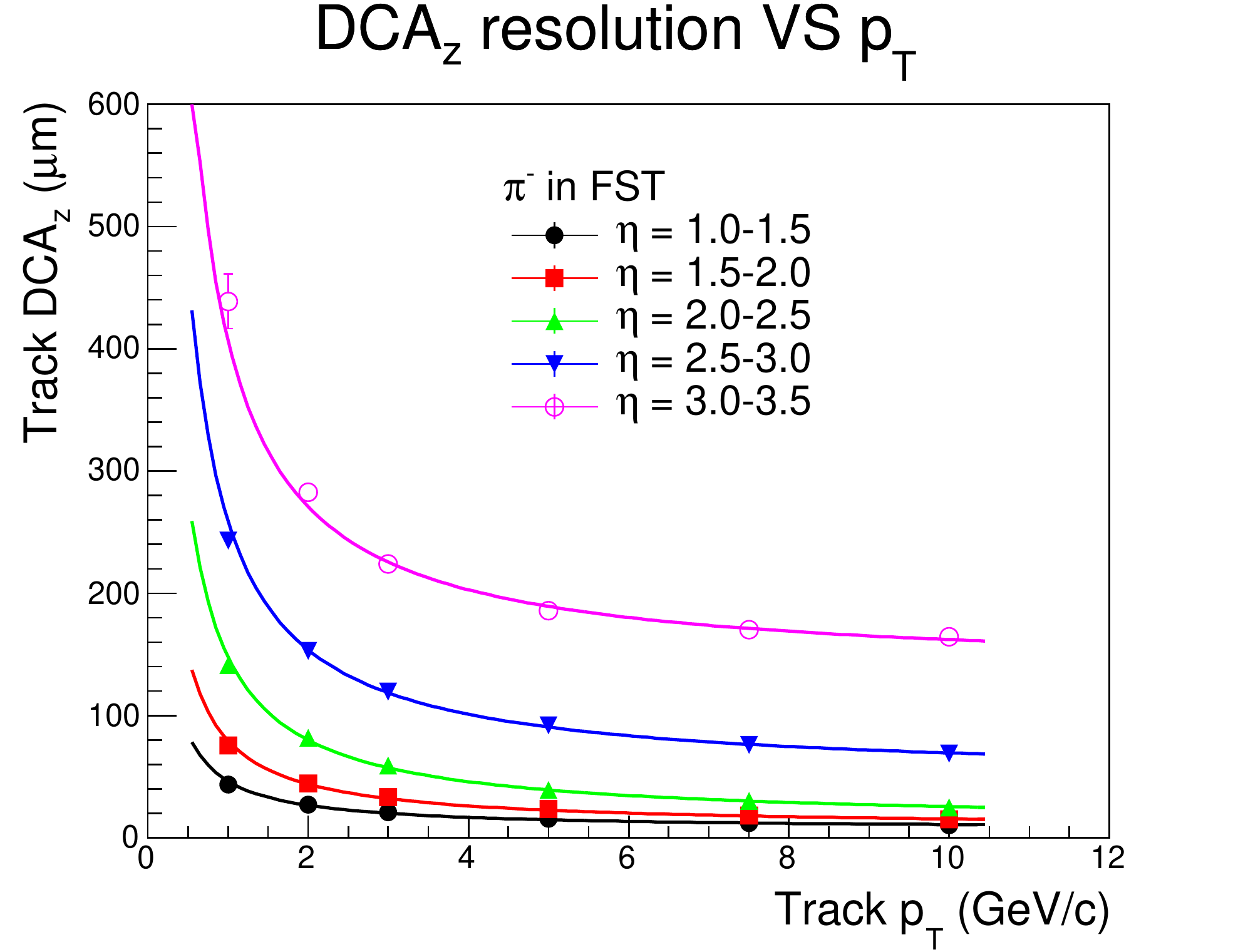}
\caption{\label{fig:fst_trk_fit0}Tracking performance for FST version 0 design with the Beast magnetic field. Pixel pitch for both barrel layers and forward planes are selected at 20 $\mu m$. Left panel shows the momentum dependent momentum resolution in different pseudorapidity regions. Middle panel shows the transverse momentum dependent $DCA_{2D}$ resolution and the right panel shows the transverse momentum dependent $DCA_{z}$ resolution in the associated pseudorapidity regions.}
\end{center}
\end{figure}

\begin{figure}[H]
\begin{center}
\includegraphics[width=0.33\textwidth]{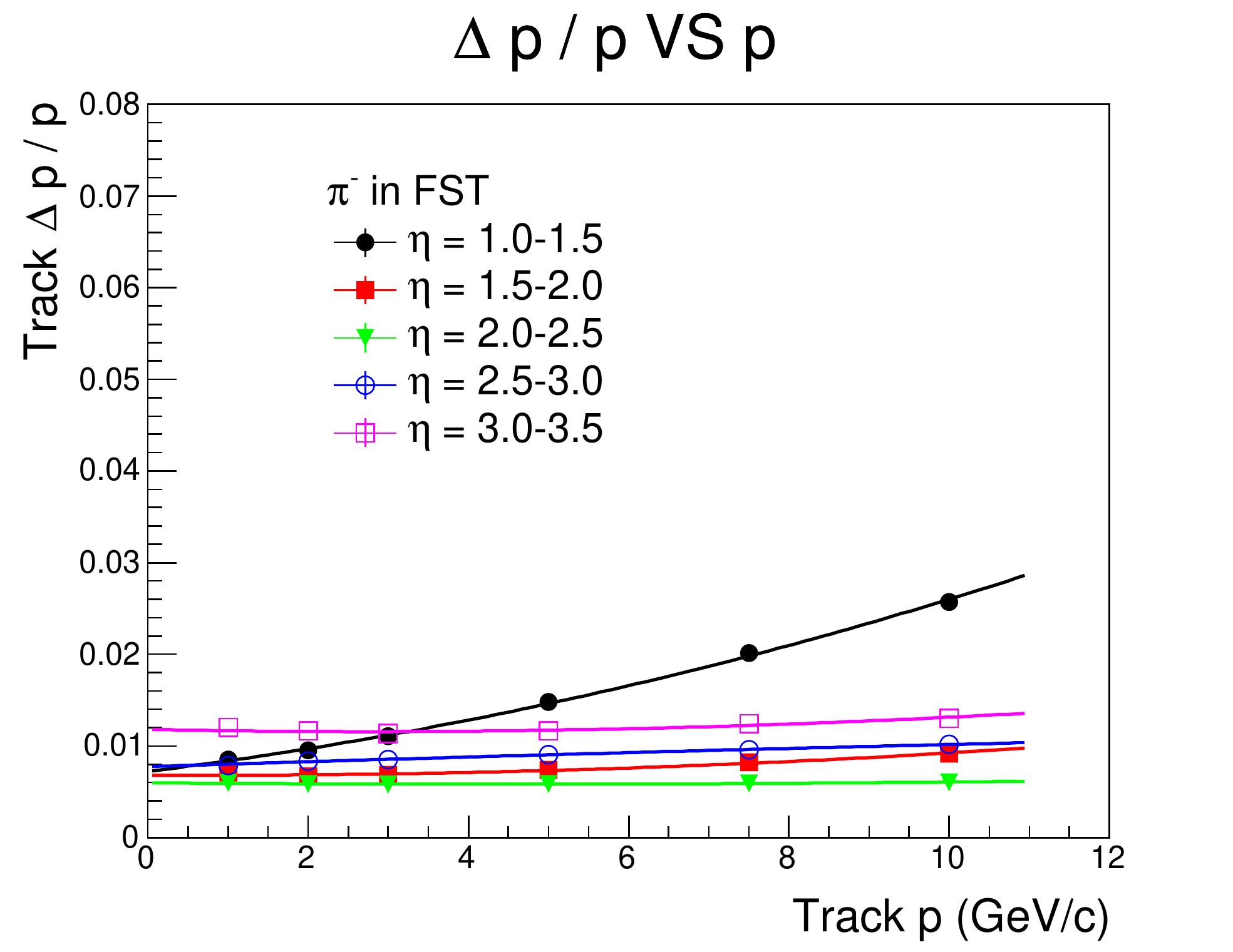}
\includegraphics[width=0.33\textwidth]{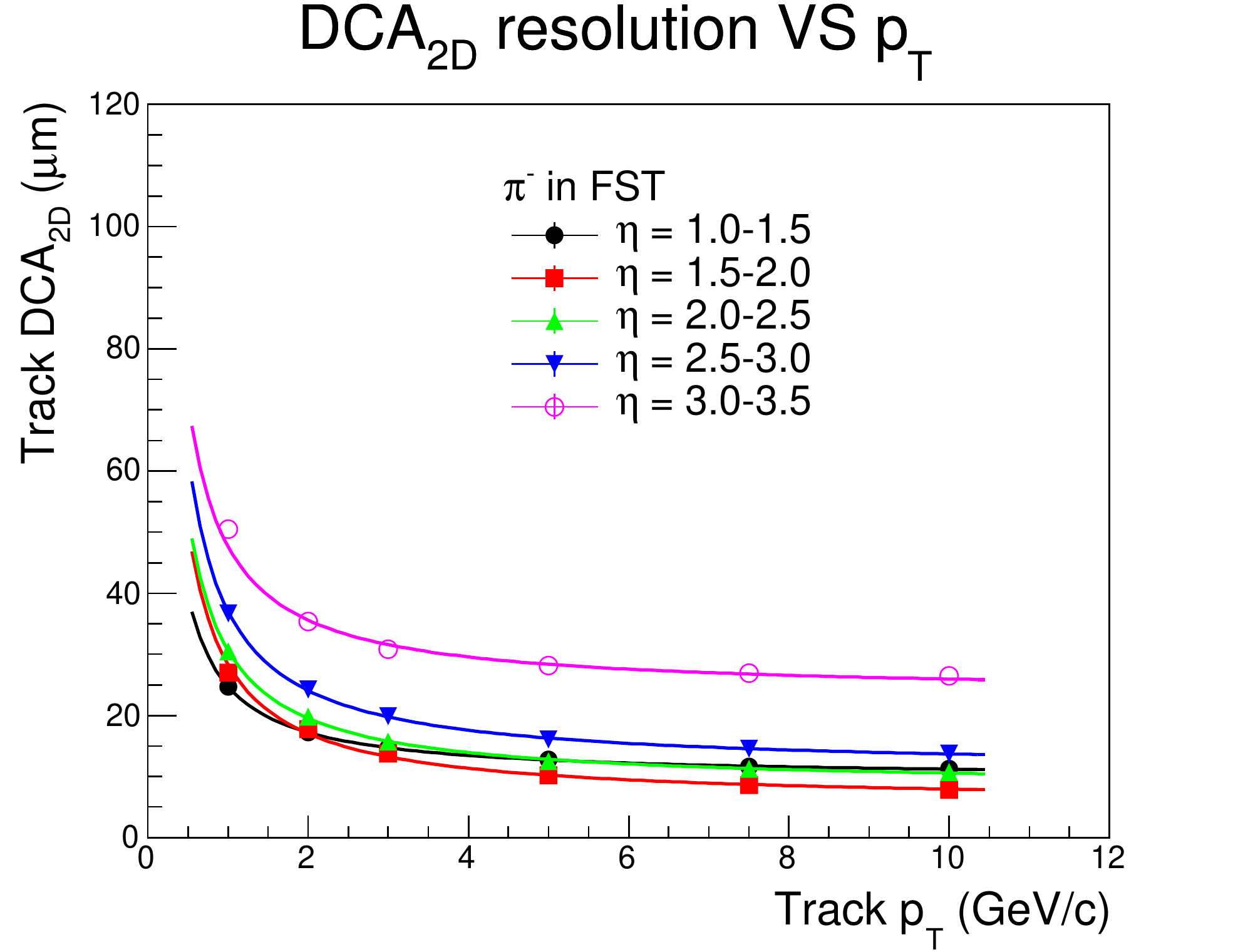}
\includegraphics[width=0.33\textwidth]{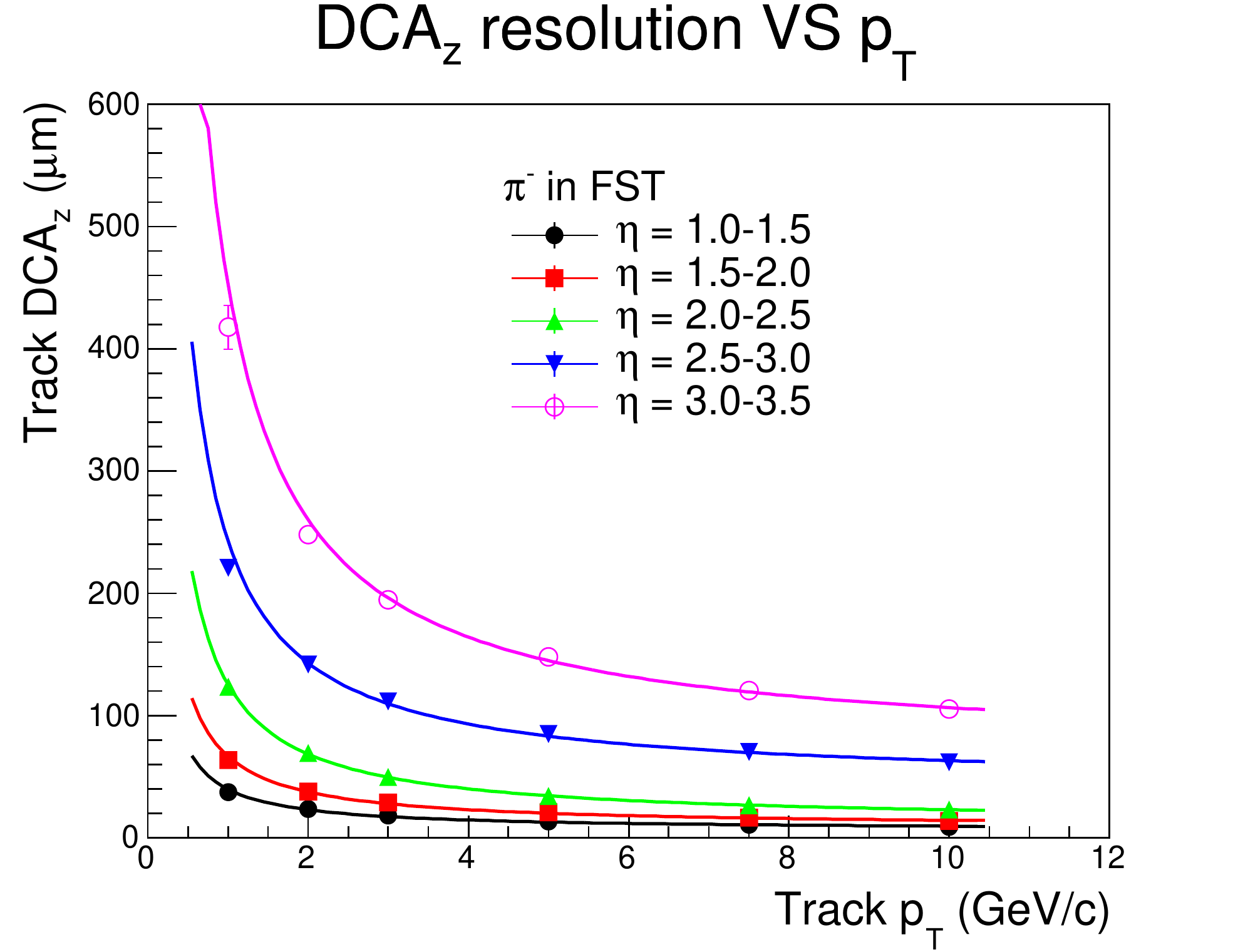}
\caption{\label{fig:fst_trk_fit}Tracking performance for FST version 4 design with the Beast magnetic field. Pixel pitch for both barrel layers and forward planes are selected at 20 $\mu m$. Left panel shows the momentum dependent momentum resolution in different pseudorapidity regions. Middle panel shows the transverse momentum dependent $DCA_{2D}$ resolution and the right panel shows the transverse momentum dependent $DCA_{z}$ resolution in the associated pseudorapidity regions.}
\end{center}
\end{figure}

\subsection{Simulation setup}
\label{sec: phy_sim}
In our current studies, we use PYTHIA8 event generator for our heavy flavor and jet studies in $e+p$ collisions. A $N_{coll}$ scaling is applied for the $e+A$ collisions. In additional to the physics $e+p$ collisions, a 12 kHz $p+p$ collision background is embedded in to the simulated events with the same proton beam energy. The evaluated tracking performance of the barrel and foward silicon vertex/tracking detector (see different design parameters in section \ref{sec:geo}) in both the central and forward pseudorapidity regions are used to smear charged tracks in generated events. In addition to the tracking performance provided by the silicon vertex/tracking detectors, 20-35 micron primary vertex resolution which depends on the track multiplity, 95$\%$ K/$\pi$/p separation and 95$\%$ electron identification efficiency are included.

\subsection{Open heavy flavor hadron reconstruction and physics projection}
\label{sec: OHF_had}
Heavy flavor products have strong a discriminating power between different model predictions of nuclear transport coefficients. In the factorizaton frame, the covolution function of the heavy flavor hadron cross-section includes the hadronization process. This results in that heavy flavor hadron production will be an ideal probe to map out the hadronization process in vacuum and nuclear medium by comparing to the other channels such as jet production.

In these simulation studies, the open heavy flavor hadrons including D-meson and B-meson are reconstructed by matching the charged tracking transverse Distance Closest Approach ($DCA_{2D}$) within a certain value. The cuts are varied depending on the reconstructed particle species, the average value is set at 100 $\mu m$. The simulation analysis chain is the following:

\begin{enumerate}
  \item event generation in PYTHIA.
  \item Detector performance smearing of generated particles.
  \item Heavy flavor hadron and jet reconstruction.
  \item Physics projection scaled with the integrated luminosity.
\end{enumerate}

\begin{figure}[H]
\centering
\includegraphics[width=0.99\textwidth]{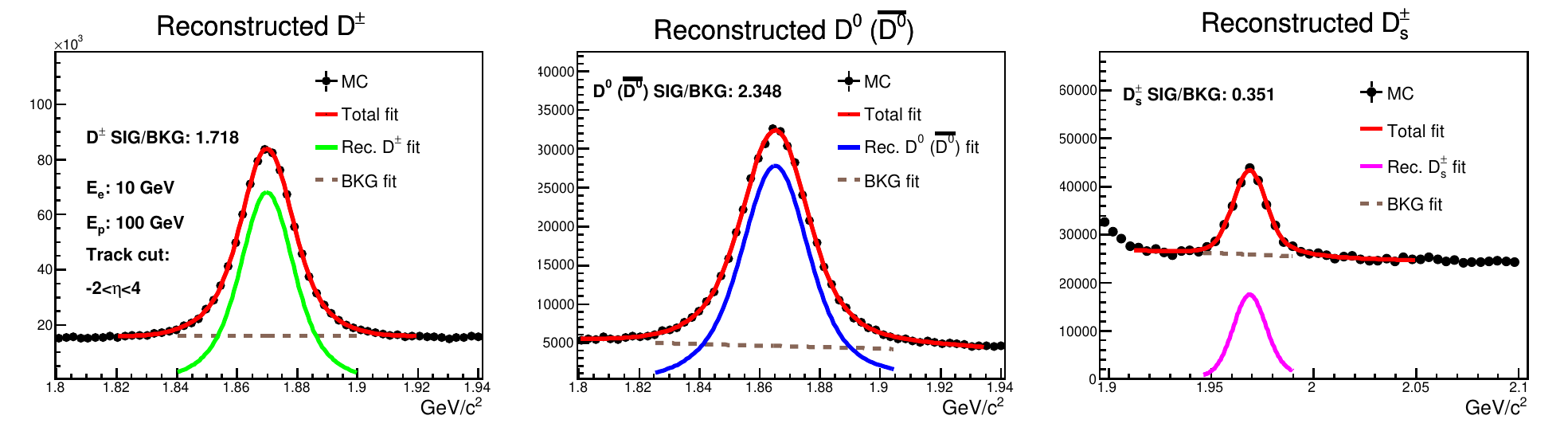}
\includegraphics[width=0.66\textwidth]{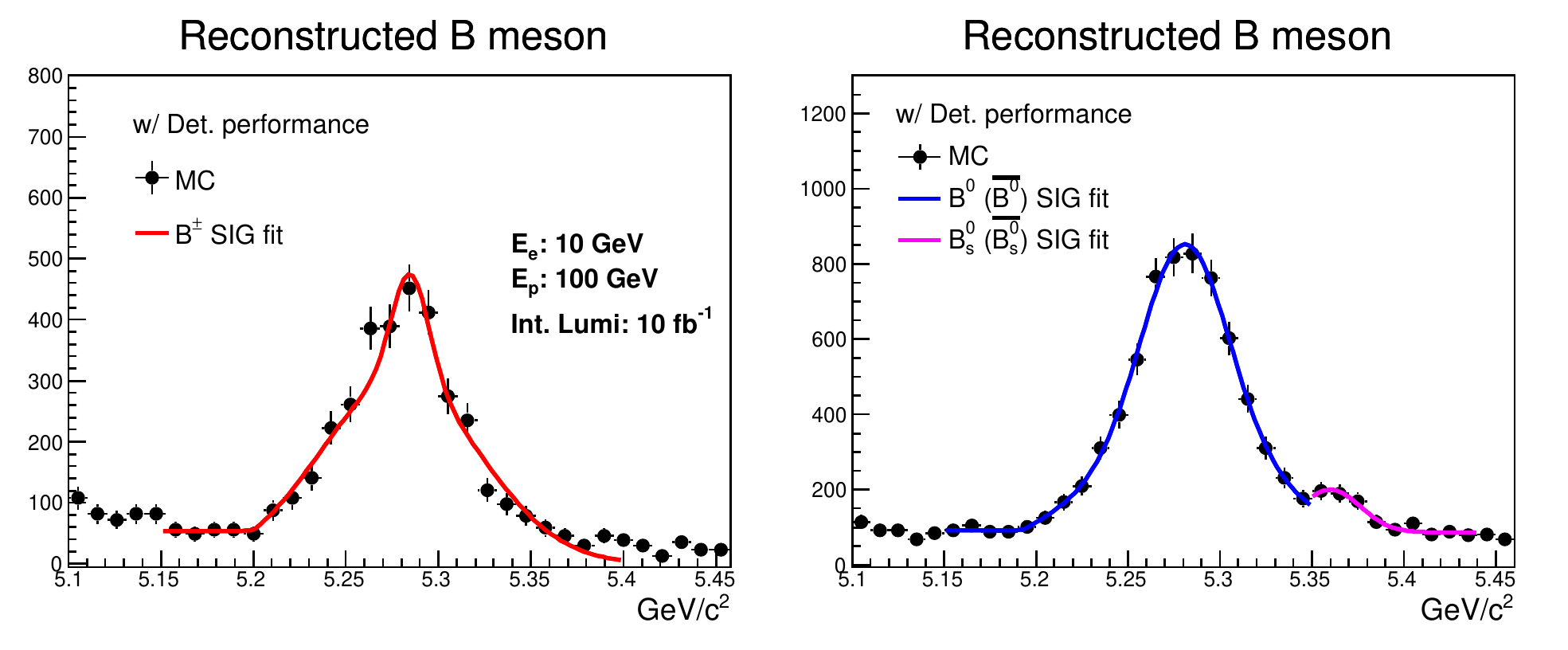}
\caption{\label{fig:HF_had_1} Reconstructed D-mesons and B-mesons using the FST version 0 design with the Barbar magnetic field. Pixel pitch for both barrel layers and forward planes are selected at 20 $\mu m$. The integrated luminosity of $e+p$ collisions at $\sqrt{s} = 63$ GeV is 10 $fb^{-1}$. }
\end{figure}

Different geometries of the proposed Forward Silicon Tracker (FST) have been studies in section \ref{sec: fun4all}, impacts on the open heavy flavor hadron by their tracking performances are studied. We start with the parameterization of the tracking performance for one version of the barrel and forward silicon vertex/tracking detector design. Figure~\ref{fig:fst_trk_fit0} shows the tracking performance with the FST version 0 design using the Beast magnetic field. Figure~\ref{fig:fst_trk_fit} shows the tracking performance with the FST version 4 design using the Beast magnetic field. Better tracking momentum and DCA resolutions are achieved by the FST version 4 design.

\newpage
\begin{figure}[H]
\centering
\includegraphics[width=0.8\textwidth]{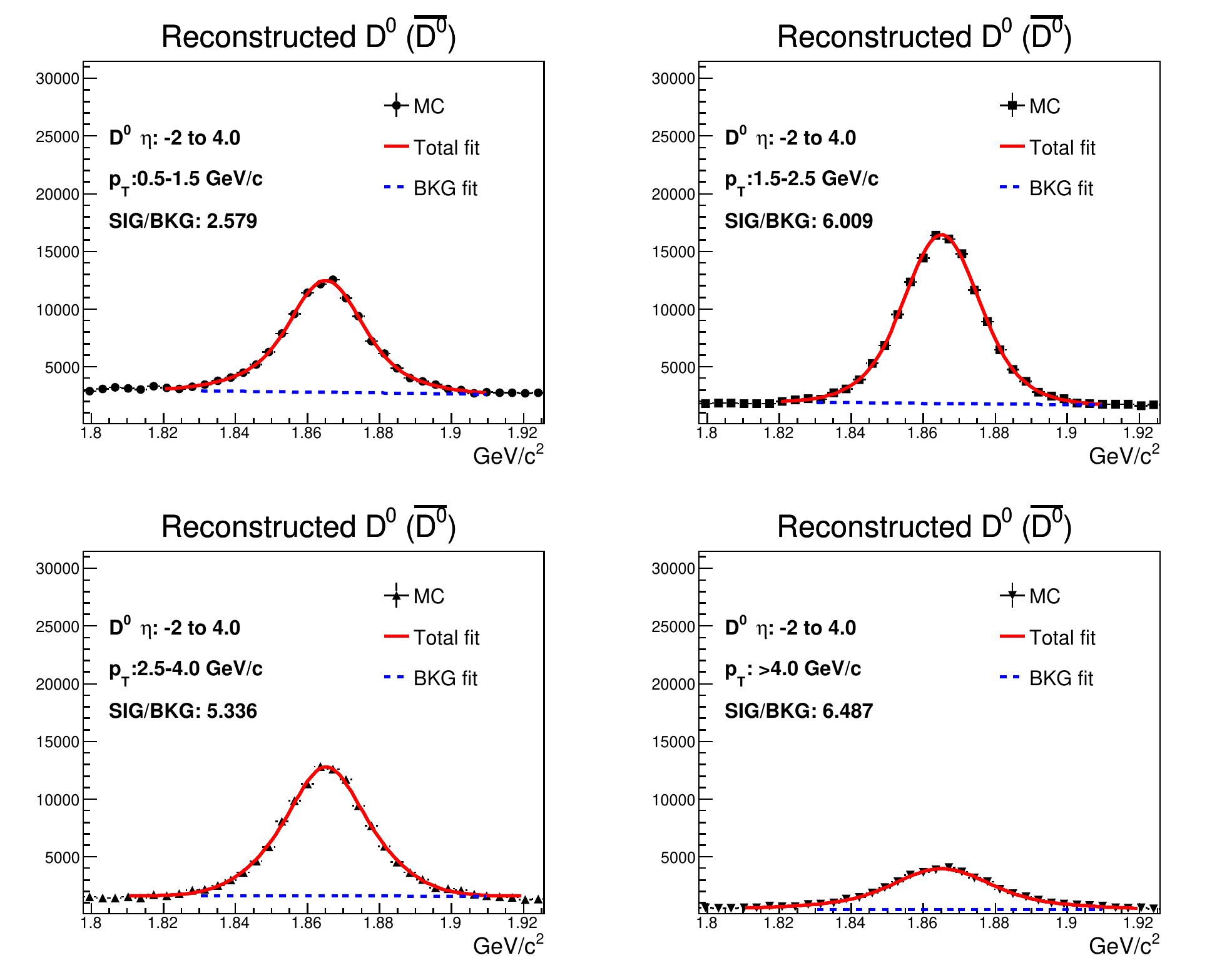}
\includegraphics[width=0.8\textwidth]{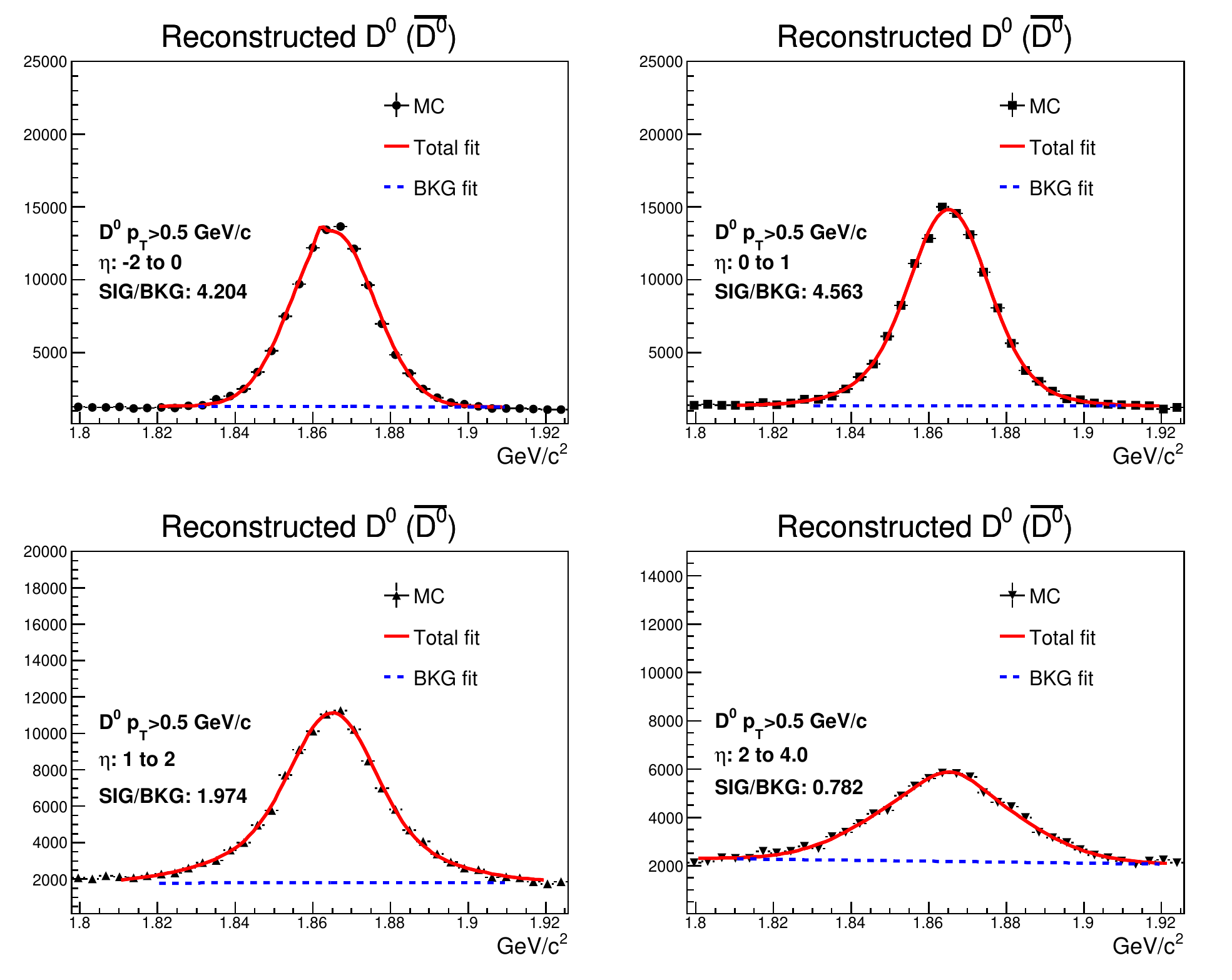}
\caption{\label{fig:HF_had_2} Reconstructed $D^{0}$ ($\bar{D^{0}}$) mesons using the FST version 0 design with the Barbar magnetic field. Pixel pitch for both barrel layers and forward planes are selected at 20 $\mu m$. The integrated luminosity of $e+p$ collisions at $\sqrt{s} = 63$ GeV is 10 $fb^{-1}$. Top four panels present the invariant mass distributions of reconstructed $D^{0}$ ($\bar{D^{0}}$) in different $p_{T}$ bins. The bottom four panels show the }
\end{figure}

 Figure~\ref{fig:HF_had_1} shows the mass spectrum of fully reconstructed $D^{\pm}$, $D^{0}$ ($\bar{D^{0}}$), $D^{\pm}_{s}$, $B^{\pm}$, $B^{0}$ ($\bar{B^{0}}$) and $B^{0}_{s}$ ($\bar{B^{0}_{s}}$) based on the tracking performance shown in Figure~\ref{fig:fst_trk_fit}. For these heavy flavor hadron reconstructions, charged tracks are required to have pseudorapidity within -2 to 4. Clear D-meson signals have been obtained on top of the combinatorial backgrounds. The signal over background ratios and the reconstruction efficiency are listed in the associated panels. Clean B-mesons can be reconstructed with this FST design, while the width of reconstructed $B^{0}_{s}$ ($bar{B^{0}_{s}}$) is a bit wide and could not get separation from the reconstructed $B^{0}$ ($\bar{B^{0}}$) mass peak.
 
 The kinematic dependence of the reconstructed D-meson has been studied. The top four panels of Figure~\ref{fig:HF_had_2} present the mass distributions of reconstructed $D^{0}$ ($\bar{D^{0}}$) with $-2< \eta < 4$ in different D-meson $p_{T}$ regions. The $p_{T}$ bins are 0.5-1.5 GeV/c, 1.5-2.5 GeV/c, 2.5-4.0 GeV/c, $>$ 4.0 GeV/c. The bottom four panels of Figure~\ref{fig:HF_had_2} present the mass distributions of reconstructed $D^{0}$ ($\bar{D^{0}}$) with $p_{T} > 0.5$ GeV/c in different D-meson pseudorapdity $\eta$ regions. The pseudorapidty $\eta$ bins are -2 to 0, 0 to 1, 1 to 2 and 2 to 4.0.
 
\begin{figure}[H]
\centering
\includegraphics[width=0.6\textwidth]{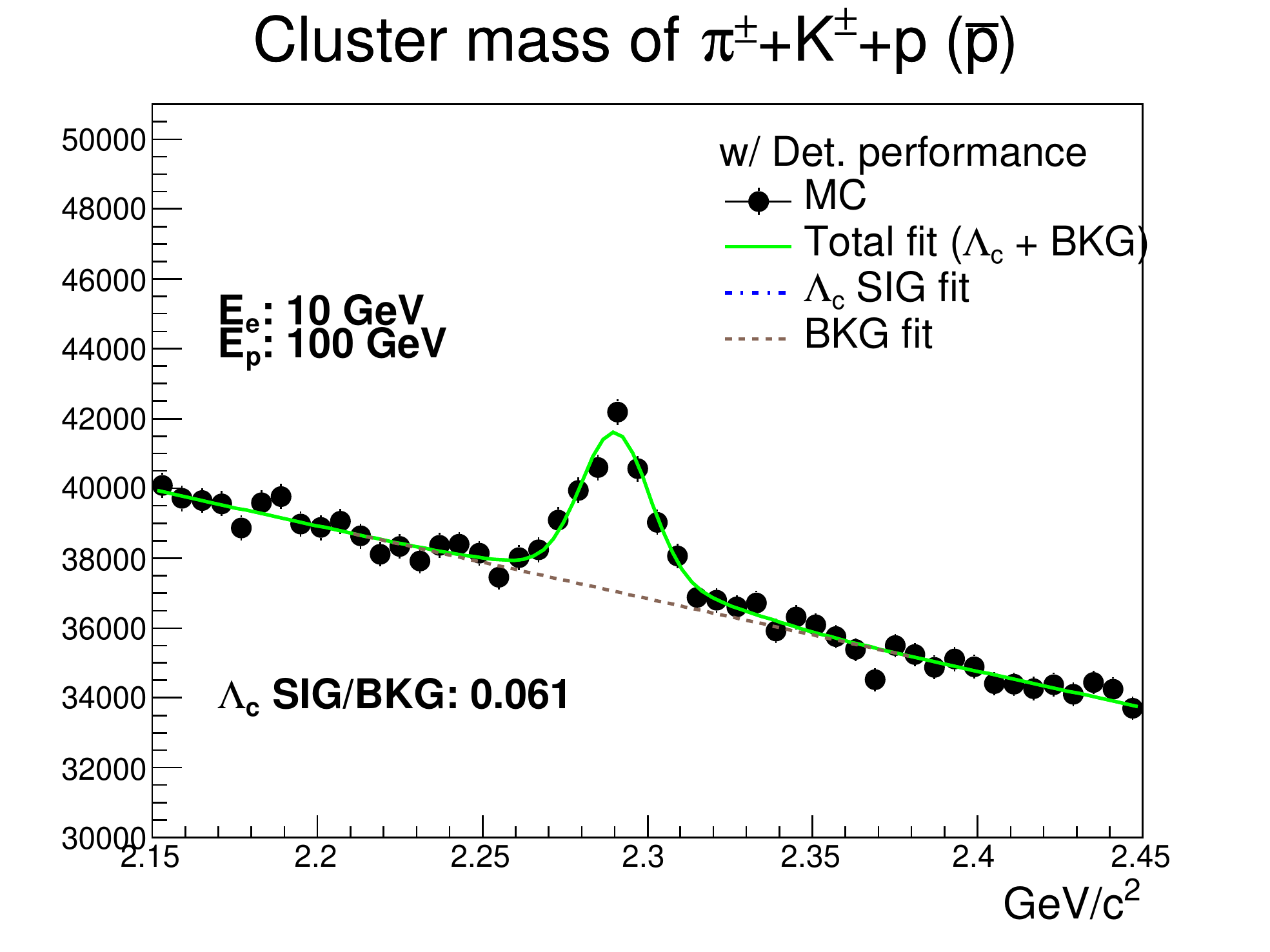}
\caption{\label{fig:HF_had_3} Reconstructed $\Lambda_{c}$ mass spectrum using the FST version 0 design with the Babar magnetic field. Pixel pitch for both barrel layers and forward planes are selected at 20 $\mu m$. The integrated luminosity of $e+p$ collisions at $\sqrt{s} = 63$ GeV is 10 $fb^{-1}$.}
\end{figure}

 In addition to heavy flavor meson reconstruction, we also looked for the heavy flavor hadron reconstruction (e.g. $\Lambda_{c}$). Although the combinatorial background is significantly higher than the D-meson mass spectrum, clear $\Lambda_{c}$ signal can be obtained as shown in Figure~\ref{fig:HF_had_3}. 
 
\begin{figure}[H]
\centering
\includegraphics[width=0.99\textwidth]{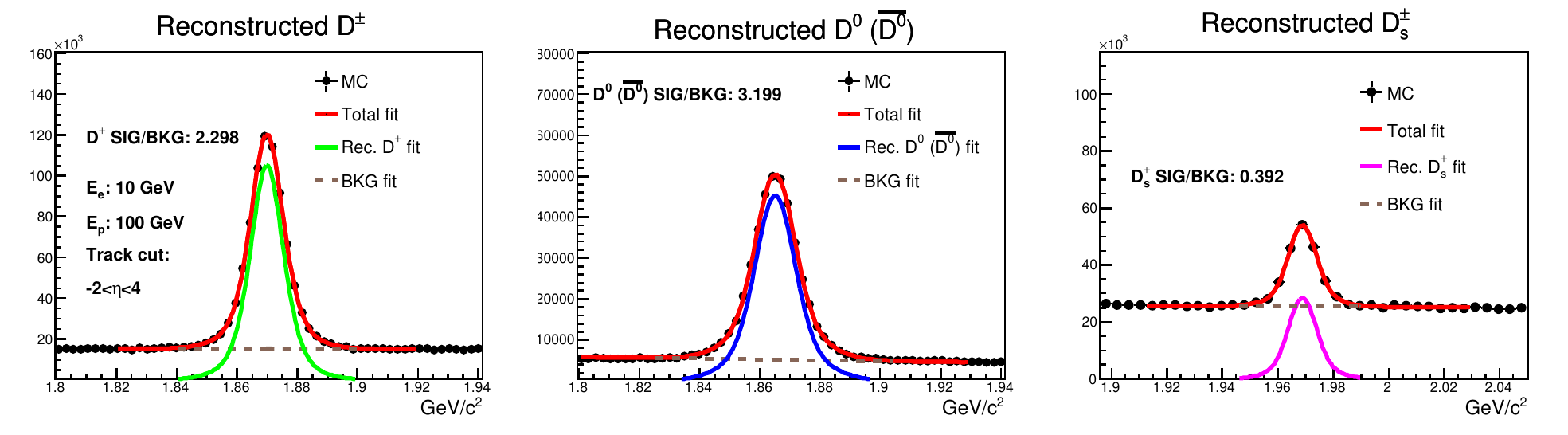}
\includegraphics[width=0.99\textwidth]{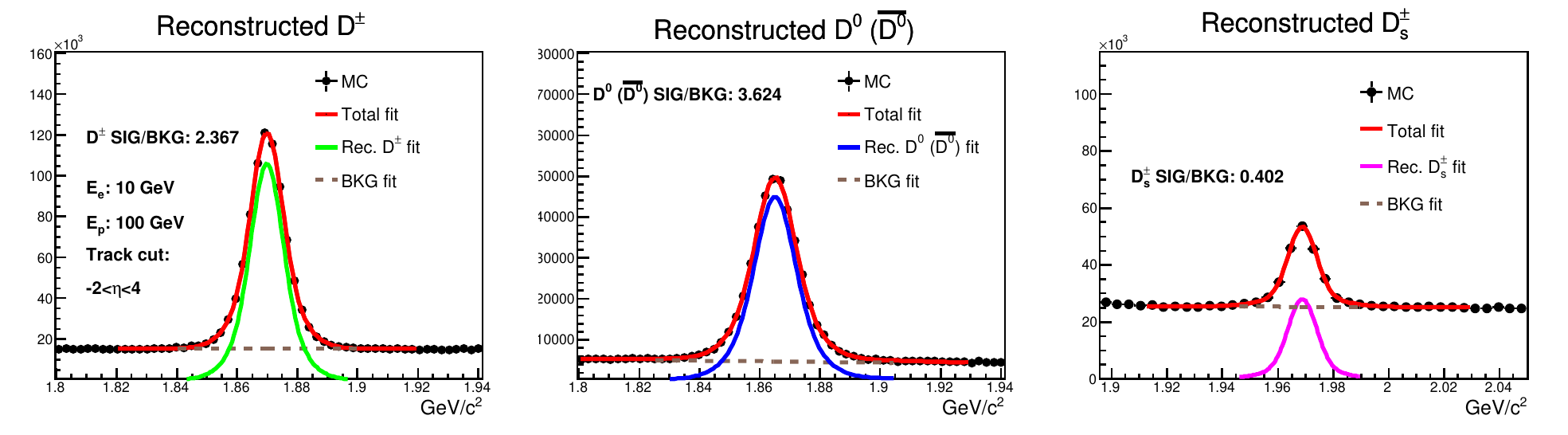}
\caption{\label{fig:HF_had_diff_1} Reconstructed D-meson mass spectrum using the FST version 0 and version 4 designs with the Beast magnetic field. Pixel pitch for both barrel layers and forward planes are selected at 20 $\mu m$. The integrated luminosity of $e+p$ collisions at $\sqrt{s} = 63$ GeV is 10 $fb^{-1}$.}
\end{figure}

\begin{figure}[H]
\centering
\includegraphics[width=0.7\textwidth]{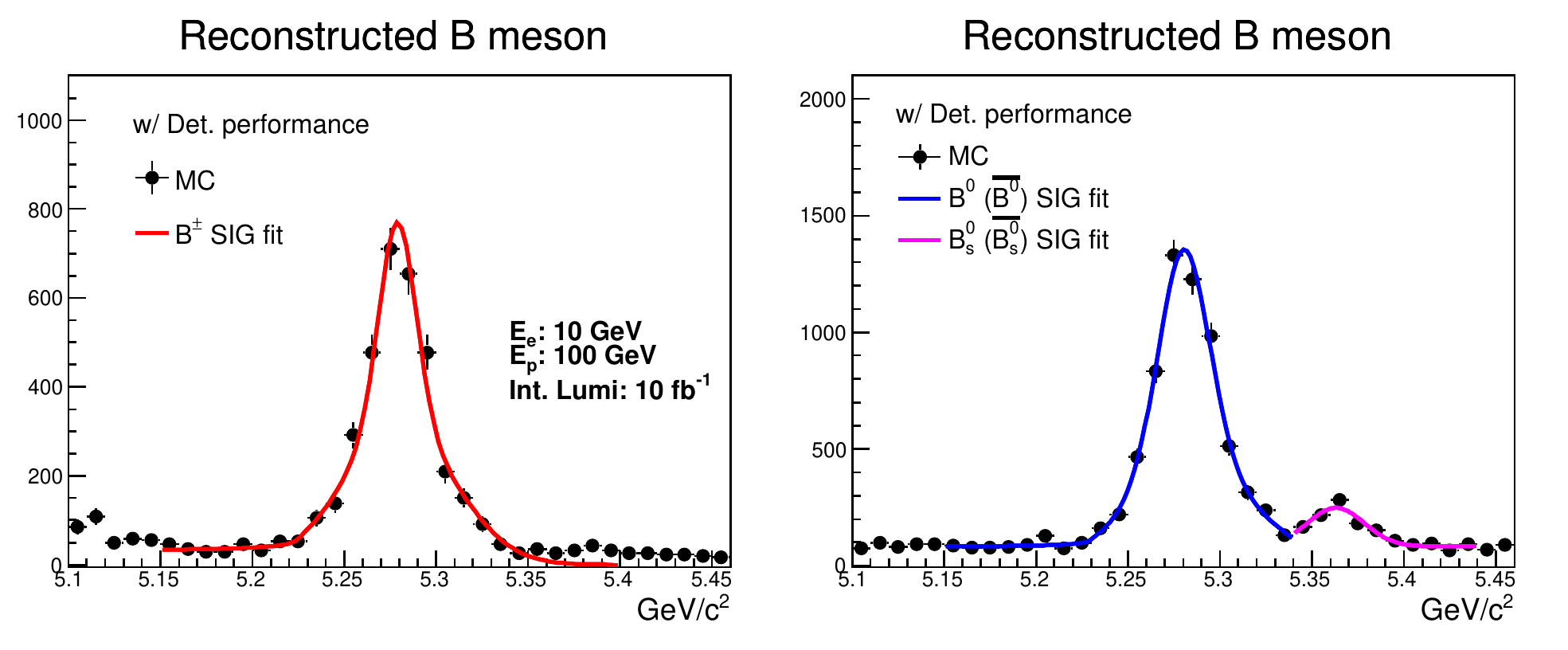}
\includegraphics[width=0.7\textwidth]{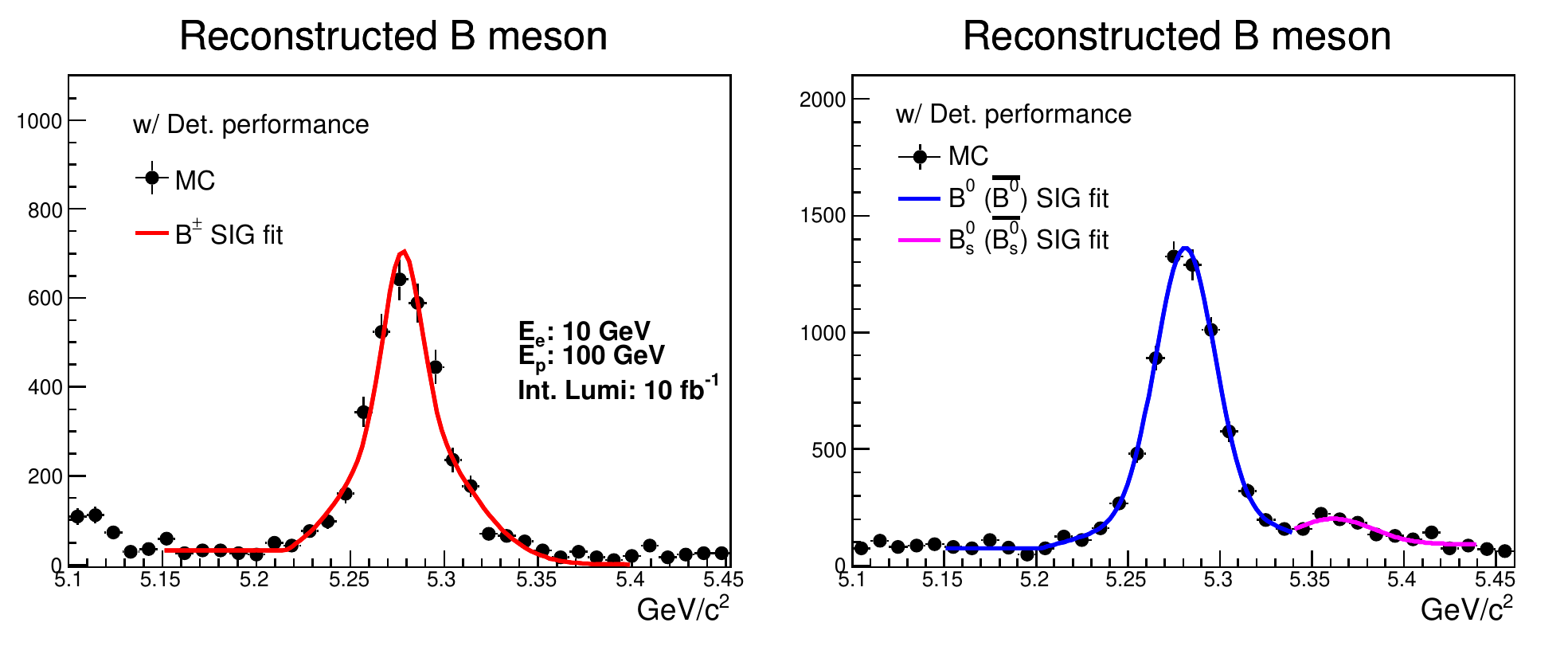}
\caption{\label{fig:HF_had_diff_2} Reconstructed B-meson mass spectrum using the FST version 0 and version 4 designs with the Beast magnetic field. Pixel pitch for both barrel layers and forward planes are selected at 20 $\mu m$. The integrated luminosity of $e+p$ collisions at $\sqrt{s} = 63$ GeV is 10 $fb^{-1}$.}
\end{figure}

\begin{figure}[H]
\centering
\includegraphics[width=0.9\textwidth]{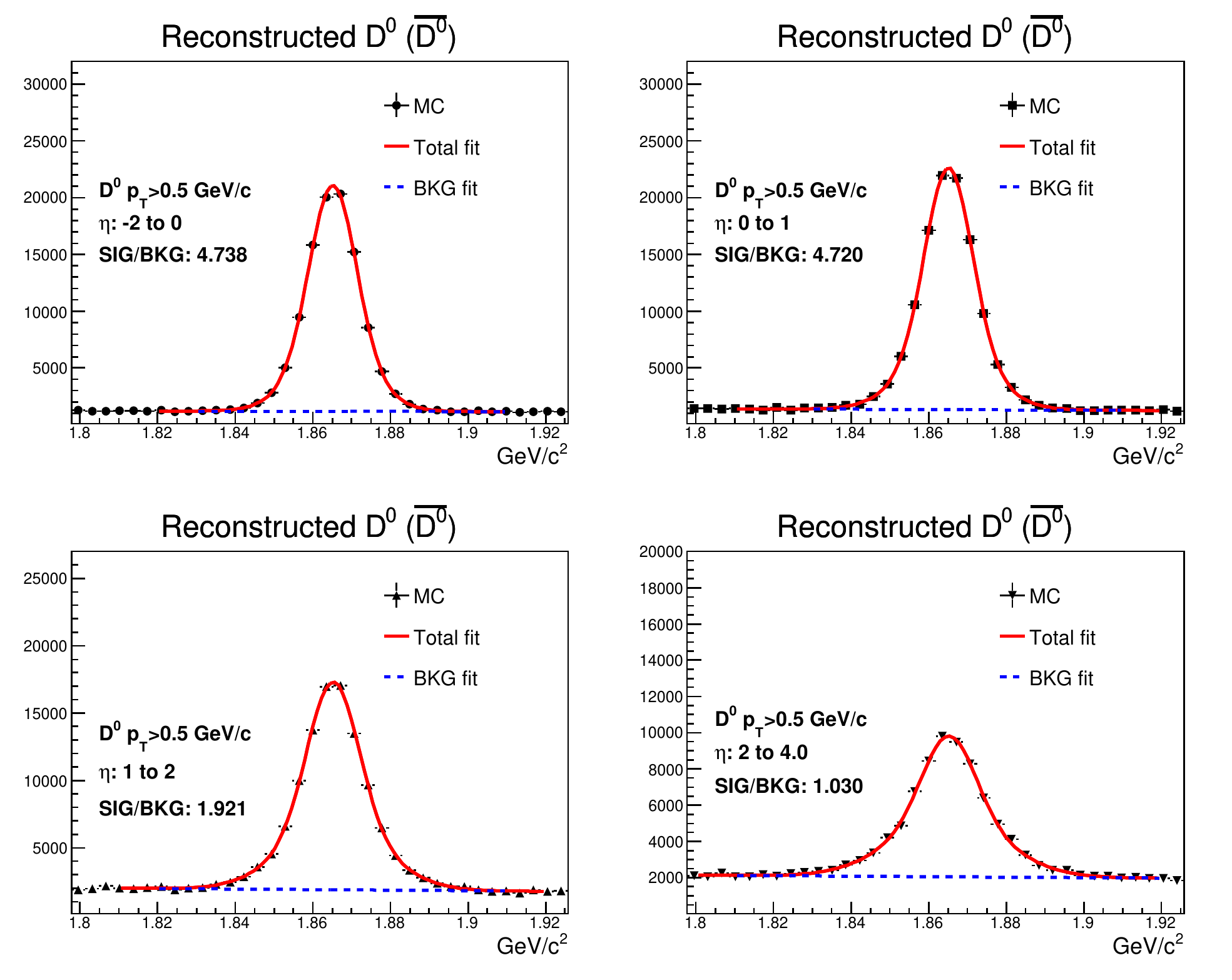}
\includegraphics[width=0.9\textwidth]{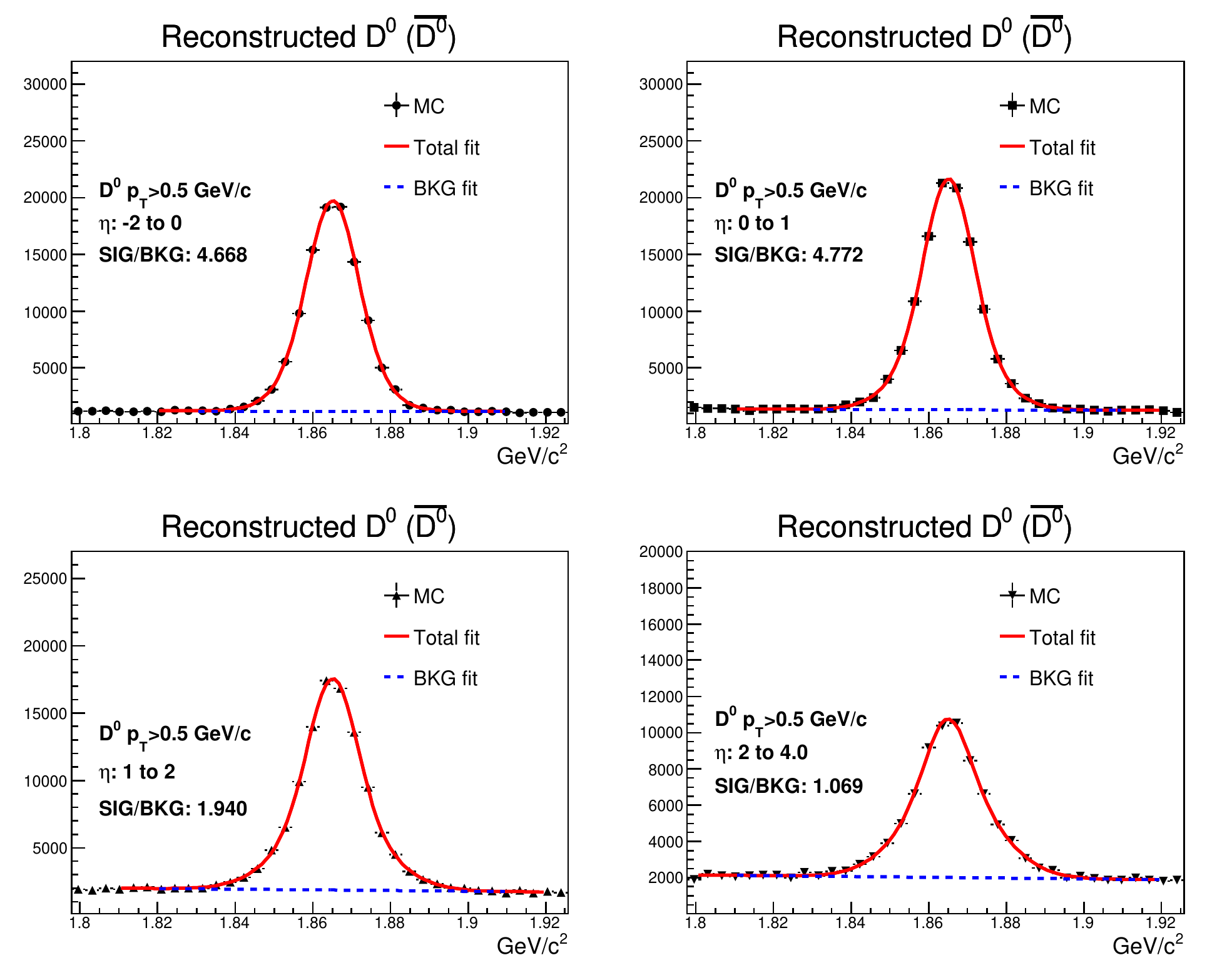}
\caption{\label{fig:HF_had_diff_3} Reconstructed $D^{0}$ ($\bar{D^{0}}$) meson mass spectrum in different pseudorapidity regions using the FST version 0 and version 4 designs with the Beast magnetic field. Pixel pitch for both barrel layers and forward planes are selected at 20 $\mu m$. The integrated luminosity of $e+p$ collisions at $\sqrt{s} = 63$ GeV is 10 $fb^{-1}$.}
\end{figure}

\begin{figure}[H]
\centering
\includegraphics[width=0.49\textwidth]{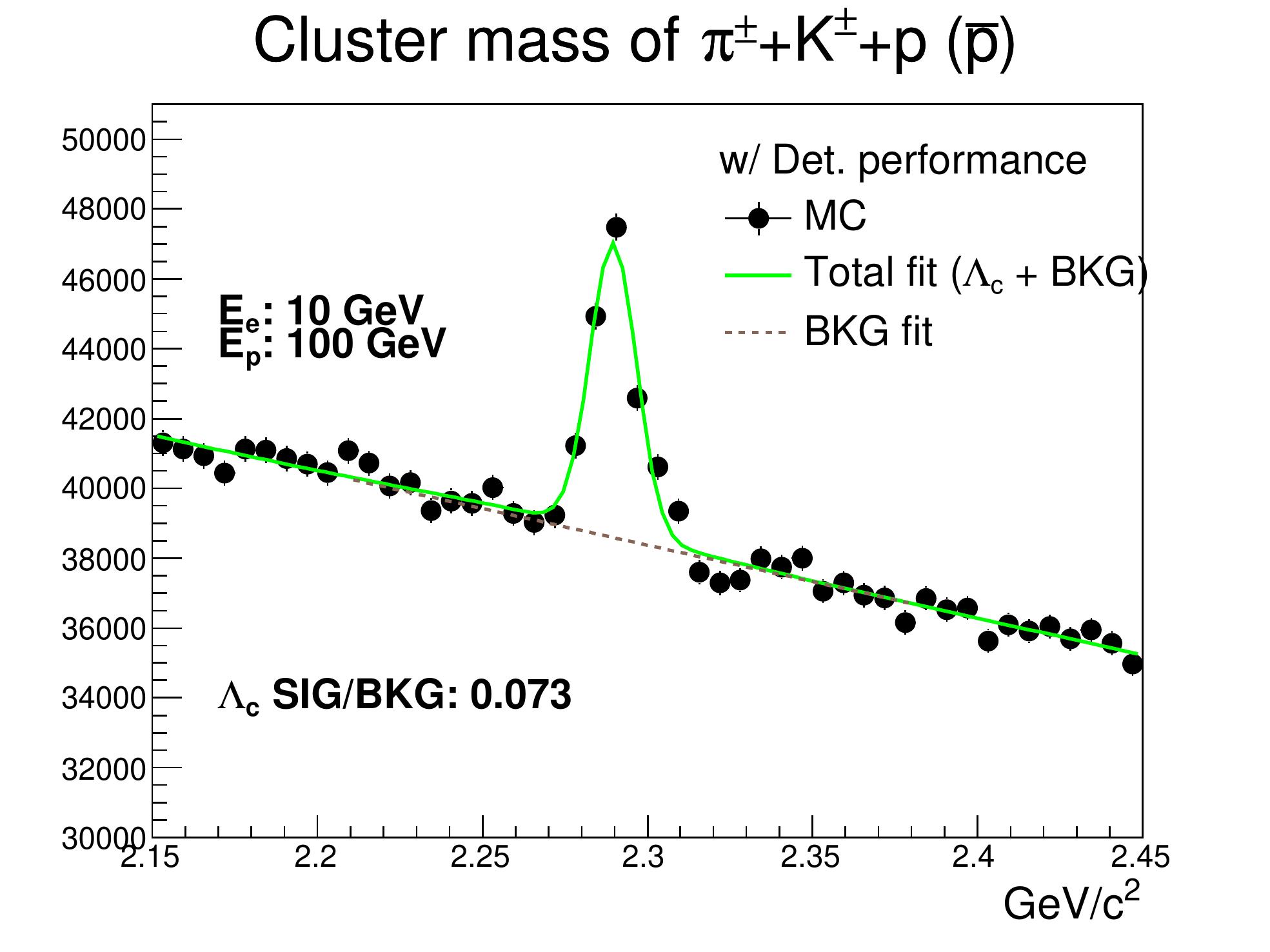}
\includegraphics[width=0.49\textwidth]{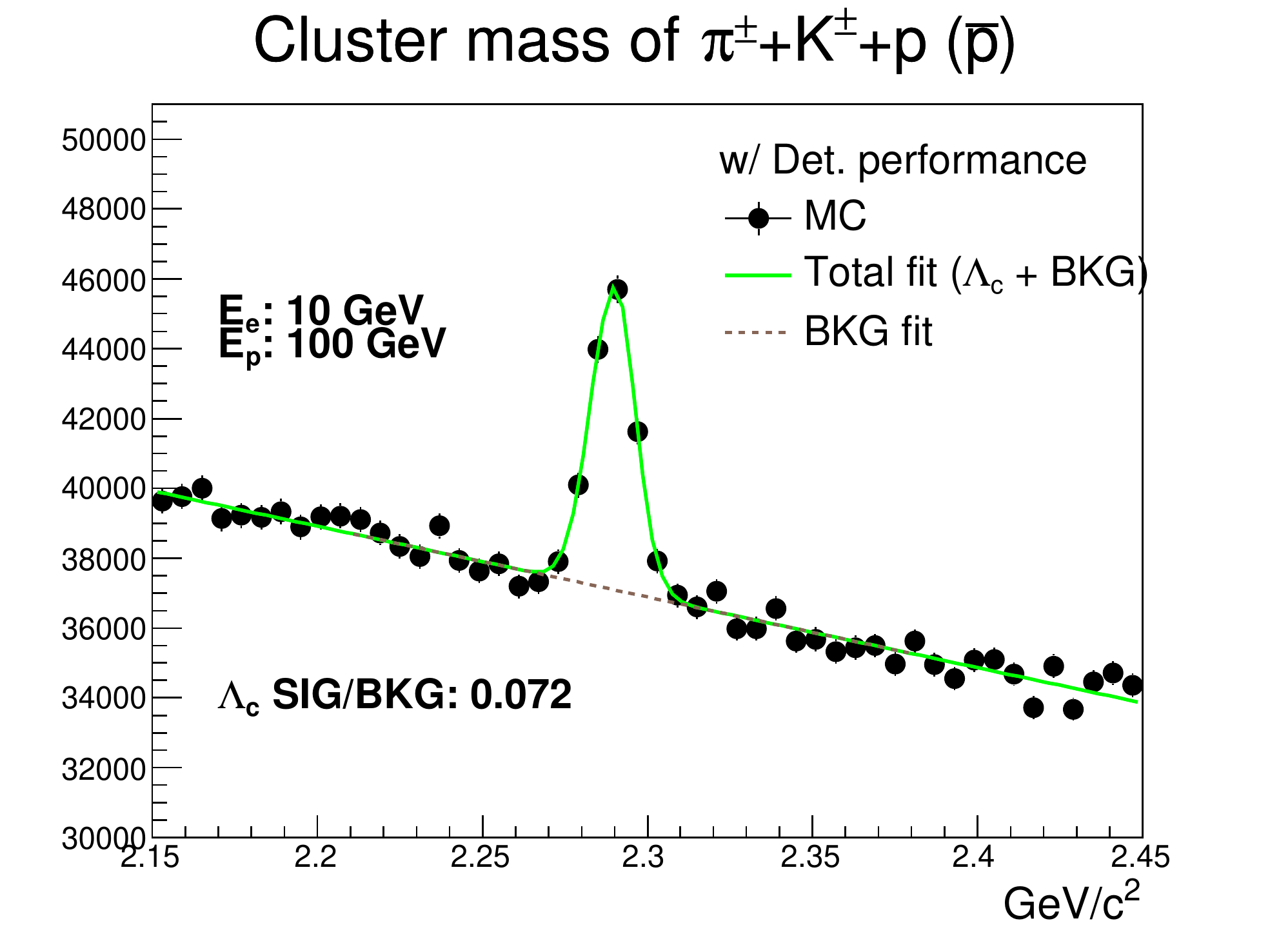}
\caption{\label{fig:HF_had_diff_4} Reconstructed $\Lambda_{c}$ mass spectrum using the FST version 0 and version 4 designs with the Beast magnetic field. Pixel pitch for both barrel layers and forward planes are selected at 20 $\mu m$. The integrated luminosity of $e+p$ collisions at $\sqrt{s} = 63$ GeV is 10 $fb^{-1}$.}

\end{figure}
\begin{figure}[H]
\centering
\includegraphics[width=0.96\textwidth]{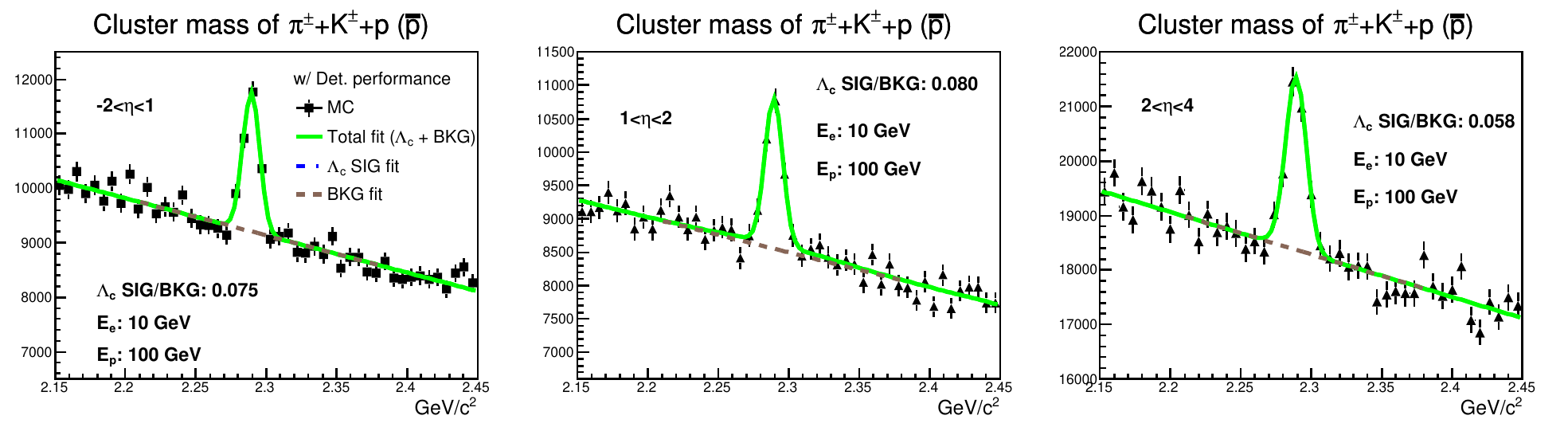}
\caption{\label{fig:HF_had_diff_5} Reconstructed $\Lambda_{c}$ mass spectrum in different pseudorapidity regions using the FST version 4 designs with the Beast magnetic field. Pixel pitch for both barrel layers and forward planes are selected at 20 $\mu m$. The integrated luminosity of $e+p$ collisions at $\sqrt{s} = 63$ GeV is 10 $fb^{-1}$.}
\end{figure}

Different FST detector geometries and magnetic field maps are used for the open heavy flavor hadron reconstruction. Figure~\ref{fig:HF_had_diff_1} shows the comparison of the reconstructed D-meson with using the FST version 0 and version 4 designs with the Beast magnetic field in 10 $fb^{-1}$ $e+p$ collisions at $\sqrt{s} = 63$ GeV. With the same simulation sample, comparison for reconstructed B-mesons are shown in Figure~\ref{fig:HF_had_diff_2}, comparison for reconstructed D-mesons in different pseudorapidity regions are shown in Figure~\ref{fig:HF_had_diff_3} and comparison for reconstructed $\Lambda_{c}$ are shown in Figure~\ref{fig:HF_had_diff_4}. Pseudorapidity dependent reconstructed $\Lambda_{c}$ mass spectrum with the FST version 4 design and the Beast magnetic filed have been shown in Figure~\ref{fig:HF_had_diff_5}. 
 
 These results indicate adding one outer barrel layer and one outer forward plane on top of the 5 barrel layer and 5 forward plane silicon vertex/tracker detector does not significantly change the signal over background ratio for reconstructed D-mesons, B-mesons and $\Lambda_{c}$ hadrons. Including a more forward and low material budget tracking detector such as a GEM tracker could further improve the tracking momentum resolution and provide better mass resolutions in the more forward pseudoradity region. These studies will be carried out once the detector design and performance evaluation is done.
 
\begin{figure}[H]
\centering
\includegraphics[width=0.6\textwidth]{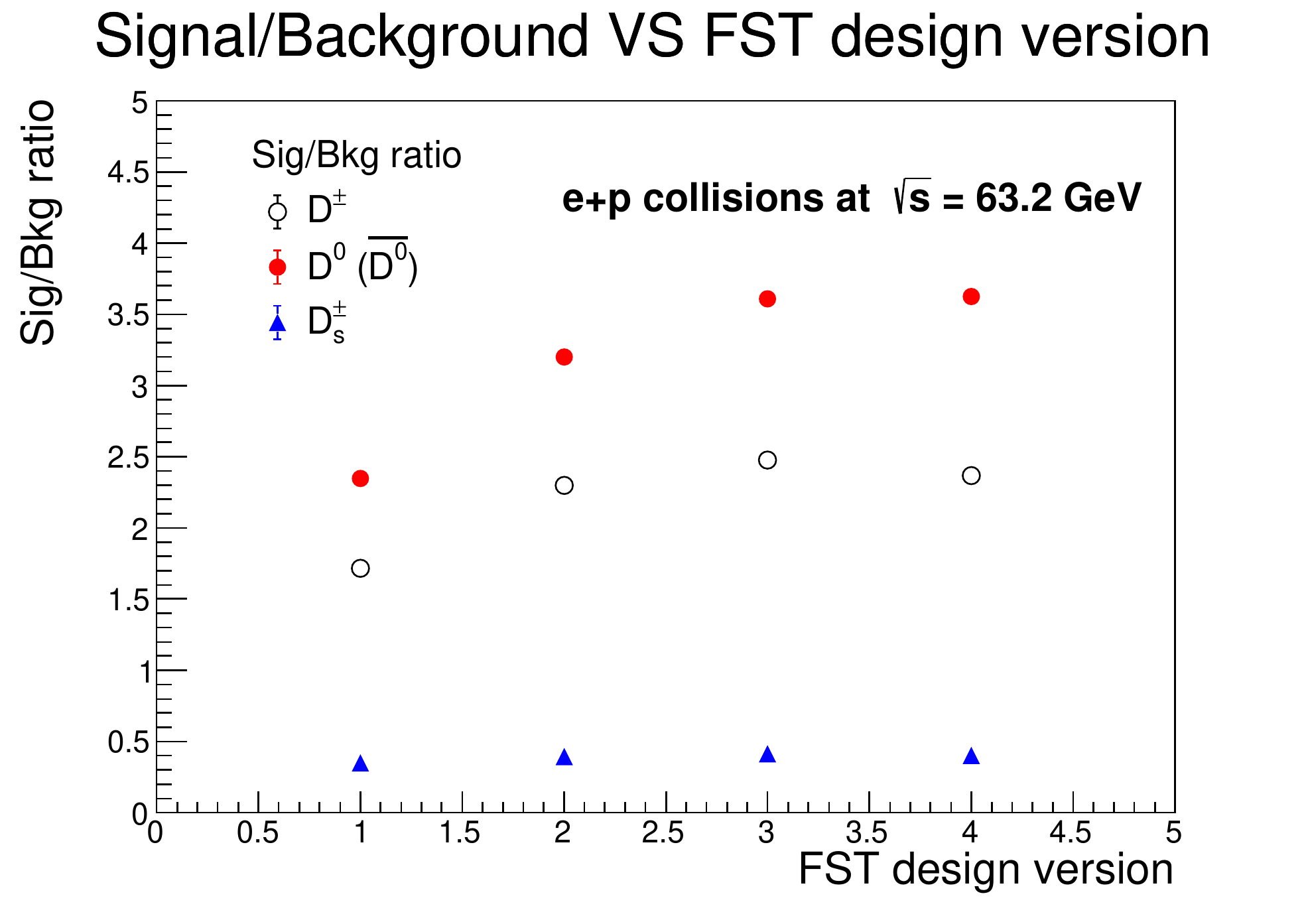}
\caption{\label{fig:D_sigbkg_1} Comparison of signal/background ratios for reconstructed D-mesons with different FST designs. These values are determined in simulation of $e+p$ collisions at $\sqrt{s} = 63$ GeV with integrated luminosity of 10 $fb^{-1}$.}
\end{figure}

\begin{figure}[H]
\centering
\includegraphics[width=0.6\textwidth]{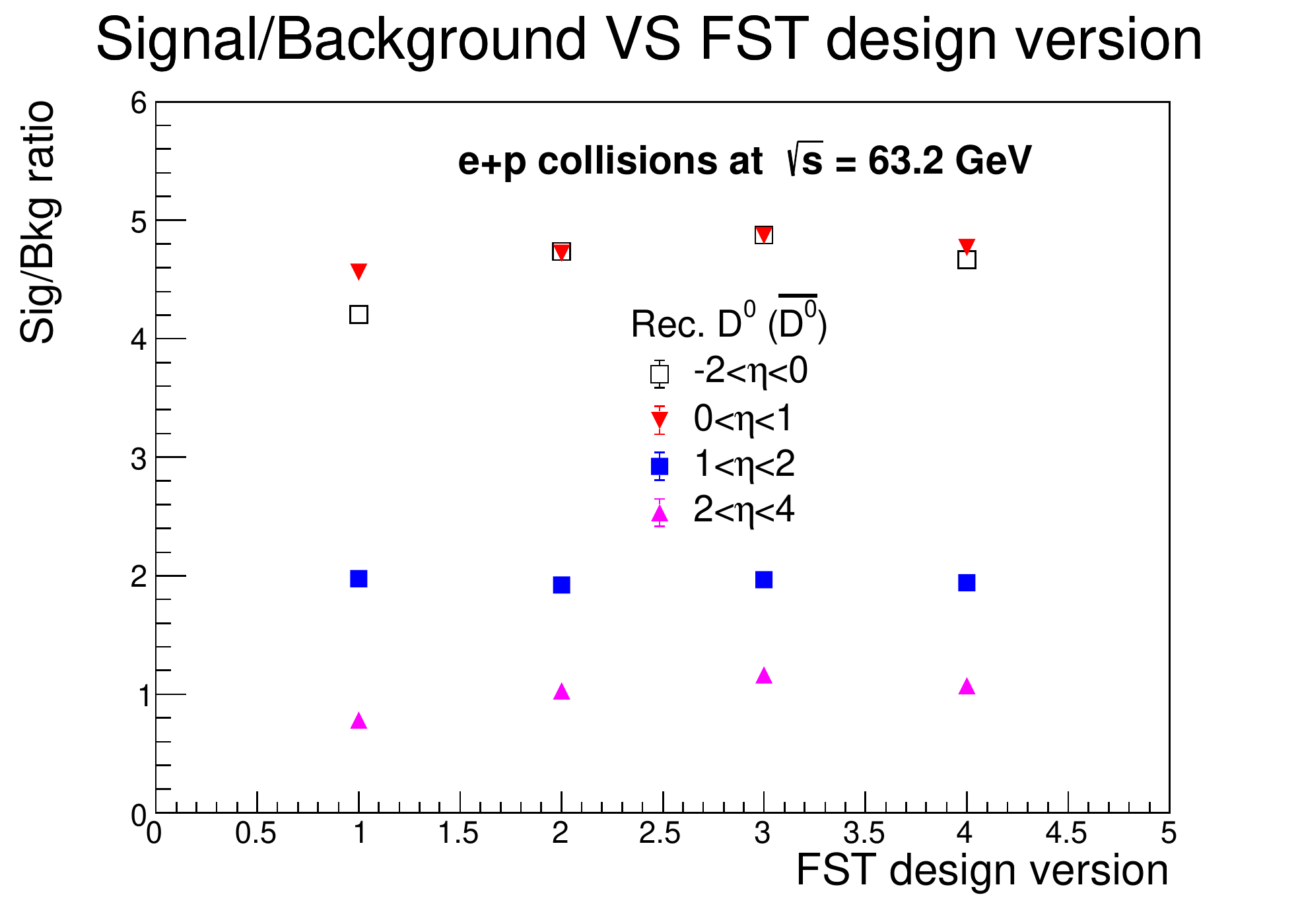}
\caption{\label{fig:D_sigbkg_2} Comparison of reconstructed $D^{0}$ ($\bar{D^{0}}$) in different pseudorapidity regions with different FST designs. These values are determined in simulation of $e+p$ collisions at $\sqrt{s} = 63$ GeV with integrated luminosity of 10 $fb^{-1}$.}
\end{figure}

\begin{figure}[H]
\centering
\includegraphics[width=0.6\textwidth]{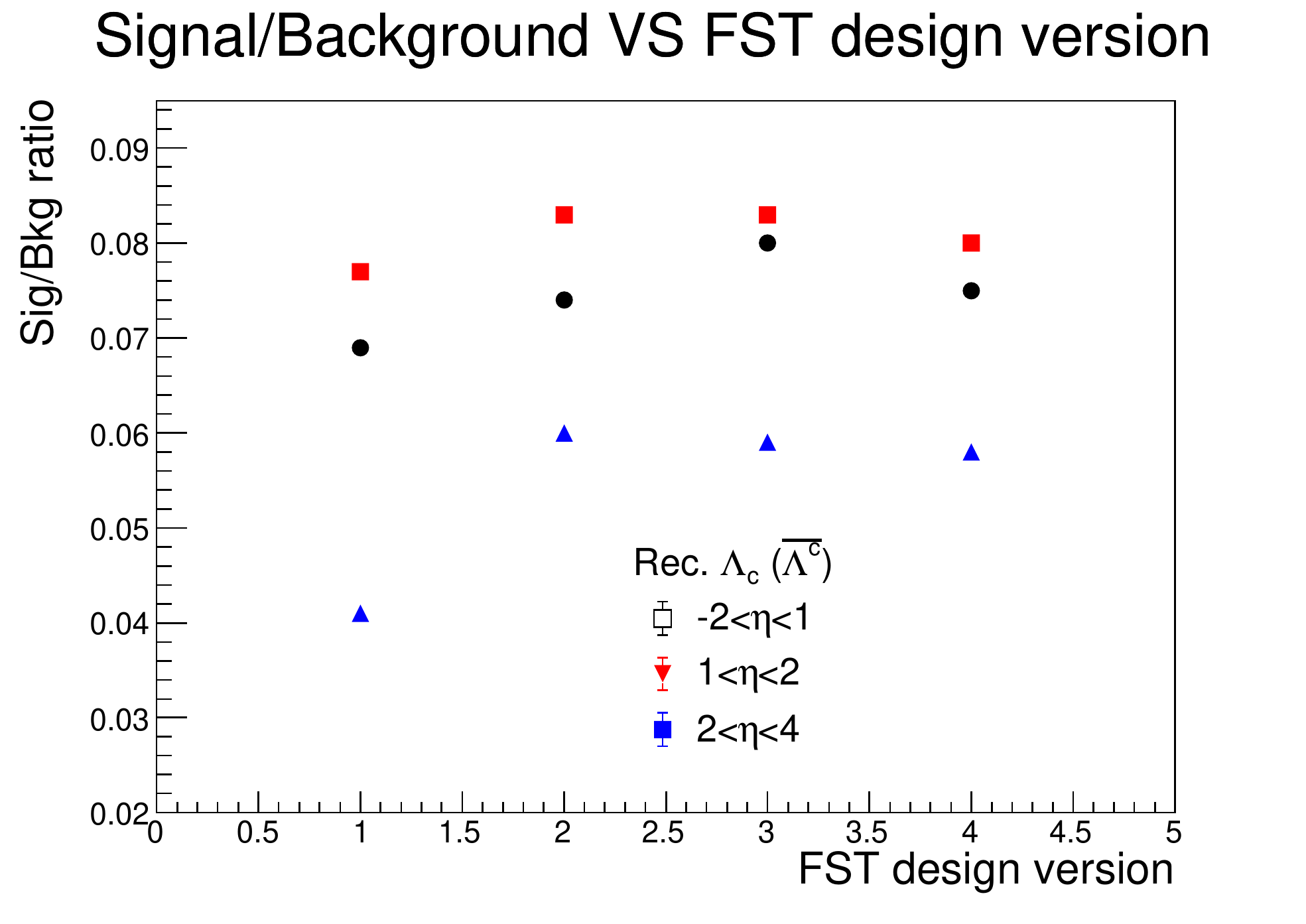}
\caption{\label{fig:Lam_sigbkg_1} Comparison of reconstructed $\Lambda_{c}$ ($\bar{\Lambda_{c}}$) in different pseudorapidity regions with different FST designs. These values are determined in simulation of $e+p$ collisions at $\sqrt{s} = 63$ GeV with integrated luminosity of 10 $fb^{-1}$.}
\end{figure}

\begin{table}[h]
    \centering
    \caption{\label{tab:fst_ver} FST version corresponding geometries and magnet options}
    \begin{tabular}{c c c c c}
    \hline
    \hline
    %-------------------------------%
    Name & FST index 1 & FST index 2 & FST index 3 & FST index4 \\ 
    \hline
    geometry version in Table \ref{tab:simFST_Geo} & version 0  & version 0 & version 1 & version 4 \\
    
    \hline
    Magnet options & Babar & Beast  & Beast  & Beast \\
    \hline
    \hline
\end{tabular}
\end{table}

\begin{figure}[H]
\centering
\includegraphics[width=0.48\textwidth]{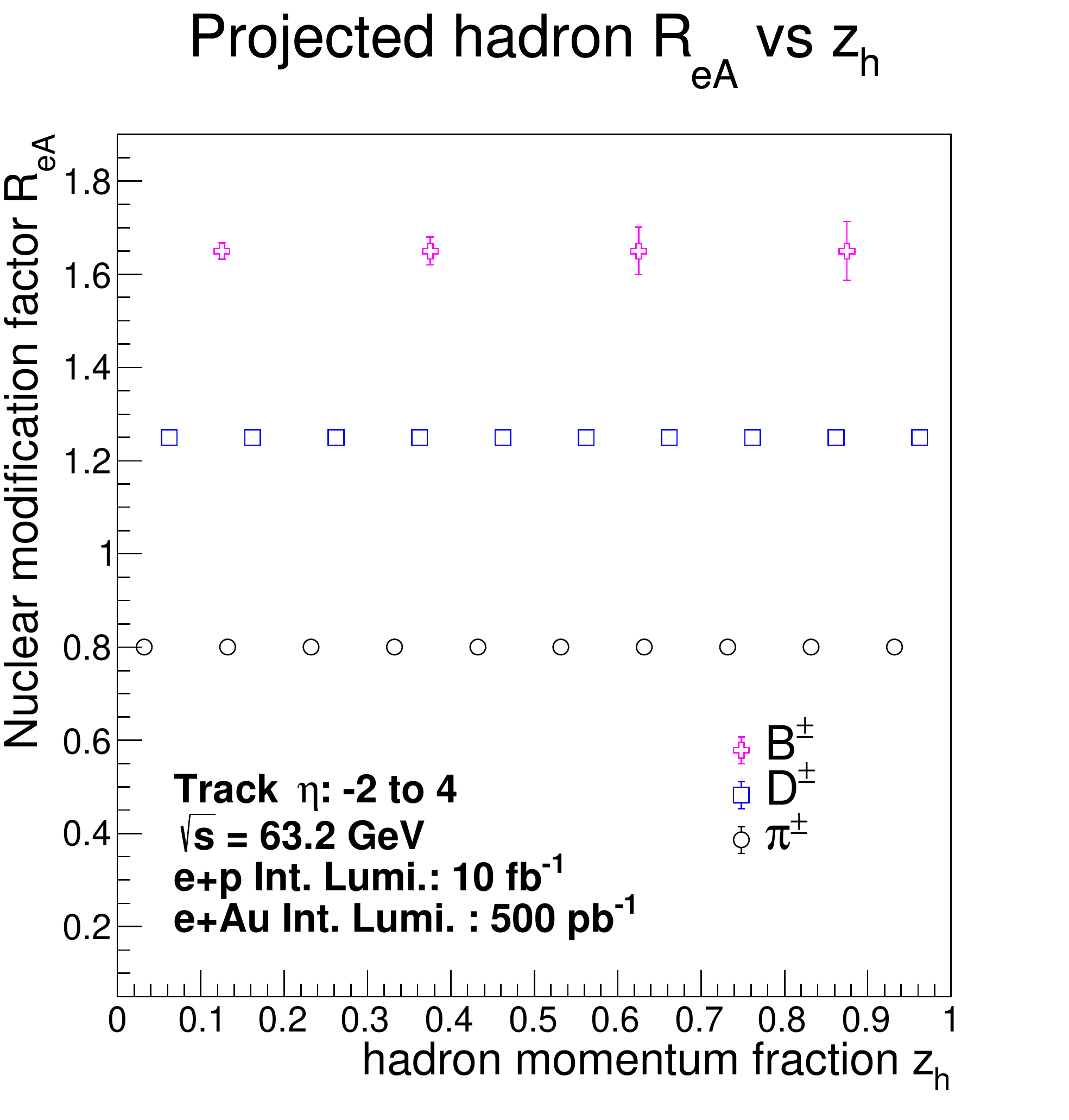}
\includegraphics[width=0.48\textwidth]{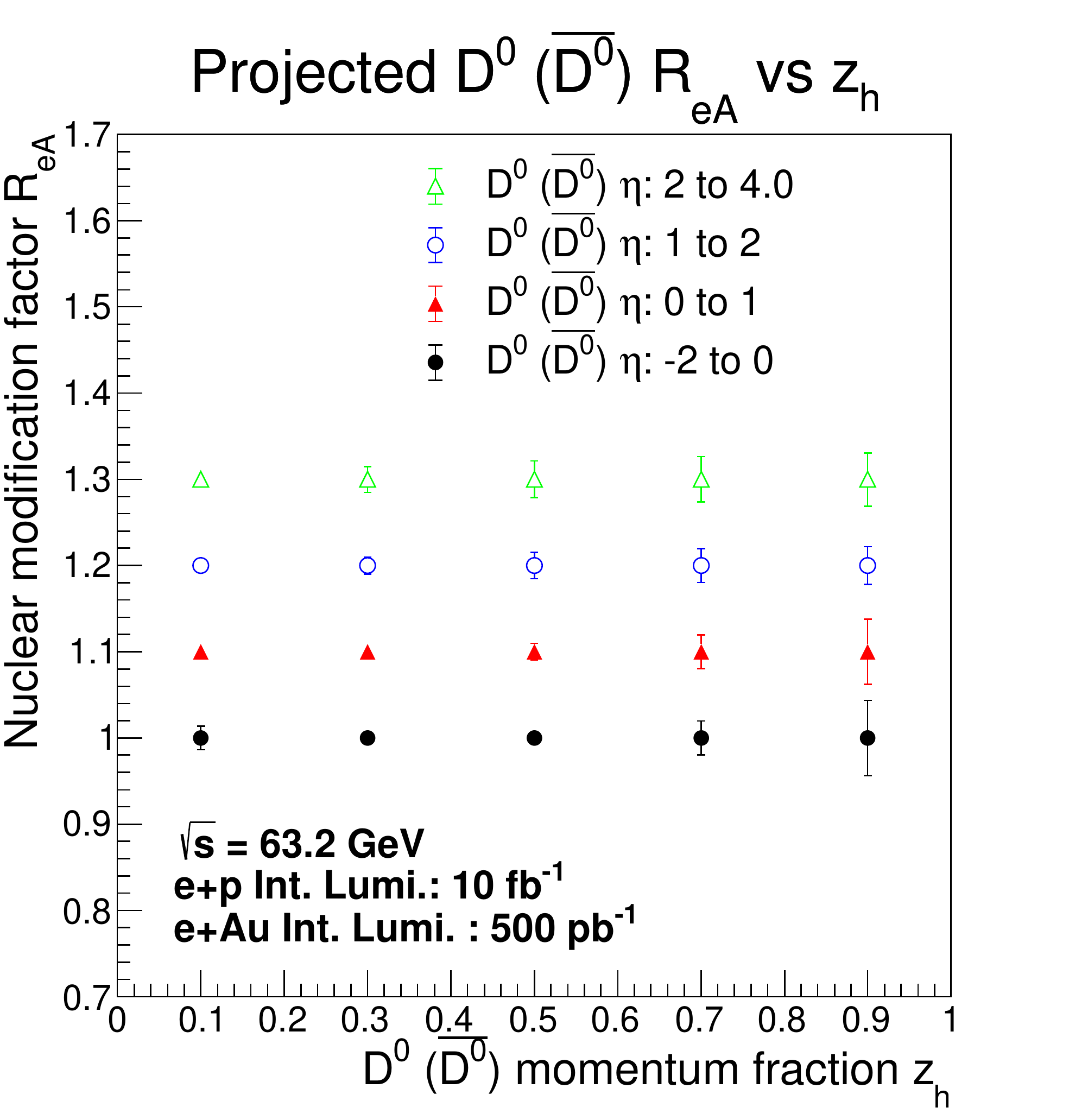}
\caption{\label{fig:ReA_1} Projections of nuclear modification factor $R_{eAu}$ for reconstructed flavor dependent hadron versus the hadron momentum fraction $z_{h}$ (left panel). $R_{eAu}$ projections of reconstructed $D^{0}$ ($\bar{D^{0}}$) in different pseudorapidity bins are shown in the right panel. Detector performance from FST version 0 design in Babar magnetic filed is used. The statistical uncertainties are projected with signal yields in $e+p$ and $e+Au$ collisions at $\sqrt{s} = 63$ GeV.}
\end{figure}

Summary of the signal over background ratio for the reconstructed D-meson within pseudorapidity of -2 to 4 is shown in Figure~\ref{fig:D_sigbkg_1}. The pseudorapidity separated reconstructed $D^{0}$ ($\bar{D^{0}}$) signal over background ratio with different FST designs are shown in  Figure~\ref{fig:D_sigbkg_2}. Pseudorapidity dependent reconstructed $\Lambda_{c}$ signal over background ratios are shown in Figure~\ref{fig:Lam_sigbkg_1}. The corresponding FST geometries and the magnet selections are listed in Table~\ref{tab:fst_ver}. The signal over background ratios for reconstructed D-mesons have dominant impacts by the tracking momentum resolutions which is associated with the magnetic filed options.

\begin{figure}[H]
\centering
\includegraphics[width=0.48\textwidth]{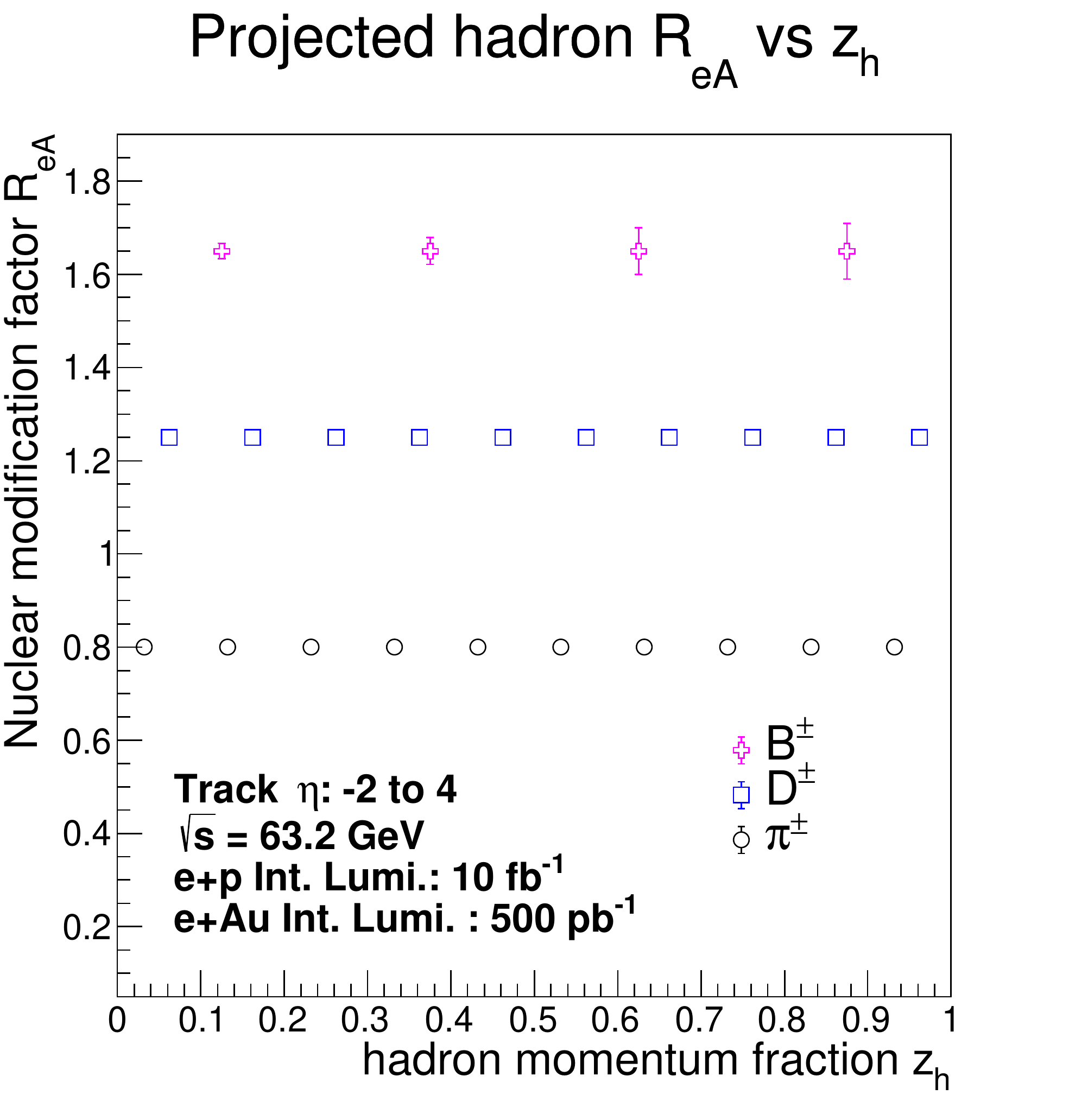}
\includegraphics[width=0.48\textwidth]{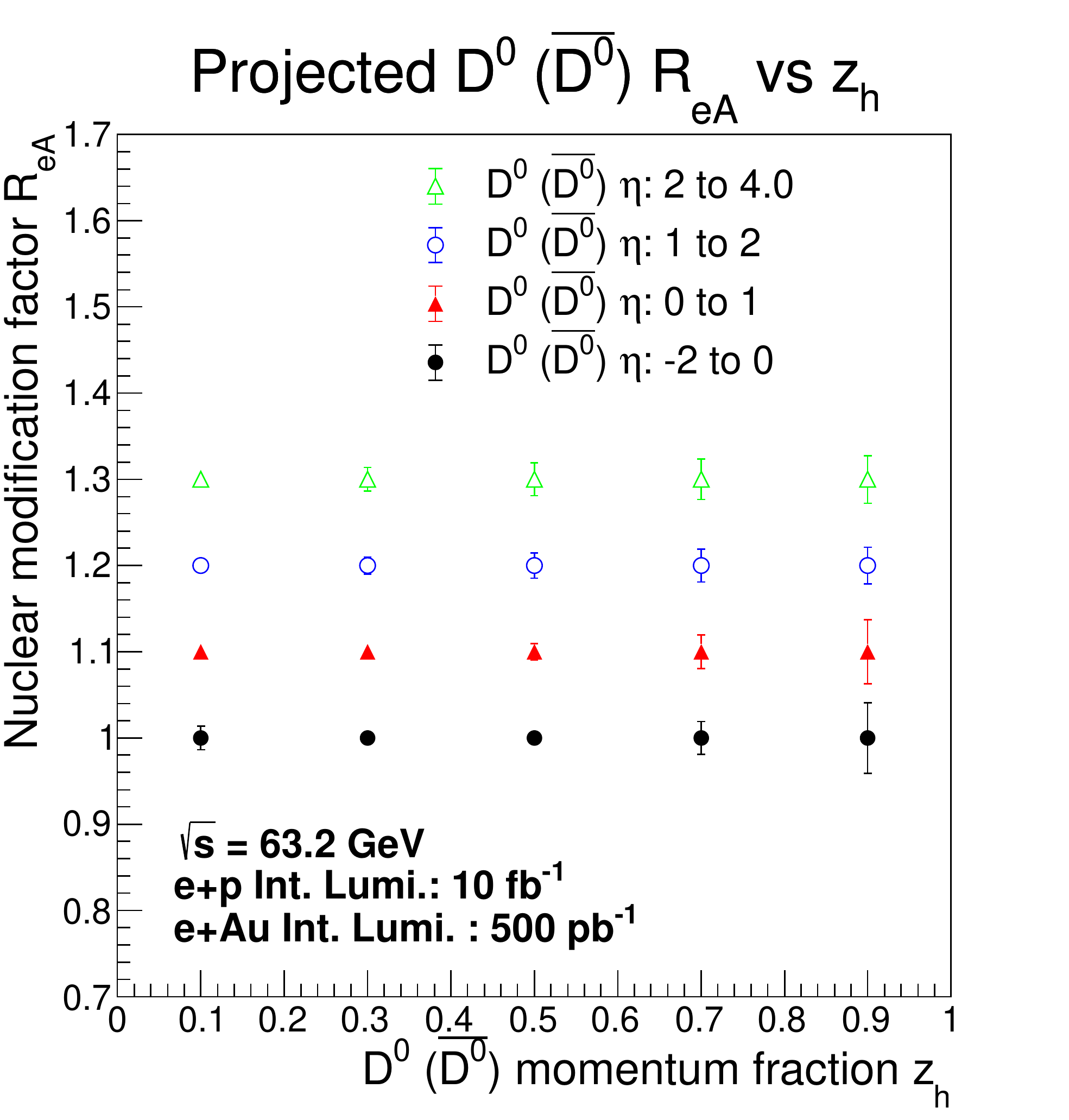}
\caption{\label{fig:ReA_2} Projections of nuclear modification factor $R_{eAu}$ for reconstructed flavor dependent hadron versus the hadron momentum fraction $z_{h}$ (left panel). $R_{eAu}$ projections of reconstructed $D^{0}$ ($\bar{D^{0}}$) in different pseudorapidity bins are shown in the right panel. Detector performance from FST version 0 design in Beast magnetic filed is used. The statistical uncertainties are projected with signal yields in $e+p$ and $e+Au$ collisions at $\sqrt{s} = 63$ GeV.}
\end{figure}

\begin{figure}[H]
\centering
\includegraphics[width=0.48\textwidth]{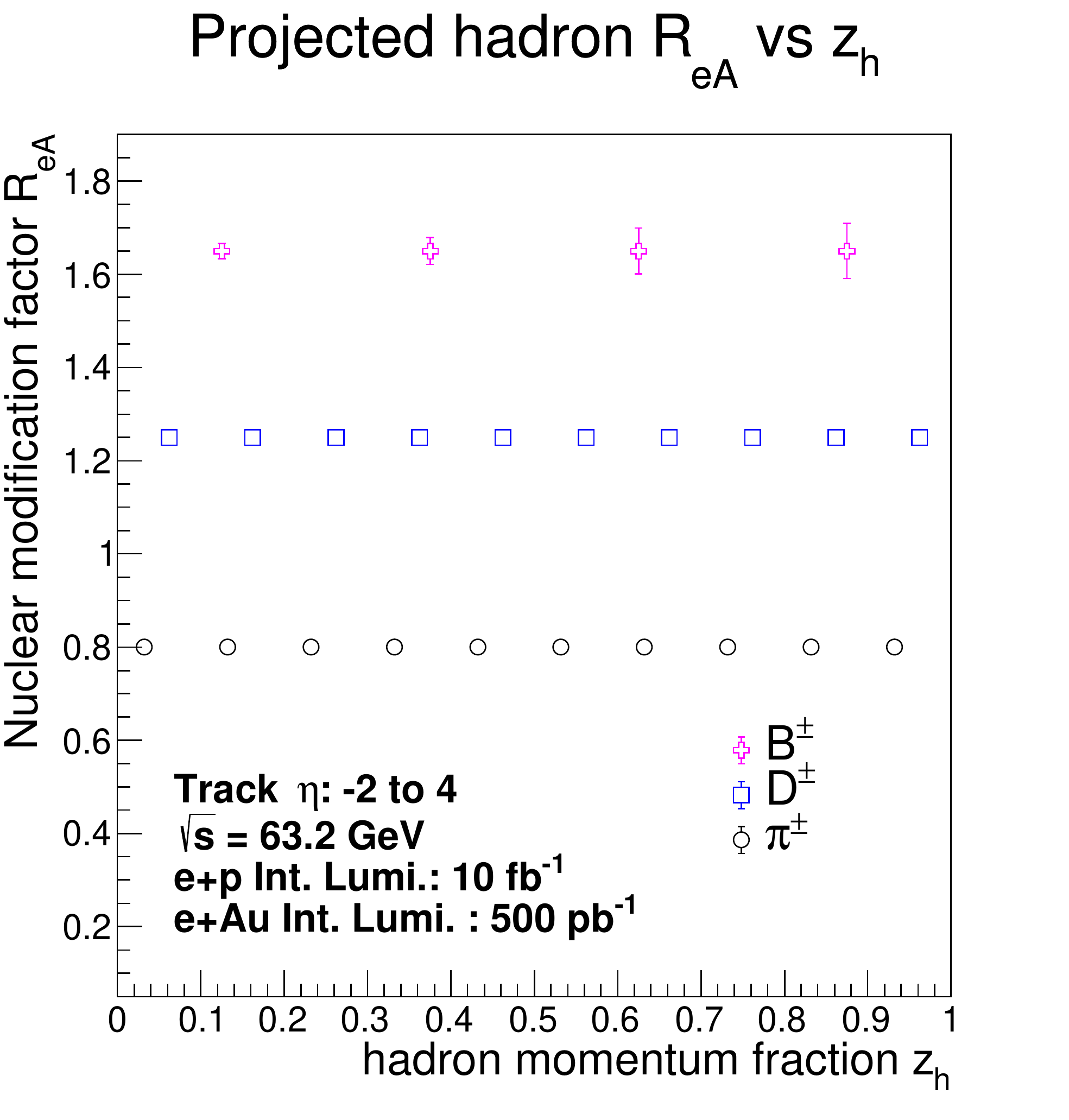}
\includegraphics[width=0.48\textwidth]{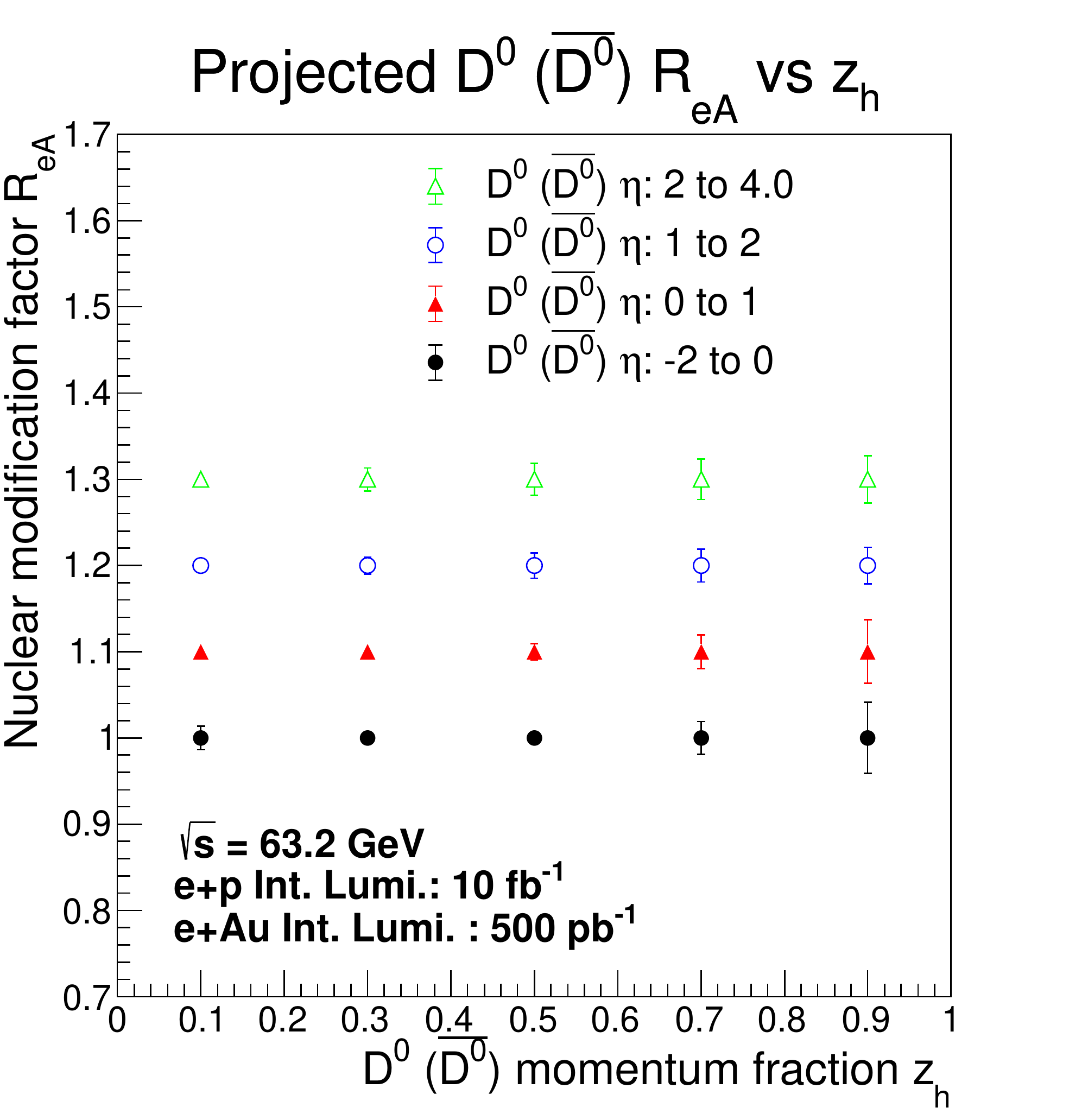}
\caption{\label{fig:ReA_3} Projections of nuclear modification factor $R_{eAu}$ for reconstructed flavor dependent hadron versus the hadron momentum fraction $z_{h}$ (left panel). $R_{eAu}$ projections of reconstructed $D^{0}$ ($\bar{D^{0}}$) in different pseudorapidity bins are shown in the right panel. Detector performance from FST version 4 design in Beast magnetic filed is used. The statistical uncertainties are projected with signal yields in $e+p$ and $e+Au$ collisions at $\sqrt{s} = 63$ GeV.}
\end{figure}

Nuclear modification factor $R_{eA}$ measurements for different flavor hadrons at the future EIC will not only explore both initial and final state effects on hadron production in nuclear medium \cite{Chudakov_2016, PhysRevD.96.114005, Vitev:2019zau} but also provide further information on hadronization process and its flavor dependence \cite{eic_hf_ivan}. Figure~\ref{fig:ReA_1} to \ref{fig:ReA_3} present the projected nuclear modification factor for reconstructed flavor dependent hadron versus the hadron momentum fraction $z_{h}$ with detector performance from different FST designs and different magnetic filed options in $e+p$ and $e+Au$ collisions at $\sqrt{s} = 63$ GeV. Projected $R_{eAu}$ for reconstructed $D^{0}$ ($\bar{D^{0}}$) in different pseudorapidity bins are also shown in these figures. 

Comparing Figure~\ref{fig:ReA_1} and Figure~\ref{fig:ReA_2}, smaller projection uncertainties have been achieved for more forward $D^{0}$ ($\bar{D^{0}}$) measurements with the same FST design by using the Beast magnetic field which produces a better signal over background ratio. Good statistics can be achieved for inclusive D-meson measurements at the future EIC, and different FST design and different magnet options have little impacts on their projected statistical uncertainties. These reconstructed heavy flavor hadrons provide a good discriminating power to separate different theoretical predictions on the nuclear transport coefficients. Forward heavy flavor measurements to be carried out at the EIC can provide better constraints on the hadron fragmentation processes in medium as discussed in \cite{eic_hf_ivan}.

\subsection{Open heavy flavor jet reconstruction and physics projection}
\label{sec: OHF_jet}
\begin{figure}[H]
\centering
\includegraphics[width=0.57\textwidth]{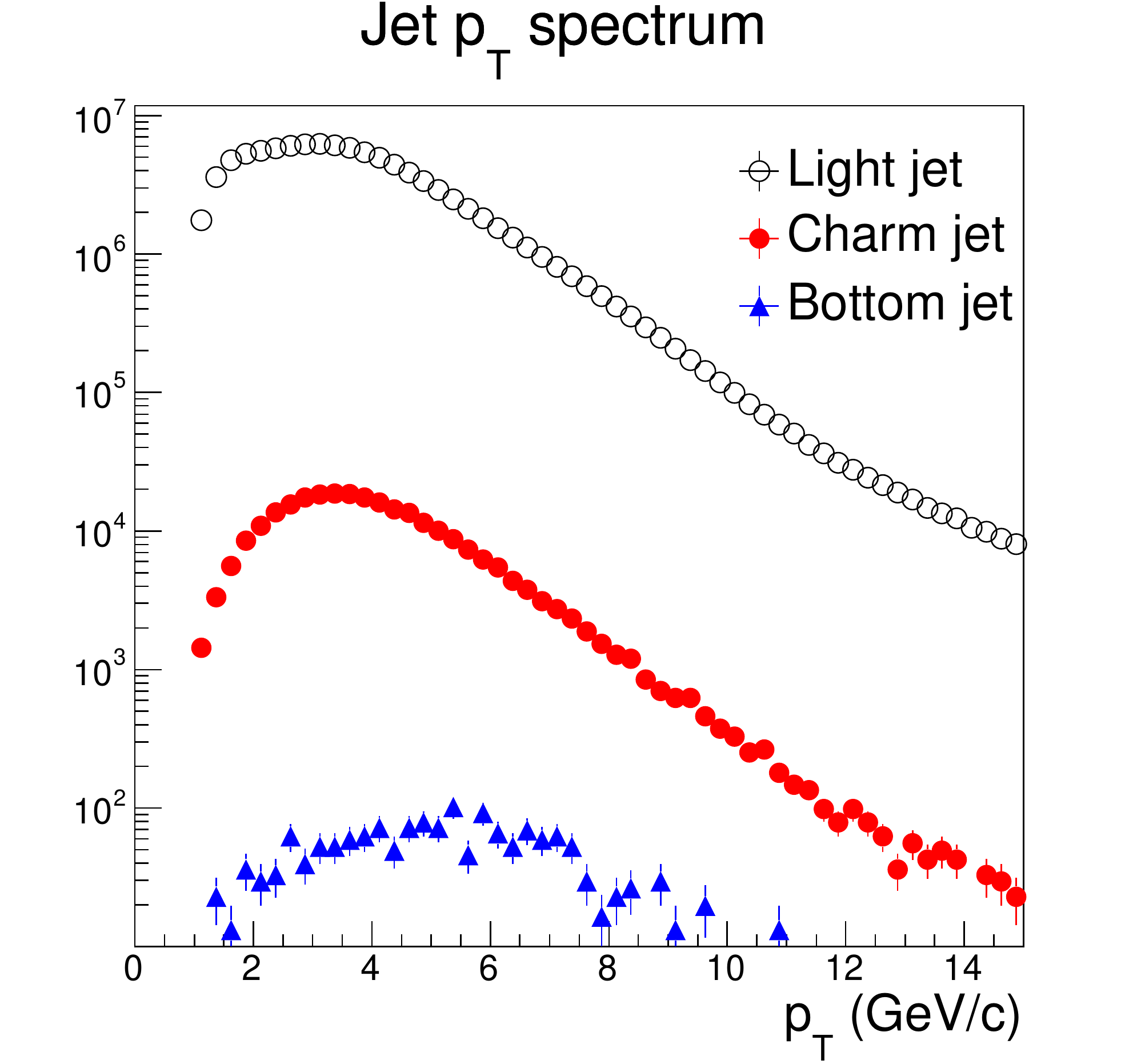}
\caption{\label{fig:Rjet_1} $p_{T}$ spectrum of reconstructed light flavor jets (black open circles), charm jets (red closed triangles) and bottom jets (blue closed triangles). The statistical uncertainties are projected with 10 $fb^{-1}$ $e+p$ at $\sqrt{s} = 63$ GeV.}
\end{figure}

The future EIC will provide a clean environment for jet studies as well. Initial jet reconstructions have been achieved based on true particle information. Inclusive jets are reconstructed with the anti-$k_{T}$ jet algorithm with cone radius at 1.0. Then jets are tagged with fully reconstructed heavy flavor meson by requiring these reconstructed heavy flavor hadrons be within the associated jet cone \cite{Li:2020hp}. If there is not a reconstructed heavy flavor hadron can be found within the jet cone, this jet is labeled as light flavor jet. Figure~\ref{fig:Rjet_1} show the spectrum of reconstructed light flavor jets and heavy flavor (charm and bottom) jets. These distributions are not corrected by the corresponding reconstruction efficiencies.

\begin{figure}[H]
\centering
\includegraphics[width=0.99\textwidth]{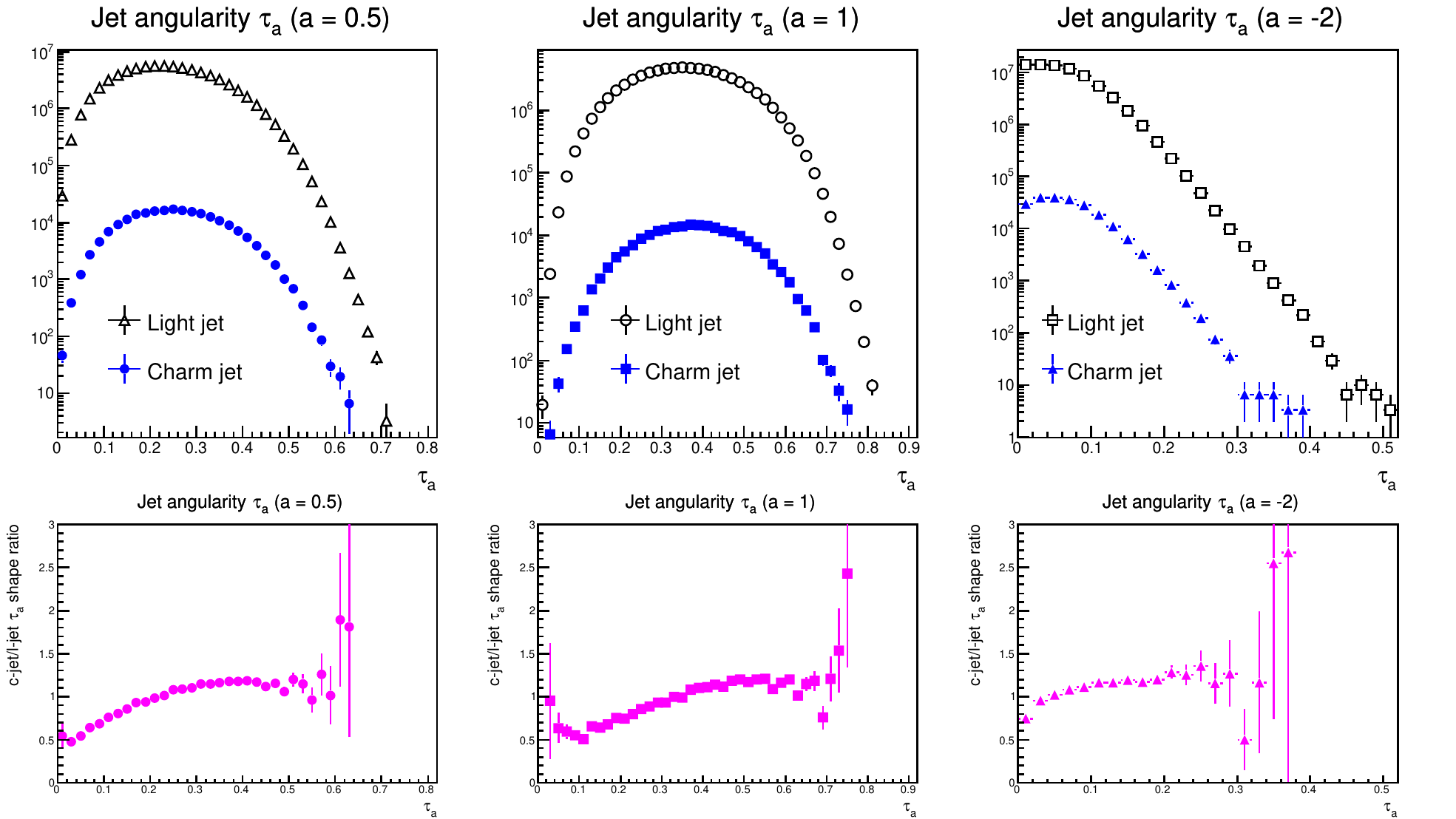}
\caption{\label{fig:Rjet_2} Jet angularity distributions for light flavor jets (black open points) and charm tagged jets (blue closed points) with different power order $a$ value selections are shown in the top panel. Distributions with $a = 0.5$ are shown in the left, with $a = 1$ are shown in the middle and $a = -2$ are shown in the right. Bottom panel shows the ratio of normalized charm jet angularity distribution over normalized light flavor jet angularity distributions in the top panel with the corresponding $a$ value selection. The statistical uncertainties are projected with 10 $fb^{-1}$ $e+p$ at $\sqrt{s} = 63$ GeV.}
\end{figure}

Jet substructure observable can image the nucleon/nuclei 3D structure and help map out the hadronization process in vacuum and nuclear medium. Recent theoretical developments \cite{jet_ana} suggest the jet angularity observable has a better discriminating power to distinguish quark or gluon originated jets. Following the same definition in \cite{jet_ana}, we have studied the jet angularity for light flavor jets and charm tagged jets with different power order $a$ value selections. Figure~\ref{fig:Rjet_2} shows the jet angularity distributions of light flavor jets and charm tagged jets and ratio distributions of their shapes in 10 $fb^{-1}$ $e+p$ at $\sqrt{s} = 63$ GeV. Charm jets have a broader jet shape which causes a increasing trends in the jet angularity ratio distributions presented in the bottom panels of Figure~\ref{fig:Rjet_2}. Nuclear modification effects for different flavor jets are under study. 

\subsection{Quarkonia and exotic state studies and physics projection}
\label{sec: exotic}

Hadronization inside the nucleus is expected to play an important role on particle production in $e$A collisions at the EIC \cite{Vitev:2019zau}.  Hadrons that propagate through the nucleus are subject to interactions with partons that lead to energy loss and, in the case of bound quarkonium states, dissociation via breakup.  This will lead to reduction in the nuclear modification factor $R_{eA}$.

Quarkonium production has been studied extensively in fixed-target $p$A collisions, where hadronization inside the nucleus also occurs. Experiments at Fermilab \cite{E866} and the SPS \cite{NA50} have measured differences between the suppression of the $\psi(2S)$ state compared to the $J/\psi$.  Since these two states have the same quark content, the interactions of the primordial $c\bar{c}$ pair prior to hadronization are identical, regardless of the final state the pair eventually projects onto.  Therefore, the suppression mechanism must occur after the pair has hadronized into a final state.  This effect can be quantitatively explained when considering the different radii of the final states:  the relatively large $\psi(2S)$ state samples a larger volume of nuclear matter while passing through the nucleus, and has a correspondingly higher probability of encountering a partons an undergoing beakup than the relatively tightly bound $J/\psi$ \cite{Arleo:1999af}.  Similar effects are expected to occur in $e$A collisions.

While the spectrum of charmonia states is well understood \cite{HigherCharmonia}, the ever-expanding list of $XYZ$ exotic states remain a mystery \cite{ExoticReview}.  Various explanations of these unexpected states have been put forth, including tetraquarks, hadroncharmonium, hadronic molecules, and other exotic hybrid states.  Given the large number of states that have been found, and their various properties, it is unlikely that a single model will be able to describe them all.  Additional data is needed to discriminate between various model calculations.

\begin{figure}[H]
\centering
\includegraphics[width=0.7\textwidth]{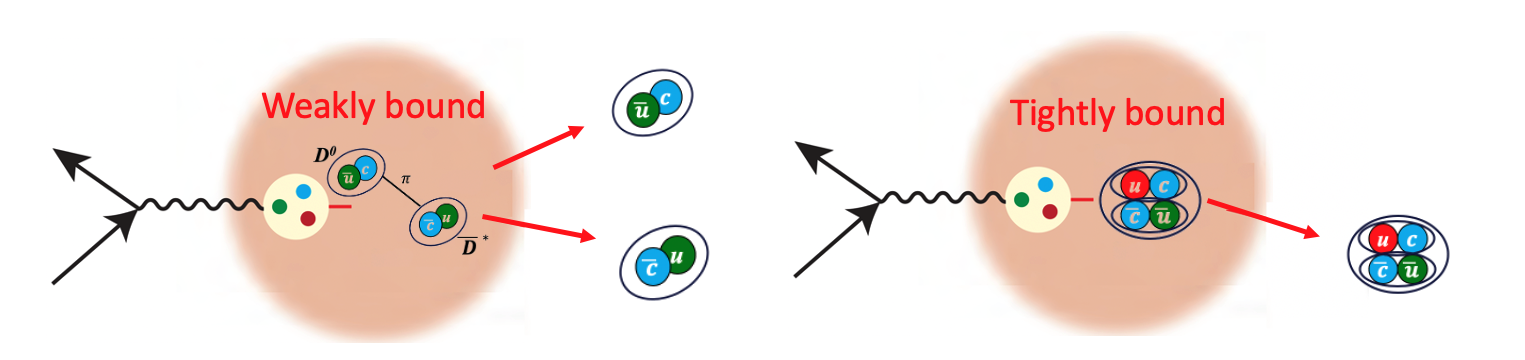}
\caption{\label{fig:exotic_1} The ratio of nuclear modification factors $R_{eA}$ for X(3872) to $psi(2S)$, for two different assumptions of the X(3872) structure.}
\end{figure}

Similar to conventional charmonia, exotics produced in $e$A collisions will also undergo interactions inside the nucleus, which can lead to disruption.  From our experience in fixed target $p$A experiments discussed above, we expect that the magnitude of these disruption effects will depend on the size of the final state.  A weakly-bound hadronic molecule with a large radius would suffer significantly more disruption that a tightly bound tetraquark (see Fig. \ref{fig:exotic_1} for a conceptual drawing).  Therefore measurements of the nuclear modification of exotic states can provide discrimination between models of their structure.

We extend the charmonium breakup model from Ref. \cite{Arleo:1999af} to also consider the exotic tetraquark candidate X(3872).  A Glauber model of the nucleus is prepared in simulation, and the starting point of the X(3872) is randomly selected inside the nucleus.  The state expands as it crosses the nucleus and considered disrupted if it approaches a nucleon within a distance of less than $\sqrt{\sigma_{c\bar{c}}\pi}$, where $\sigma_{c\bar{c}}$ is a breakup cross section that depends on the size of the state.  These simulations are run for two different models of the X(3872), one where it is considered a compact tetraquark with a final radius of 1 fm, and one where the X(3872) is modeled as a weakly bound hadronic molecule with a radius of 7 fm.  Since the X(3872) and the well-known conventional charmonium state $\psi(2S)$ are both measureable through their decays to $J/\psi\pi^{+}\pi^{-}$, the $\psi(2S)$ is also modeled, with a radius of 0.84 fm.
 
\begin{figure}[H]
\centering
\includegraphics[width=0.7\textwidth]{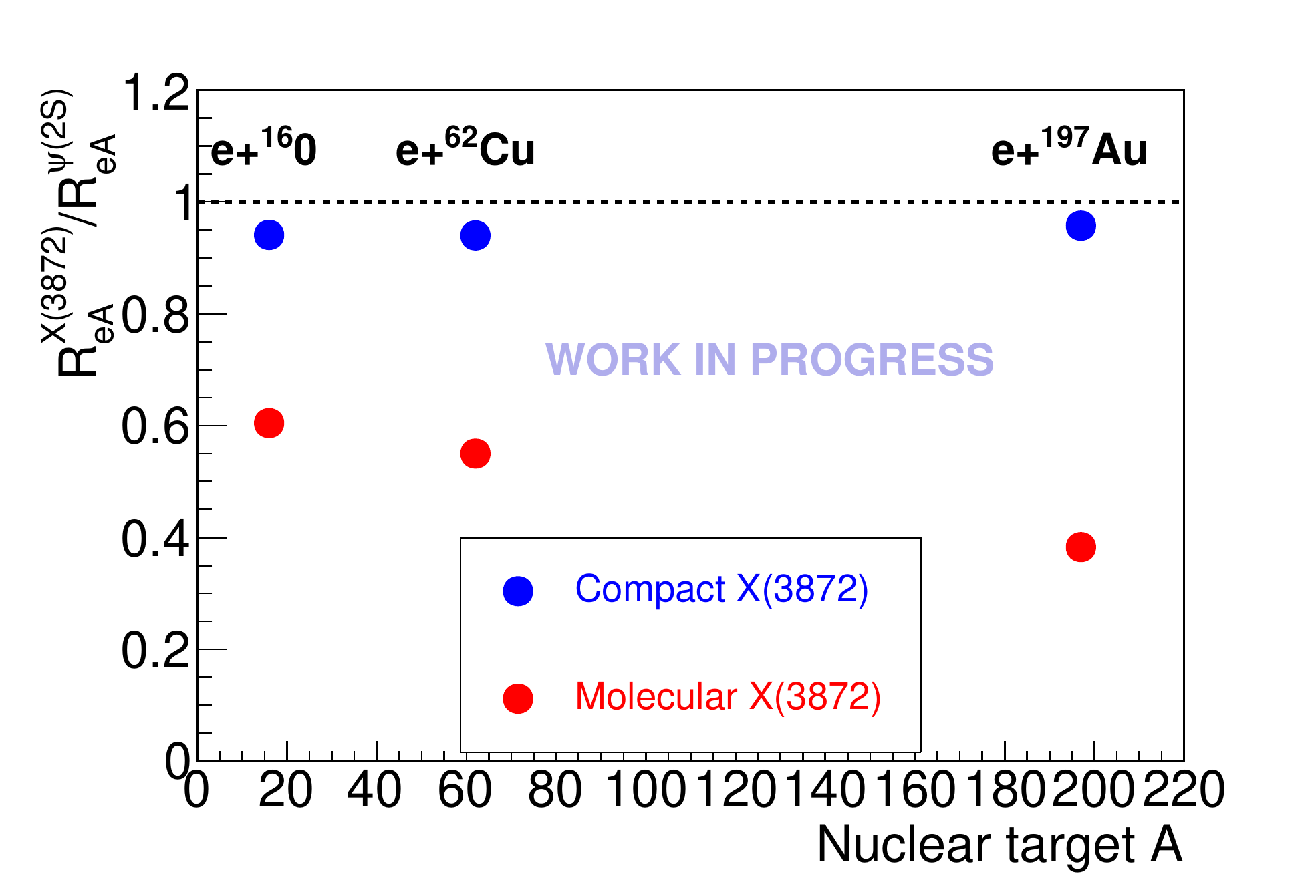}
\caption{\label{fig:HF_exotic_2} The ratio of nuclear modification factors $R_{eA}$ for X(3872) to $psi(2S)$, for two different assumptions of the X(3872) structure.}
\end{figure}

The ratio of nuclear modification factors for the X(3872) and the $\psi(2S)$ taken from this model is shown in Fig. \ref{fig:HF_exotic_2}, for three different nuclear targets.  We see that, for a compact X(3872) with a radius similar to the $psi(2S)$, there is approximately a 10$\%$ difference between the species $R_{eA}$.  However, for the molecular X(3872), the nuclear modification is different by a factor of approximately 2, showing that measurements of this ratio at the EIC can be used to discriminate between various structure models.  We note here that X(3872) is only used as an example; the technique is equally applicable to the charged $Z^{\pm}$ states (which are strong candidates for compact tetraquarks) and the $P_{c}^{\pm}$ pentaquark states (which are baryon-meson hadronic molecule candidates).

\section{Summary}
\label{sec: phy_sum}
Detailed studies have been performed for the proposed forward silicon tracker detector and its associated heavy flavor and jet measurements. Different conceptual detector designs and their tracking performance with different silicon sensor options and magnet options have been implemented. Full analysis framework has been developed for open heavy flavor and jet reconstruction. New physics observables such as the flavor dependent nuclear modification factor, flavor dependent jet angularity and exotic structure have been explored. These studies significantly enrich the ongoing EIC physics developments and will help provide guidance on the detector technology down selection and associated detector design.

\section{Acknowledgements}
\label{sec: ack}
This work is supported by the Los Alamos National Laboratory LDRD office 20200022DR project.

\newpage

%--------------------------------------------------------------------%
\bibliographystyle{elsarticle-num}
\bibliography{ref}

\begin{thebibliography}{10}
\expandafter\ifx\csname url\endcsname\relax
  \def\url#1{\texttt{#1}}\fi
\expandafter\ifx\csname urlprefix\endcsname\relax\def\urlprefix{URL }\fi
\expandafter\ifx\csname href\endcsname\relax
  \def\href#1#2{#2} \def\path#1{#1}\fi

\bibitem{eic_WP}
A.~Accardi, et~al., {Electron-Ion Collider: The next QCD frontier}, The
  European Physical Journal A 52 (2016) 268.
\newblock \href {https://doi.org/10.1140/epja/i2016-16268-9}
  {\path{doi:10.1140/epja/i2016-16268-9}}.

\bibitem{bib:pythia6}
T.~Sj\"{o}strand, S.~Mrenna, P.~Skands, {PYTHIA} 6.4 physics and manual,
  Journal of High Energy Physics 2006~(05) (2006) 026--026.
\newblock \href {https://doi.org/10.1088/1126-6708/2006/05/026}
  {\path{doi:10.1088/1126-6708/2006/05/026}}.

\bibitem{bib:pythiaeRHIC}
K.~Kauder, E.~Aschenauer, Pythia6 with radiative corrections,
  \href{https://eic.github.io/software/pythia6.html#running-pythiaerhic}{EIC
  software} {[Online]} (July 2020).

\bibitem{Li:2020sru}
X.~Li, et~al., {A New Heavy Flavor Program for the Future Electron-Ion
  Collider}, EPJ Web Conf. 235 (2020) 04002.
\newblock \href {http://arxiv.org/abs/2002.05880} {\path{arXiv:2002.05880}},
  \href {https://doi.org/10.1051/epjconf/202023504002}
  {\path{doi:10.1051/epjconf/202023504002}}.

\bibitem{gems:SAULI20162}
F.~Sauli, The gas electron multiplier (gem): Operating principles and
  applications, Nuclear Instruments and Methods in Physics Research Section A:
  Accelerators, Spectrometers, Detectors and Associated Equipment 805 (2016) 2
  -- 24, special Issue in memory of Glenn F. Knoll.
\newblock \href {https://doi.org/https://doi.org/10.1016/j.nima.2015.07.060}
  {\path{doi:https://doi.org/10.1016/j.nima.2015.07.060}}.

\bibitem{Li:2020hp}
X.~Li, {Heavy flavor and jet studies for the future Electron-Ion Collider}
  (2020).
\newblock \href {http://arxiv.org/abs/2007.14417} {\path{arXiv:2007.14417}}.

\bibitem{Chudakov_2016}
E.~Chudakov, D.~Higinbotham, C.~Hyde, S.~Furletov, Y.~Furletova, D.~Nguyen,
  M.~Stratmann, M.~Strikman, C.~Weiss,
  \href{https://doi.org/10.1088%2F1742-6596%2F770%2F1%2F012042}{Heavy quark
  production at an electron-ion collider}, Journal of Physics: Conference
  Series 770 (2016) 012042.
\newblock \href {https://doi.org/10.1088/1742-6596/770/1/012042}
  {\path{doi:10.1088/1742-6596/770/1/012042}}.
\newline\urlprefix\url{https://doi.org/10.1088%2F1742-6596%2F770%2F1%2F012042}

\bibitem{PhysRevD.96.114005}
E.~C. Aschenauer, S.~Fazio, M.~A.~C. Lamont, H.~Paukkunen, P.~Zurita,
  \href{https://link.aps.org/doi/10.1103/PhysRevD.96.114005}{Nuclear structure
  functions at a future electron-ion collider}, Phys. Rev. D 96 (2017) 114005.
\newblock \href {https://doi.org/10.1103/PhysRevD.96.114005}
  {\path{doi:10.1103/PhysRevD.96.114005}}.
\newline\urlprefix\url{https://link.aps.org/doi/10.1103/PhysRevD.96.114005}

\bibitem{Vitev:2019zau}
I.~Vitev, {Radiative processes and jet modification at the EIC}, in: {Probing
  Nucleons and Nuclei in High Energy Collisions}: {Dedicated to the Physics of
  the Electron Ion Collider}, 2020, pp. 244--247.
\newblock \href {http://arxiv.org/abs/1912.10965} {\path{arXiv:1912.10965}},
  \href {https://doi.org/10.1142/9789811214950\_0050}
  {\path{doi:10.1142/9789811214950\_0050}}.

\bibitem{eic_hf_ivan}
H.~T. Li, Z.~L. Liu, V.~I, {Heavy meson tomography of cold nuclear matter at
  the electron-ion collider} (2020).
\newblock \href {http://arxiv.org/abs/2007.10994} {\path{arXiv:2007.10994}}.

\bibitem{jet_ana}
Z.-B. Kang, K.~Lee, F.~Ringer, {Jet angularity measurements for single
  inclusive jet production}, J. High Energ. Phys. 04 (2018).

\bibitem{E866}
M.~J. Leitch, et~al., {Measurement of differences between $J/\psi$ and $\psi'$
  suppression in p-A collisions at 800-GeV/c}, Phys.\ Rev.\ Lett. 84 (2000)
  3256--3260.
\newblock \href {http://arxiv.org/abs/nucl-ex/9909007}
  {\path{arXiv:nucl-ex/9909007}}, \href
  {https://doi.org/10.1103/PhysRevLett.84.3256}
  {\path{doi:10.1103/PhysRevLett.84.3256}}.

\bibitem{NA50}
B.~Alessandro, et~al., {J/psi and psi-prime production and their normal nuclear
  absorption in proton-nucleus collisions at 400-GeV}, Eur. Phys. J. C 48
  (2006) 329.
\newblock \href {http://arxiv.org/abs/nucl-ex/0612012}
  {\path{arXiv:nucl-ex/0612012}}, \href
  {https://doi.org/10.1140/epjc/s10052-006-0079-4}
  {\path{doi:10.1140/epjc/s10052-006-0079-4}}.

\bibitem{Arleo:1999af}
F.~Arleo, P.~Gossiaux, T.~Gousset, J.~Aichelin, {Charmonium suppression in p-A
  collisions}, Phys. Rev. C 61 (2000) 054906.
\newblock \href {http://arxiv.org/abs/hep-ph/9907286}
  {\path{arXiv:hep-ph/9907286}}, \href
  {https://doi.org/10.1103/PhysRevC.61.054906}
  {\path{doi:10.1103/PhysRevC.61.054906}}.

\bibitem{HigherCharmonia}
T.~Barnes, S.~Godfrey, E.~Swanson, {Higher charmonia}, Phys. Rev. D 72 (2005)
  054026.
\newblock \href {http://arxiv.org/abs/hep-ph/0505002}
  {\path{arXiv:hep-ph/0505002}}, \href
  {https://doi.org/10.1103/PhysRevD.72.054026}
  {\path{doi:10.1103/PhysRevD.72.054026}}.

\bibitem{ExoticReview}
S.~L. Olsen, T.~Skwarnicki, D.~Zieminska,
  \href{https://link.aps.org/doi/10.1103/RevModPhys.90.015003}{{Nonstandard
  heavy mesons and baryons: experimental evidence}}, Rev. Mod. Phys. 90 (2018)
  015003.
\newblock \href {https://doi.org/10.1103/RevModPhys.90.015003}
  {\path{doi:10.1103/RevModPhys.90.015003}}.
\newline\urlprefix\url{https://link.aps.org/doi/10.1103/RevModPhys.90.015003}

\end{thebibliography}

\end{document}